\documentclass[12pt]{article} 
\usepackage[a4paper,textwidth=490.0pt,textheight=703.1pt]{geometry}
\usepackage{url}
\usepackage{booktabs}

\usepackage{epsfig} %
\usepackage{float} %
\usepackage{amssymb} %
\usepackage{amsmath} %
\usepackage{verbatim} %
\usepackage{cite} %
\usepackage{color} %
\usepackage{eufrak}
\usepackage[english]{babel}
\usepackage[latin1]{inputenc}
\usepackage[T1]{fontenc}
\usepackage{ae}
         
\usepackage[colorinlistoftodos]{todonotes}

\newcommand{\ep}{\varepsilon}

\newcommand{\beq}{\begin{equation}}
\newcommand{\eeq}{\end{equation}}
\newcommand{\bea}{\begin{eqnarray}}
\newcommand{\eea}{\end{eqnarray}}

\newcommand*\pFqskip{8mu}
\catcode`,\active
\newcommand*\pFq{\begingroup
        \catcode`\,\active
        \def ,{\mskip\pFqskip\relax}%
        \dopFq
}
\catcode`\,12
\def\dopFq#1#2#3#4#5{%
        {}_{#1}F_{#2}\biggl[\genfrac..{0pt}{}{#3}{#4};#5\biggr]%
        \endgroup
}

\newcommand{\KK}{{\mathbb K}}

\allowdisplaybreaks[4]

\begin{document} 
\setlength{\baselineskip}{0.515cm}

\sloppy 
\thispagestyle{empty} 
\begin{flushleft} 
DESY 21--071
\\ 
DO--TH 21/16
\\ 
RISC Report Series 21--17
\\ 
SAGEX-21-10-E
\\ 
November 2021 
\end{flushleft}

\mbox{} \vspace*{\fill} \begin{center}

{\LARGE\bf Hypergeometric Structures in}

\vspace*{3mm}
{\LARGE\bf Feynman Integrals}

\vspace{3cm} 
\large 
{\large J.~Bl\"umlein$^a$, M.~Saragnese$^a$ and C.~Schneider$^b$}

\normalsize 

\vspace{1.cm} 
{\it $^a$~Deutsches Elektronen--Synchrotron DESY,}\\ {\it Platanenallee 6, 
15738 Zeuthen, Germany}

\vspace*{2mm} 
{\it $^b$~Johannes Kepler University Linz, Research Institute for Symbolic Computation (RISC), Altenberger Stra\ss{}e 69, A-4040 Linz, 
Austria}


\end{center} 
\normalsize 
\vspace{\fill} 
\begin{abstract} 
\noindent
Hypergeometric structures in single and multiscale Feynman integrals emerge in a wide class of topologies. 
Using integration-by-parts relations, associated master or scalar integrals have to be calculated. For this 
purpose it appears useful to devise an automated method which recognizes the respective (partial) differential 
equations related to the corresponding higher transcendental functions. We solve these equations through 
associated recursions of the expansion coefficient of the multivalued formal Taylor series. The expansion 
coefficients can be determined using either the package {\tt Sigma} in the case of linear difference equations 
or by applying heuristic methods in the case of partial linear difference equations. In the present context a 
new type of sums occurs, the Hurwitz harmonic sums, and generalized versions of them. The code {\tt HypSeries} 
transforming classes of differential equations into analytic series expansions is described. Also partial 
difference equations having rational solutions and rational function solutions of Pochhammer symbols are 
considered, for which the code {\tt solvePartialLDE} is designed. Generalized hypergeometric 
functions, Appell-,~Kamp\'e de F\'eriet-, Horn-, Lauricella-Saran-, Srivasta-, and Exton--type functions 
are 
considered. We illustrate the algorithms by examples.
\end{abstract}

\vspace*{\fill} \noindent
\newpage

\section{Introduction} 
\label{sec:Intro}

\vspace*{1mm} 
\noindent 
It is a general observation, that the Feynman parameter integrals for certain classes of topologies can be expressed
in terms of higher transcendental functions of the hypergeometric type, cf.~e.g.~\cite{HAMBERG,Davydychev:2003mv,
Bierenbaum:2007qe,Kalmykov:2020cqz}. 
This concerns their representation before expanding in the dimensional parameter $\ep = D - 4$, with $D$ the dimension of 
space--time.   
Here we consider the generalization of the Euler integrals to the generalized
hypergeometric functions $_pF_q$, and the multiple hypergeometric functions of the Appell-, Kamp\'e de F\'eriet-,  Horn-,
and Lauricella-, Saran-,  
Srivastava-, and Exton type 
\cite{HYPKLEIN,HYPBAILEY,SLATER1,APPELL1,APPELL2,KAMPE1,KAMPE2,HORN,EXTON72a,
EXTON1,EXTON2,SCHLOSSER,
Anastasiou:1999ui,Anastasiou:1999cx,SRIKARL,Lauricella:1893, Saran:1954,Saran:1955,ERDELYI,Kalmykov:2020cqz}. 
In physical applications a standard integration 
method is that of solving systems of ordinary and partial differential equation systems \cite{DEQ} generated by the 
integration by parts (IBP) relations \cite{IBP}. In this context it is important to recognize the differential equations of
the classes of the aforementioned functions, since their mathematical structure is widely known. This allows the direct analytic
solution of at least this part of the physical problem. Starting with certain topologies, more general differential equations
will contribute, requiring different solution technologies. Structures of the above kind have been obtained 
in general
off--shell representations at the one--loop level for multi--leg diagrams, 
cf.~e.g.~\cite{Boos:1990rg,Fleischer:2003rm,
Watanabe:2013ova,Bluemlein:2017rbi,Phan:2018cnz}. At the two- and 
three--loop level for various scattering processes related structures are found, 
cf.~e.g.~\cite{Anastasiou:1999ui,Anastasiou:1999cx,      
Bauberger:1994nk,Ablinger:2012qm}.

For all the above quantities the (partial) differential equations are known and they partly turn out to be of rather
high order. On the other hand, one may consider the formal multiple Taylor expansion of these higher transcendental functions,
which allows one to obtain difference equations for the corresponding expansion coefficients. This is advised, since these are 
remarkably simpler.

In this paper we describe a systematic classification of partial differential equations for scalar or master integrals 
of one or more scales w.r.t.\ known solutions in the hypergeometric classes. These differential equations have multivariate
multiple series solutions for parameters $x_1, ..., x_n$ in the vicinity of $\{0, ..., 0\}$ as formal Taylor series.
We determine the expansion coefficients, which are obtained as rational product solutions. In identifying the associated 
non--linear coefficient pattern for the respective case the expansion coefficients can be factorized into rational
expressions of Pochhammer symbols.

The method works for general values of the space--time dimension $D$. The (multiple) infinite series representations 
found can be finally expanded in the dimensional parameter $\ep$, which transforms the summand in terms of Pochhammer symbols and general hypergeometric products by introducing in addition (cyclotomic) harmonic sums \cite{Vermaseren:1998uu,Blumlein:1998if,Ablinger:2011te} or generalized versions, like Hurwitz
harmonic sums. One may try to simplify the obtained multiple sums in terms of hypergeometric products and indefinite nested sums to expressions that are purely given in terms of indefinite nested sums using the package \texttt{EvaluateMultiSums}~\cite{Ablinger:2010pb,Blumlein:2012hg,Schneider:2013zna,Schneider:19}. The underlying summation engine is based on the package {\tt Sigma} \cite{SIG1,SIG2} that contains 
non--trivial algorithms in the setting of difference rings~\cite{DR,TermAlgebra,LinearSolver}.

Hypergeometric structures emerge in the calculation of Feynman integrals using 
Feynman parameter representations from the simplest topologies onward, cf.~e.g.\ \cite{IZ}.
In the most simple cases they can be calculated in terms of Euler Beta functions
\begin{eqnarray}
\int_0^1 dz z^a (1-z)^b = B(a+1,b+1).
\end{eqnarray}
Here and in the following we represent the respective higher transcendental functions in their convergence region
but allow to perform analytic continuations to their whole analyticity range, cf.~\cite{WW}. The next more involved
function is ${_2F_1}$
\begin{eqnarray}
\pFq{2}{1}{a_1,a_2}{b_1}{z} = \frac{\Gamma(b_1)}{\Gamma(a_1) \Gamma(b_1-a_1)} \int_0^1 dx x^{a_1-1} (1-x)^{b_1-a_1-1}
(1-zx)^{-a_2}
\end{eqnarray}
followed by $_{p+1}F_p$ by the iterative integral
\begin{eqnarray}
\int_0^1 dx x^{a-1} (1-x)^{b-1} 
\pFq{p}{q}{a_1,...,a_p}{b_1,...b_q}{x z} = \frac{\Gamma(a) \Gamma(b)}{\Gamma(a+b)} \pFq{p+1}{q+1}
{a_1,...,a_p,a}{b_1,...b_q,a+b}{z}. 
\end{eqnarray}
Other topologies \cite{Ablinger:2012qm} lead to the integral representation of the Appell function $F_1$ 
\begin{eqnarray}
I &=& \int_0^1 dw_1 \int_0^1 dw_2 \theta(1-w_1-w_2) w_1^{b-1} w_2^{b'-1} (1-w_1-w_2)^{c-b-b'-1}(1-w_1 x - w_2 y)^{-a}
\nonumber\\ 
  &=& \frac{\Gamma(b) \gamma(b') \Gamma(c-b-b')}{\Gamma(c)} F_1\left[a;b,b';c;x,y\right]  
\end{eqnarray}
and others.
The solution of Feynman integrals through Feynman parameterizations mapping to higher transcendental functions
is not a method which can be easily made uniform. At a certain stage it will also require the use of 
Mellin--Barnes integral representations \cite{MB} to be solved by the residue theorem. Although this method 
can establish links to higher transcendental functions in principle since those have Pochhammer Umlauf-integral
representations \cite{POCHHAMMER,KF,SLATER1}, it may easily lead to non--minimal representations \cite{Blumlein:2010zv}
which are difficult to reduce analytically, if one is not only interested in numerical results \cite{MBnum}.

The advantage of all these representations lies in the fact that multiple Feynman parameter integrals are reduced
to much lower dimensional infinite sum representations, which are one--fold in the case of generalized 
hypergeometric
functions, two--fold e.g.\ for Appell and Horn functions 
\cite{APPELL1,APPELL2,Anastasiou:1999ui,Anastasiou:1999cx,HORN}, 
three--fold for the Srivastava functions
\cite{SRIKARL} and further given by multi--sum Lauricella-type functions \cite{Lauricella:1893} in more 
involved cases, with an
early application in \cite{Bauberger:1994nk}.

In this paper we will consider (partial) differential equations for master integrals w.r.t.\ their parameters. The master 
integrals 
are obtained as the result of the IBP--reduction. Usually these are first order equations. However, one will decouple the
corresponding systems, cf.~\cite{BCP13,Zuercher:94}, using algorithms which are available, e.g., in {\tt OreSys} \cite{ORESYS}. In this way  higher order 
differential equations will emerge. In the case of partial differential equations one may use, e.g., Janet bases \cite{JANET}.

These (partial) differential equations can be mapped to corresponding multi--variate difference equations expanding 
the associated ansatz in terms of the multi--variate formal Taylor series
\begin{eqnarray}
\sum_{k_1, ..., k_n = 0}^\infty f[k_1, ..., k_n] x_1^{k_1} ... x_n^{k_n}.
\label{eq:TAYLOR}
\end{eqnarray}
The recurrences obeyed by the expansion coefficients $f[k_1, ..., k_n]$
can be solved using difference ring techniques \cite{SIG1,SIG2} for the linear case. For the multivariate case we will utilize ideas from~\cite{kauers10,kauers11} that led to the new package \texttt{solvePartialLDE} that may support in parts the solving task. Finally, one may express $f[k_1, ..., k_n]$
as a rational term of (multi--indexed) Pochhammer symbols, which contain all parameters of the original differential
equations, like particle masses and kinematic invariants of the processes considered, including the dimensional parameter 
$\ep$. In the general case first product--solutions emerge, which can be factored into Pochhammer--structures by solving
algebraic equations. Alternatively, one can keep the multiplicands in non--linear form and can apply a new 
function implemented in the package \texttt{EvaluateMultiSum} that produces the $\ep$--expansions without introducing algebraic extensions.
In most cases, we will limit our consideration to the principal structure of the known classes of higher transcendental functions of the 
hypergeometric type, i.e. those where $f[k_1, ..., k_n]$ is given by Pochhammer--ratios
\begin{eqnarray}
f[k_1, ..., k_n] = \frac{\prod_{i=1}^p (a(i))_{l(i)}}{\prod_{i=j}^q (b(j))_{m(j)}},~~~l(i), m(j),p,q \in \mathbb{N},
\end{eqnarray}
and $l(i), m(j)$ are linear functions of $k_r \in \mathbb{N}$ with integer coefficients. 
Here the Pochhammer symbols are defined by
\begin{eqnarray}
(a)_n = \frac{\Gamma(a+n)}{\Gamma(a)},~~~a \in \mathbb{C} \backslash \mathbb{Z}_-,~~n \in \mathbb{N}
\end{eqnarray}
and $\mathbb{Z}_- = \mathbb{Z} \backslash \mathbb{N} \cup \{0\}$. 

The coefficients $a(i)$ and $b(j)$
will in general depend on the dimensional parameter $\ep$ and we will further
consider the expansion of the higher transcendental functions in this parameter.

The paper is organized as follows. In Section~\ref{sec:delist} we list the differential equations of the 
multi--variate
generalized hypergeometric functions up to four variables in explicit form, parameterizing them linearly. There are 
also general differential equations as those for the hypergeometric functions $_pF_q$, the  Kamp\'e de F\'eriet function 
\cite{KAMPE2}, and the Lauricella--Saran functions \cite{Lauricella:1893,Saran:1954,Saran:1955}. In Section~\ref{sec:reclist} we 
derive the recursions for the multivariate expansion coefficients of these functions. An algorithm is presented in
Section~\ref{sec:recsol} to find hypergeometric product solutions for first--order linear recurrence 
systems. 
In this way the multivariate functions $f(x_1, ..., x_n)$ can be represented in the vicinity of $(\vec{0})_n$.
The parameters of the differential and difference equations depend also on the dimensional parameter $\ep$. 
Usually one would like to perform corresponding expansions in this quantity, which we describe in 
Section~\ref{sec:epExpansion}. Here the so-called Hurwitz harmonic sums and more general versions occur, the 
summation problem of which can be dealt with the packages \texttt{EvaluateMultiSums} and {\tt Sigma}. 
In Section~\ref{sec:fullmachinery} we demonstrate the full machinery, to obtain for a given system of linear 
differential equations the first coefficients of the $\ep$--expansions in terms of indefinite nested sums 
and 
products. In Section~\ref{sec:PLDEsolver} we supplement the solving tools from Section~\ref{sec:recsol} and 
turn to general partial difference equations with rational coefficients. Based on the algorithms presented 
in~\cite{kauers10,kauers11} we present different strategies to find solutions in terms of
hypergeometric products and iterative sums over such products that appear in the calculation of Feynman integrals.
Section~\ref{sec:conclusion} contains the conclusions.

In Appendix~\ref{sec:A}  we provide for convenience a list of  the main functions dealt with in the present 
paper, which are defined by their series representation, see also the file {\tt cases.m}. 
Appendix~\ref{sec:B} illustrates the matching conditions to be met to obtain from the general solutions in a direct way
the Pochhammer-type solutions. They are given in computer-readable form in the file {\tt Mconditions.m}. 
In Appendix~\ref{sec:C} a brief description of the commands of the code {\tt HypSeries} is given and 
Appendix~\ref{sec:D} provides a brief description of the code {\tt solvePartialLDE}. For both cases we will provide 
{\tt Mathematica} notebooks illustrating the corresponding operations in 
examples. In Appendix~\ref{sec:E} a special constant is evaluated, which 
appears in one of the examples. Appendix~\ref{sec:F} lists the 
{\tt Mathematica} and other software 
packages required to execute the example notebooks.
\section{The differential equations} 
\label{sec:delist}

\vspace*{1mm} 
\noindent 
Multivariate master integrals obey partial differential equations, which are obtained after the IBP reduction and,
if necessary, the decoupling of coupled systems. In this way differential equations of higher than first order are 
obtained. The first class concerns the univariate case of the generalized hypergeometric functions; for its 
definition we refer to Appendix~\ref{sec:A}.

The differential operator of Gau\ss{}' $_2F_1$ function reads, cf.~\cite{SLATER1},
\begin{eqnarray}
x(1-x) \frac{d^2}{dx^2} + (c -(a+b+1)x) \frac{d}{dx} - ab,
\end{eqnarray}
which we write more generally as 
\begin{eqnarray}
x(1-x) \frac{d^2}{dx^2} + (A_1+B_1 x) \frac{d}{dx} + C.
\label{eq:D2F1g}
\end{eqnarray}
For the function $_{3}F_2$ one obtains
\begin{eqnarray}
x^2(1-x) \frac{d^3}{dx^3} + x (A_{{2}}+B_{{2}} x) \frac{d^2}{dx^2} 
+ (A_{{1}}+B_{{1}} x) \frac{d}{dx} + C,
\label{eq:D3F2}
\end{eqnarray}
with $A_{{2}} = b_1+b_2 +1, B_{{2}} = -(3 + a_1 + a_2 +a_3), A_{{1}} = b_1b_2, 
B_{{1}} = -(a_2 a_1 + a_3 a_1 + a_2 a_3 + a_1 +a_2 +a_3 +1), C = - a_1 a_2 a_3$.
In general, we get for ${}_{p+1}F_p$ the linear differential equation
\begin{eqnarray}
x^p (1-x) \frac{d^{p+1}}{dx^{p+1}} + \sum_{k = 1}^{p} x^{k-1}(A_k + B_k x) \frac{d^k}{dx^k} + C. 
\end{eqnarray}
The $_pF_q$ function is the homogeneous solution of the differential 
operator
\begin{eqnarray}
x \frac{d}{dx} \Biggl( x \frac{d}{dx} + b_1 -1 \Biggr) ... \Biggl( x \frac{d}{dx} + b_q -1 \Biggr)
- x \Biggl( x \frac{d}{dx} + a_1  \Biggr) ... \Biggl( x \frac{d}{dx} + a_p \Biggr).
\label{eq:PFQ1}
\end{eqnarray}
The products of the differential operators in (\ref{eq:PFQ1}) $\vartheta = x (d/dx) \equiv x \partial_x$, 
can be written in the following form
\begin{eqnarray}
\vartheta &=& x \partial_x
\\
\vartheta^2 &=& x \partial_x + x^2 \partial_x^2
\\
\vartheta^3 &=& x \partial_x + 3 x^{{2}} \partial_x^2 + x^3 \partial_x^3
\\
\vartheta^4 &=& x \partial_x + 7 x^{{2}} \partial_x^2 + 6 x^3 \partial_x^3+ x^4 \partial_x^4
\\
\vartheta^5 &=& x \partial_x + 15 x^{{2}} \partial_x^2 + 25 x^3 \partial_x^3+ 10 x^4 \partial_x^4 + x^5 
\partial_x^5,~~\text{etc.}
\end{eqnarray}
Inserting this into Eq.~(\ref{eq:PFQ1}) will imply the corresponding general differential operator, which has the form
\begin{eqnarray}
\sum_{k = 0}^m P_k(x) \frac{d^k}{dx^k},
\end{eqnarray}
with the corresponding polynomials $P_k(x)$ and $m = {\rm max}\{p,q\}$. The coefficient polynomials result 
from the expansion of (\ref{eq:PFQ1}).
 
Here and in the following we will first parameterize the differential operators in general terms. In the literature
the different coefficients are usually related by algebraic equations, which is possible, but not necessary. The list 
of these equations are given in a subsidiary file to this paper. 
The differential operators for the two--variable Horn type functions \cite{APPELL1,APPELL2,KAMPE1,KAMPE2,HORN} 
$F_1$ to $F_4$, $G_1$ to $G_3$, and $H_1$ to $H_7$, including the Appell functions 
\cite{APPELL1,APPELL2}, are given by
{
\begin{eqnarray}
a
+ (b x+c) \partial_x
+x (d+e x) \partial_x^2
+f y \partial_y
+(g y+h x y) \partial_{x,y}^2
+j y^2 \partial_y^2 &=& 0
\label{APP1}
\\
a_1
+ (b_1 y+ c_1) \partial_y
+y (d_1+e_1
   y) \partial_y^2
+f_1 x \partial_x
+ (g_1  x+ h_1 x y) \partial_{x,y}^2
+j_1 x^2 \partial_x^2 &=& 0,
\label{APP2}
\end{eqnarray}
}
with the example of the Appell $F_1$ function
{
\begin{eqnarray}
F_1 &:& x(1-x)\partial_x^2 +y(1-x) \partial_{x,y}^2 + (A+B x)\partial_x + C y\partial_y + D 
\\
F_1 &:& y(1-y)\partial_y^2 +x(1-y) \partial_{x,y}^2 + (A +B' y)\partial_y + C' x\partial_x + D'. 
\end{eqnarray}
}
Here {$\partial_{x,y}^2 = \partial_x \partial_y$}, etc.

In physical applications two more differential operators appeared in the bi--variate case, 
\cite{Anastasiou:1999ui,Anastasiou:1999cx}, to which the functions
$S_1$ and $S_2$ belong. The differential operators read
{
\begin{eqnarray}
S_1 &:& 
a+(c+b x) \partial_x +x (d+e x) \partial_x^2 +x^2 (l+p x) \partial_x^3+f y \partial_y +x 
(q+r x) y \partial_x^2 \partial_y +j y^2 \partial_y^2 
\nonumber\\ &&
+s x y^2 \partial_x \partial_y^2 +(g y+h x y) 
\partial_{x,y}^2
\\
 &:& 
a_1+f_1 x \partial_x + j_1 x^2 \partial_x^2 +(c_1+b_1 y) \partial_y+y (d_1+e_1 y) \partial_y^2 +(g_1 x+h_1 x 
y) \partial_{x,y}^2
\\
S_2 &:&
a+(c+b x) \partial_x +x (d+e x) \partial_x^2 +x^2 (l+p x) \partial_x^3 +f y \partial_y +x 
(q+r x) y \partial_x^2 \partial_y
+j y^2 \partial_y^2
\nonumber\\ &&
+s x y^2 \partial_x \partial_y^2 +(g y+h x y) 
\partial_{x,y}^2
\\
 &:&
a_1+f_1 x \partial_x +(c_1+b_1 y) \partial_y +j_1 x^2 \partial_x^2 \partial_y +y (d_1+e_1 y) 
\partial_y^2+q_1 x y \partial_x \partial_y^2 +p_1 y^2 \partial_y^3 
\nonumber\\ &&
+(g_1 x+h_1 x y) \partial_{x,y}^2.
\end{eqnarray}
}
For the Kamp\'e de F\'eriet function 
\begin{eqnarray}
	F^{p;q;k}_{l;m;n} \left[ \begin{array}{cc}
	(a_p) ; (b_q) ; (c_k) & \\
	& x, y \\
	(\alpha_l) ; (\beta_m) ; (\gamma_n) &
	\end{array}
	\right]
	&=& \sum_{r,s=0}^\infty \, \frac{\prod_{j=1}^p (a_j)_{r+s} 
        \prod_{j=1}^q (b_j)_r \prod_{j=1}^k (c_{{j}})_s}{\prod_{j=1}^l (\alpha_j)_{r+s} \prod_{j=1}^m (\beta_j)_r \prod_{j=1}^n 
       (\gamma_j)_s } \frac{x^r}{r!} \frac{y^s}{s!} \\
	&=& \sum_{r,s=0}^\infty f[r,s] x^r y^s
\end{eqnarray}
one obtains the following annihilating differential operators \cite{KAMPE1,KAMPE2}
\begin{eqnarray}
	\prod_{j=1}^p (x \partial_x + y \partial_y +a_j) \prod_{j=1}^q (x \partial_x +b_j) -{\partial_x} \prod_{j=1}^l (x \partial_x + y \partial_y -1+\alpha_j) \prod_{j=1}^m ({x \partial_x} -1+\beta_j) = 0,
\\
	\prod_{j=1}^p (x \partial_x +y \partial_y +a_j) \prod_{j=1}^k (y \partial_y +c_j) -{\partial_y} \prod_{j=1}^l (x 
\partial_x + y \partial_y -1+\alpha_j) \prod_{j=1}^n (y \partial_y -1+\gamma_j) = 0.
\end{eqnarray}

The differential operators for the triple hypergeometric series \cite{SRIKARL} read
{
\begin{eqnarray}
D_{3,1} &=& A+(B_0+B_1 x) \partial_x +x (E_0+E_1 x) \partial_x^2 + C_1 y \partial_y +{F_1} y^2 \partial_y^2 +(H_0+H_1 x) y 
\partial_{x,y}^2
\nonumber\\ &&
+D_1 z \partial_z 
+G_1 z^2 \partial_z^2 +(L_0+L_1 x) z \partial_{x,z}^2 + S_1 y z \partial_{y,z}^2
\\
D_{3,2} &=& A'+{B_1'} x \partial_x +E_1' x^2 \partial_x^2 +(C_0'+C_1' y) \partial_y +y (F_0'+F_1' y) 
\partial_y^2+x 
(H_2'+H_1' y) \partial_{x,y}^2
\nonumber\\ &&
+ D_1' z \partial_z 
+ G_1' z^2 \partial_z^2 + L_1' x z \partial_{x,z}^2
+(S_0'+S_1' y) z \partial_{y,z}^2
\\
D_{3,3} &=& 
A''+ B_1'' x \partial_x + {E''_1} x^2 \partial_x^2 + C_1'' y \partial_y + {F_1''} y^2 \partial_y^2
+ H_1'' x y \partial_{x,y}^2 + (D''_0 + D_1'' z) \partial_z 
\nonumber\\ &&
+ z (G_0'' + G_1'' z) \partial_z^2
+x {(L_2'' + L''_1 z)} \partial_{x,z}^2 + y (S_2''+ S_1'' z) \partial_{y,z}^2 .
\end{eqnarray}
}

The differential operators for the quadruple versions are given by
{
\begin{eqnarray}
D_{4,1} &=&
A + E_1 t \partial_t +L_1 t^2 \partial_t^2 +(B_0+B_1 x)\partial_x +x (F_0+F_1 x) \partial_x^2+t 
(P_0+P_1 x) \partial_{t,x}^2 +C_1 y \partial_y
\nonumber\\ &&
+G_1 y^2 \partial_y^2 +R_1 t y \partial_{t,y}^2 + (M_0+M_1 x) y 
\partial_{x,y}^2 + D_1 z \partial_z + H_1 z^2 \partial_z^2 + S_1 t z \partial_{t,z}^2 
\nonumber\\ &&
+(N_0+N_1 x) z 
\partial_{x,z}^2 +Q_1 y z \partial_{y,z}^2
\\ 
D_{4,2} &=& 
A'+E_1' t \partial_t + L_1' t^2 \partial_t^2 + B_1' x \partial_x +F_1' x^2 \partial_x^2 + P_1' t x 
\partial_{t,x}^2+(C_0'+ C_1' y) \partial_y
+y (G_0' + G_1' y) \partial_y^2 
\nonumber\\ &&
+ t (R_0'+R_1' y) \partial_{t,y}^2 + x (M_2' + M_1' y) 
\partial_{x,y}^2 + D_1' z \partial_z +H_1' z^2 \partial_z^2 +S_1' t z \partial_{t,z}^2 + N_1' x z 
\partial_{x,z}^2 
\nonumber\\ &&
+ (Q_0'+Q_1' y) z \partial_{y,z}^2
\\
D_{4,3} &=& 
A''+E_1'' t \partial_t + L_1'' t^2 \partial_t^2 + B_1'' x \partial_x + F_1'' x^2 \partial_x^2
+ P_1'' t x \partial_{t,x}^2  + C_1'' y \partial_y + G_1'' y^2 \partial_y^2 + R_1'' t y \partial_{t,y}^2 
\nonumber\\ &&
+ M_1'' x 
y \partial_{x,y}^2
+(D_0'' + D_1'' z) \partial_z + z (H_0''+H_1'' z) \partial_z^2 + t (S_0''+S_1'' z) \partial_{t,z}^2+x 
(N''_2 + N_1'' z) \partial_{x,z}^2 
\nonumber\\ &&
+ y (Q_2''+Q_1'' z) \partial_{y,z}^2
\\
D_{4,4} &=& 
A''' + (E_0''' + E_1''' t) \partial_t + t (L_0''' + L_1''' t) \partial_t^2 + B_1''' x \partial_x
+ F_1''' x^2 \partial_x^2 + (P_2''' + P_1''' t) x \partial_{t,x}^2 
\nonumber\\ &&
+ C_1''' y \partial_y + G_1''' y^2 
\partial_y^2 + (R_2''' + R_1''' t) y \partial_{t,y}^2 + M_1''' x y \partial_{x,y}^2 + D_1''' z \partial_z + H_1''' z^2 
\partial_z^2 
\nonumber\\ &&
+ (S_2'''+ S_1''' t) z \partial_{t,z}^2 + N_1''' x z \partial_{x,z}^2 + Q_1''' y z \partial_{y,z}^2.
\end{eqnarray}
}
They cover the functions $K_i,~~i = 1 ... 21$ of Refs.~\cite{EXTON72a,EXTON1}.

\section{The Recursions} 
\label{sec:reclist} 

\vspace*{1mm} 
\noindent 
The formal power series ansatz (\ref{eq:TAYLOR}) allows to obtain difference equations for the expansion 
coefficient from the differential equations given in Section~\ref{sec:delist}.\footnote{There are also contiguous relations
for the corresponding functions, cf.~\cite{CONTIG,Kalmykov:2020cqz}.}

The following recursions are obtained:
\begin{eqnarray}
_2F_1 &:& 
(C + n (1 - n + B_1)) f[n] + (1 + n) (n + A_1) f[1 + n] = 0
\label{eq:R2F1}
\\
_3F_2 &:& 
{
\big( n B_1 + (n-1)n B_2 +C-(n-2) (n-1) n \big) f[n]
    }
\nonumber\\ &&
{
+ \big( (n+1) A_1 + n (n+1) A_2 +  (n-1)n(n+1)\big) f[n+1] = 0
}    
\label{eq:R3F2}
\\
_{p+1}F_p &:& 
{
\Biggl[\frac{C}{n!} - \frac{1}{(n-p-1)!} + \sum_{k=1}^p \frac{B_k}{(n-k)!} \Biggr] f[n]
}
\nonumber\\ && 
{
+ (n+1) \Biggl[ \frac{1}{(n-p)!} +  \sum_{k=1}^p \frac{A_k}{(n-k+1)!} \Biggr] f[n+1] = 0.
}
\end{eqnarray}

In the two--variable cases the expansion coefficients of the  Horn--type functions obey
\begin{eqnarray}
&&\Big[a+b m+e (m-1) m+n \big(f+h m+j (n-1)\big)\Big] f[m,n]
\nonumber\\&&\qquad\qquad\qquad\qquad\qquad\qquad\qquad\qquad
+(1+m) (c+d m+g n) f[1+m,n]=0,
\\
&&\Big[a_1+f_1 m+j_1 (m-1) m+n \big(b_1+h_1 m+e_1 (n-1)\big)\Big] f[m,n]
\nonumber\\&&\qquad\qquad\qquad\qquad\qquad\qquad\qquad\qquad
+(1+n) (c_1+g_1 m+d_1 n) f[m,1+n]=0,
\end{eqnarray}
and for the $S_1$-functions one has
\begin{eqnarray}
&&\Big[a+b m+e (m-1) m+f n+h m n+j (n-1) n+(m-2) (m-1) m p+(m-1) m n r
\nonumber\\&&
+m (n-1) n s\Big] f[m,n]
+(1+m) \Big[c+g n+m \big(d+l (m-1)+n q\big)\Big] f[1+m,n]=0,
\\
&&\Big[a_1+f_1 m+j_1 (m-1) m+n \big(b_1+h_1 m+e_1 (n-1)\big)\Big] f[m,n]
\nonumber\\&&
+(1+n) (c_1+g_1 m+d_1 n) f[m,1+n]=0,
\end{eqnarray}
as, likewise, for the 
$S_2$-functions
\begin{eqnarray}
&&\Big[a+b m+e (m-1) m+f n+h m n+j (n-1) n+(m-2) (m-1) m p+(m-1) m n r+
\nonumber\\&&
m (n-1) n s\Big] f[m,n]+(1+m) \Big[c+g n+m \big(d+l (-1+m)+n q\big)\Big] f[1+m,n]=0,
\\
&&\Big[a_1+f_1 m+n \big(b_1+h_1 m+e_1 {(n-1)}\big)\Big] f[m,n]+(1+n) \Big[c_1+n (d_1+(-1+n) p_1)
\nonumber\\&&
+m \big(g_1+j_1 (m-1)+n q_1\big)\Big] f[m,1+n]=0.
\end{eqnarray}
For the expansion coefficients $f[r,s]$ of the Kamp\'e de F\'eriet functions the recurrences read
\begin{eqnarray}
	\prod_{j=1}^p (r+s+a_j) \prod_{j=1}^q (r+b_j) f[r,s] - (r+1) \prod_{j=1}^l (r+s+\alpha_j) \prod_{j=1}^m (r+\beta_j) 
f[r+1,s] = 0,
\\
	\prod_{j=1}^p (r+s+a_j) \prod_{j=1}^k (s+c_j) f[r,s] - (s+1) \prod_{j=1}^l (r+s+\alpha_j) \prod_{j=1}^n (s+\gamma_j) 
f[r,s+1] =0.
\end{eqnarray}

The coefficients in the 3-variable cases obey
\begin{eqnarray}
&&\Big[A+ B_1  m+ E_1  (m-1) m+ C_1  n+ H_1  m n+ F_1  (n-1) n+ D_1  p+ L_1  m p+ G_1  (p-1) p
\nonumber\\&&
+n p  S_1 \Big] f[m,n,p]+(1+m) ( B_0 + E_0  m+ H_0  n+ L_0  p) f[1+m,n,p]=0,
\\
&&\Big[ A' + B'_1  m+ E'_1  (m-1) m+ C'_1  n+ H'_1  m n+ F'_1  (n-1) n+ D'_1  p+ L'_1  m p+ G'_1  (p-1) p
\nonumber\\&&
+n p  S'_1 \Big] f[m,n,p]+(1+n) ( C'_0 + H'_2  m+ F'_0  n+p  S'_0 ) f[m,1+n,p]=0,
\\
&&\Big[ A'' + B''_1  m+ E''_1  (m-1) m+ C''_1  n+ H''_1  m n+ F''_1  (n-1) n+ D''_1  p+ L''_1  m p+ G''_1  (p-1) p
\nonumber\\&&
+n p  S''_1 \Big] f[m,n,p]+(1+p) ( D''_0 + L''_2  m+ G''_0  p+n  S''_2 ) f[m,n,1+p]=0.
\end{eqnarray}

For the 4-variable systems one has
\begin{eqnarray}
&&\Big[A+ B_1  m+ F_1  (m-1) m+ C_1  n+m  M_1  n+ G_1  (n-1) n+ D_1  p+m  N_1  p+ H_1  (p-1) p
\nonumber\\&&
+ E_1  q+m  P_1  q+ L_1  (q-1) q+n p  Q_1 +n q  R_1 +p q  S_1 \Big] f[m,n,p,q]+(1+m) ( B_0 + F_0  m
\nonumber\\&&
+ M_0  n+ N_0  p+ P_0  q) f[1+m,n,p,q]=0,
\\[2mm]
&&\Big[ A' + B'_1  m+ F'_1  (m-1) m+ C'_1  n+m  M'_1  n+ G'_1  (n-1) n+ D'_1  p+m  N'_1  p+ H'_1  (p-1) p
\nonumber\\&&
+ E'_1  q+m  P'_1  q+ L'_1  (q-1) q+n p  Q'_1 +n q  R'_1 +p q  S'_1 \Big] f[m,n,p,q]+(1+n) ( C'_0 +m  M'_2 
\nonumber\\&&
+ G'_0  n+p  Q'_0 +q  R'_0 ) f[m,1+n,p,q]=0,
\\[2mm]
&&\Big[ A'' + B''_1  m+ F''_1  (m-1) m+ C''_1  n+m  M''_1  n+ G''_1  (n-1) n+ D''_1  p+m  N''_1  p+ H''_1  (p-1) p
\nonumber\\&&
+ E''_1  q+m  P''_1  q+ L''_1  (q-1) q+n p  Q''_1 +n q  R''_1 +p q  S''_1 \Big] f[m,n,p,q]+(1+p) ( D''_0 +m  N''_2 
\nonumber\\&&
+ H''_0  p+n  Q''_2 +q  S''_0 ) f[m,n,1+p,q]=0,
\\[2mm]
&&\Big[ A''' + B'''_1  m+ F'''_1  (m-1) m+ C'''_1  n+m  M'''_1  n+ G'''_1  (n-1) n+ D'''_1  p+m  N'''_1  p
\nonumber\\&&
+ H'''_1  (p-1) p+ E'''_1  q+m  P'''_1  q+ L'''_1  (q-1) q+n p  Q'''_1 +n q  R'''_1 +p q  S'''_1 \Big] f[m,n,p,q]
\nonumber\\&&
+(1+q) ( E'''_0 +m  P'''_2 + L'''_0  q+n  R'''_2 +p  S'''_2 ) f[m,n,p,1+q]=0.
\end{eqnarray}
\section{The Solution of the Recursions} 
\label{sec:recsol}

Let $\KK$ be a field of characteristic $0$ (i.e., a field that contains $\mathbb Q$ as subfield).
A power series in $r$ variables
\begin{equation}
f(x_1,\ldots,x_r) = \sum_{n_i\ge0} A(n_1,\ldots,n_r) x_1^{n_1}\cdots x_r^{n_r}
\label{eq:hyp-series}
\end{equation}
is called a multiple hypergeometric series if the multivariate sequence $A:\mathbb{N}^r\to\KK$ is hypergeometric, i.e., we have
\begin{equation}
s_i\,A(n_1,\ldots,n_i,\ldots,n_r) = t_i\,A(n_1,\ldots,n_i+1,\ldots,n_r) , \qquad i=1,\ldots,r
\label{eq:hyp-ratio}
\end{equation}
for polynomials $s_i,t_i\in\KK[n_1,\dots,n_r]$ being coprime.

Often the hypergeometric sequence $A$ is given in terms of binomial coefficients, Pochhammer symbols, 
$\Gamma$--functions and related special functions. However, in concrete applications one often starts with 
a 
given system of partial linear differential equations and searches for a hypergeometric series solution as 
specified above. Plugging this ansatz into the equations and doing coefficient comparison w.r.t.\ $x_1^{n_1}\cdots x_r^{n_r}$ yield a system of partial linear difference equations; for concrete examples see Section~\ref{sec:reclist}. In the general case not too many methods are known that can support the use to solve these difference equations; for some first steps in this direction we refer the reader to Section~\ref{sec:PLDEsolver}. In the following we concentrate on first--order systems 
of the form~\eqref{eq:hyp-ratio}.

We remark that in many concrete applications such a system~\eqref{eq:hyp-ratio} can be found. In particular, this is the case if the underlying system of linear differential equations is of the form
\begin{equation}
\Big[
s_i \Bigl(x_1\frac{\partial}{\partial x_i}, \ldots, x_i \frac{\partial}{\partial x_i}, \ldots, x_r \frac{\partial}{\partial x_r} \Bigr) 
- \frac{1}{x_i} t_i \Bigl(x_1 \frac{\partial}{\partial x_1}, \ldots, x_i\frac{\partial}{\partial x_i}-1, \ldots, x_r \frac{\partial}{\partial x_r} \Bigr) 
\Big] f(x_1,\ldots,x_r)
= 0.
\label{eq:DEsystem}
\end{equation}
To show this, we utilize the crucial property 
\begin{equation}
x_i \frac{\partial}{\partial x_i} x_1^{n_1}\cdots x_r^{n_r} = n_i x_1^{n_1}\cdots x_r^{n_r}
\end{equation}
which implies that for a polynomial $p(n_1,\ldots,n_r)$ we have
\begin{equation}
p \Big(x_1\frac{\partial}{\partial x_1}, \ldots, x_i \frac{\partial}{\partial x_i}, \ldots, x_r \frac{\partial}{\partial x_r} \Bigr) x_1^{n_1}\cdots x_r^{n_r} = p(n_1,\ldots,n_r) x_1^{n_1}\cdots x_r^{n_r} .
\end{equation}
Thus
\begin{align}
\label{eq:t-term}
&\Big[ s_i \Bigl(x_1\frac{\partial}{\partial x_1}, \ldots, x_i \frac{\partial}{\partial x_i}, \ldots, x_r \frac{\partial}{\partial x_r} \Bigr) \Big] f(x_1,\ldots,x_r)
\nonumber\\
&\qquad \qquad \qquad \qquad \qquad = \sum_{n_i\ge 0} s_i(n_1,\ldots,n_i,\ldots,n_r)  A(n_1,\ldots,n_i,\ldots,n_r) x_1^{n_1}\cdots x_r^{n_r}	\\ 
&\Big[ t_i \Bigl(x_1 \frac{\partial}{\partial x_1}, \ldots, x_i\frac{\partial}{\partial x_i}-1, \ldots, x_r \frac{\partial}{\partial x_r} \Bigr) \Big] f(x_1,\ldots,x_r)
\nonumber\\
&\qquad \qquad \qquad \qquad \qquad = \sum_{n_i\ge 0} t_i(n_1,\ldots,n_i-1,\ldots,n_r)  A(n_1,\ldots,n_i,\ldots,n_r) x_1^{n_1}\cdots x_r^{n_r} 
\end{align}
and therefore, dividing the second equation by $x_i$ from the left,
\begin{align}
\label{eq:s-term}
&\Big[ \frac{1}{x_i} t_i \Bigl(x_1 \frac{\partial}{\partial x_1}, \ldots, x_i\frac{\partial}{\partial x_i}-1, \ldots, x_r \frac{\partial}{\partial x_r} \Bigr) \Big] f(x_1,\ldots,x_r)
\nonumber\\
& \qquad \qquad \qquad \qquad \qquad = \sum_{n_i\ge 0} 
t_i(n_1,\ldots,n_i-1,\ldots,n_r)  A(n_1,\ldots,n_i,\ldots,n_r) 
x_1^{n_1}\cdots x_i^{n_i-1} \cdots x_r^{n_r}. 
\end{align}
The coefficient of the term $x_1^{n_1}\cdots x_i^{n_i}\cdots x_r^{n_r}$ in 
\eqref{eq:t-term} and in 
\eqref{eq:s-term} is respectively
\begin{equation}
s_i(n_1,\ldots,n_i,\ldots,n_r) A(n_1,\ldots,n_i,\ldots,n_r)
\end{equation}
and
\begin{equation}
t_i(n_1,\ldots,n_i,\ldots,n_r) A(n_1,\ldots,n_i+1,\ldots,n_r).
\end{equation}
This shows, due to \eqref{eq:hyp-ratio}, that~\eqref{eq:DEsystem} holds.

For example, for the case of the Gauss hypergeometric function
\begin{equation}
_2F_1(a,b;c;x) = \sum_{n\ge 0} \frac{(a)_n (b)_n}{(c)_n n!}x^n
\end{equation}
one has
\begin{eqnarray}
A(n) &=& \frac{(a)_n (b)_n}{(c)_n n!} \\
s(n) &=& (a+n)(b+n) \\
t(n) &=& (n+1)(c+n) 
\end{eqnarray}
and the differential equation obeyed by $_2F_1(a,b;c;x)$ is, from \eqref{eq:DEsystem},
\begin{equation}
\Big[ \Big(a+ x \frac{\partial}{\partial x} \Big) \Big(b+ x \frac{\partial}{\partial x} \Big) - \frac{1}{x} \Big(x \frac{\partial}{\partial x}\Big)	\Big( x \frac{\partial}{\partial x} -1 +c \Big) \Big] {}_2F_1(a,b;c;x) = 0.
\end{equation}

\vspace*{1mm} 
\noindent 

\subsection{An algorithm for hypergeometric products}\label{sec:solveProd}

Given a hypergeometric sequence $A$ with~\eqref{eq:hyp-ratio} we seek for a representation in terms of indefinite nested products that can be modeled, e.g., within the summation package \texttt{Sigma}.

For the univariate case $r=1$ this task is immediate. Since $s_1,t_1\in\KK[n_1]$ have only finitely many roots, there is a $\lambda_1\in\mathbb{N}$ such that $s_1(k)\neq0\neq t_1(k)$ for all $k\geq\lambda_1$. Thus for $n_1\geq\lambda_1$ we get
\begin{equation}\label{Equ:Prod1}
\begin{split}
A(n_1)&=\frac{s_1(n_1-1)}{t_1(n_1-1)}A(n_1-1)\\
&=\frac{s_1(n_1-1)}{t_1(n_1-1)}\frac{s_1(n_1-2)}{t_1(n_1-2)}A(n_1-2)=\dots=\left(\prod_{k=\lambda_1+1}^{n_1}\frac{s_1(k-1)}{t_1(k-1)}\right)A(\lambda_1)
\end{split}
\end{equation}
where $A(n_1)$ can be written in terms of the hypergeometric product $\prod_{k=\lambda_1+1}^{n_1}\frac{s_1(k-1)}{t_1(k-1)}$ 
which is nonzero for each $n_1\geq0$. In other words, a hypergeometric sequence is either trivial, i.e., it is the $0$ sequence from a certain point on (if $A(\lambda_1)=0$) or is nonzero for all $n\geq\lambda_1$. 

Next, we turn to the multivariate case.
As introduced in~\cite{AP:02} we call a sequence non--trivial if the zero points vanish on a nonzero 
polynomial from $\KK[n_1,\dots,n_r]$. In other words $A$ is almost everywhere a nonzero sequence. An important consequence of~\cite[Prop~4]{AP:02} is that for such a hypergeometric sequence with~\eqref{eq:hyp-ratio} the following compatibility property holds for $R_i=\frac{s_i}{t_i}\in\KK(n_1,\dots,n_r)$: for $1\leq i\leq j\leq r$,
\begin{equation}\label{Equ:CompatibilityProp}
\frac{R_i(n_1,\dots,n_j+1,\dots,n_r)}{R_i(n_1,\dots,n_j,\dots,n_r)}=\frac{R_j(n_1,\dots,n_i+1,\dots,n_r)}{R_j(n_1,\dots,n_i,\dots,n_r)}.
\end{equation}
In particular, the Ore-Sato Theorem~\cite{OS:90} holds: $A$ can be written as a product in terms of geometric products and factorial terms; for a rigorous (and rather involved) proof see~\cite{AP:02} and for further generalizations see~\cite{OSGeneral}.

In the following we will introduce a special case of the Ore-Sato theorem that deals with the problem to represent $A$ in terms of hypergeometric products which are valid for all $(n_1,\dots,n_r)\in\mathbb N$ where the $n_i$ are chosen sufficiently large.
This is precisely the situation that we require for hypergeometric power series as given in~\eqref{eq:hyp-series}.

As it turns out, such a representation is always possible if we require the following additional assumptions (which hold in the univariate case automatically): we can choose\footnote{In general there is no algorithm by the Davis--Matiyasevich--Putnam--Robinson
theorem~\cite{Hilbert10} that can decide if there is an integer root (or even infinitely many integer roots). However, in our applications the polynomials are usually small, mostly even linear and thus such integers $\lambda_i$ can be determined.} $\lambda_i\in\mathbb N$ such that for all $(n_1,\dots,n_r)\in\mathbb N^r$ with $n_i\geq\lambda_i$  we have
$$s_i(n_1,\dots,n_r)\neq0\neq t_i(n_1,\dots,n_r).$$
Therefore~\eqref{eq:hyp-ratio} is equivalent to
\begin{equation}
A(n_1,\ldots,n_i+1,\ldots,n_r) = R_i(n_1,\dots,n_r)\,A(n_1,\ldots,n_i,\ldots,n_r) , \qquad i=1,\ldots,r
\end{equation}
with $R_i(n_1,\dots,n_r)\neq0$ for all $n_i\geq\lambda_i$.
Applying these relations iteratively shows that for any  $(n_1,\dots,n_r)\in\mathbb N^r$ with $n_i\geq\lambda_i$ 
there is a $c\in\KK\setminus\{0\}$ such that
$$A(n_1,\dots,n_r)=c\,A(\lambda_1,\dots,\lambda_r).$$
Similarly to the univariate case we get the following consequence: $A(n_1,\dots,n_r)$ is the zero sequence (for all $n_i\geq\lambda_i$) if $A(\lambda_1,\dots,\lambda_r)=0$ or it is nonzero for all $n_i\geq\lambda_i$ otherwise. 

\medskip

\textit{Remark:} In the second case all zeroes of $A$ are finite and thus vanish on a particular chosen nonzero polynomial. Thus we can apply~\cite[Prop~4]{AP:02} as above. If the compatibility criteria~\eqref{Equ:CompatibilityProp} does not hold, then $A$ must be the zero sequence or the hypergeometric system is inconsistent.

\medskip

With these properties there is a simple algorithm which finds a product representation of $A$ of the form
\begin{equation}\label{Equ:FinalProdForm}
c\,\left(\prod_{k=\lambda_1}^{n_1}h_1(k,n_2,\dots,n_r)\right)\left(\prod_{k=\lambda_2}^{n_2}h_2(k,n_3\dots,n_r)\right)\dots \left(\prod_{k=\lambda_r}^{n_r}h_r(k)\right)
\end{equation}
with $c=A(\lambda_1,\dots,\lambda_r)\in\KK\setminus\{0\}$ and $h_i(x,n_{i+1},\dots,n_r)\in\KK(x,n_{i},\dots,n_r)$ with $1\leq i\leq r$. In particular, we have that $\prod_{k=\lambda_i}^{n_i}h_i(k,n_{i+1}\dots,n_r)\neq0$ for all $n_i\geq\lambda_i$ with $i=1,\dots,r$.

If $r=1$, such a product can be derived immediately with~\eqref{Equ:Prod1} and $c=A(\lambda_1)$. Otherwise, the algorithm works by recursion (induction on $r>1$). As in the case $r=1$ it follows that we can write
$$A(n_1,\dots,n_r)=A(\lambda_1,n_2,\dots,n_r)\prod_{k=\lambda_1+1}^{n_1}h_1(k,n_2,\dots,n_r)$$
with $h_1(k,n_2,\dots,n_r)=R_1(k-1,n_2,\dots,n_r)=\frac{s_1(k-1,n_2,\dots,n_r)}{t_1(k-1,n_2,\dots,n_r)}$; note that
$A(\lambda_1,n_2,\dots,n_r)\neq0$ for all $(n_2,\dots,n_r)\in\mathbb N$ with $n_i\geq\lambda_i$.  

Now consider the multivariate sequence
$$A'(n_2,\dots,n_r):=A(\lambda_1,n_2,\dots,n_r)
$$
which satisfies
$$A'(n_2,\ldots,n_i+1,\ldots,n_r) = R_i(\lambda_1,n_2,\dots,n_r)\,A'(n_2,\ldots,n_i,\ldots,n_r) , \qquad i=2,\ldots,r$$
with $R_i(\lambda_1,n_2,\dots,n_r)\in\KK[n_2,\dots,n_r]$, where 
$R_i(\lambda_1,n_2,\dots,n_r)\neq0$ for all $n_i\geq\lambda_i$.
Obviously, $A'$ is again hypergeometric with all the assumptions (in particular satisfying the compatibility criteria in~\eqref{Equ:CompatibilityProp}) and we can proceed by induction/recursion. Thus we get
$$A'(n_2,\dots,n_r)=c\,\left(\prod_{k=\lambda_1}^{n_2}h_2(k,n_3\dots,n_r)\right)
\dots \left(\prod_{k=\lambda_r}^{n_r}h_r(k)\right),$$
with 
$c=A'(\lambda_2,\dots,\lambda_r)=A(\lambda_1,\lambda_2,\dots,\lambda_r)\in\KK\setminus\{0\}$ 
and $h_i(x,n_{i+1},\dots,n_r)\in\KK(x,n_{i},\dots,n_r)$ with $2\leq i\leq r$. 
This finally shows~\eqref{Equ:FinalProdForm}. 

We remark that in all the examples of this article we can set $\lambda_i=0$ for $1\leq i\leq r$.

\subsection{Examples}\label{sec:exProdExp}

Let us illustrate the solution of some of the recursions in explicit form. Here we refer first to the general
representation of the corresponding differential and difference equations. We consider the differential equation
(\ref{eq:D2F1g}) which leads to the recurrence (\ref{eq:R2F1}) for the expansion coefficient $f[n]$. The recursion is 
of order one and is solved for $f[n] = 0$. {\tt Sigma} obtains the following product solution
\begin{eqnarray}
f[n] = 
\frac{\prod_{i_1=1}^n \big(
        2
        +B_1
        -C
        -3 i_1
        -B_1 i_1
        +i_1^2
\big)}{n! (A_1)_n}
\equiv \frac{\prod_{i_1=1}^n \big[-C +B_1(1-i_1) + (1-i_1)(2-i_1)
\big]}{n! (A_1)_n},
\label{eq:T1}
\nonumber\\
\end{eqnarray}
which is not yet expressed by Pochhammer symbols.

{\tt Mathematica} allows to obtain the factorization of the 
product in (\ref{eq:T1}) in terms of Pochhammer symbols by      
\begin{eqnarray}
f[n] = \frac{(\alpha_1)_n (\alpha_2)_n}{(A_1)_n n!},
\end{eqnarray}
with 
\begin{eqnarray}
\alpha_{1(2)} = -\frac{1}{2}(1+B_1) \mp  \frac{1}{2} \sqrt{(1+B_1)^2 + 4 C}.
\end{eqnarray}
By replacing $A_1,B_1$ and $C$ directly to
\begin{eqnarray}
C \rightarrow -a b,~~A_1 \rightarrow c,~~B_1 \rightarrow -1 - a - b
\end{eqnarray}
one obtains
\begin{eqnarray}
f[n] = \frac{(a)_n (b)_n}{(c)_n n!}.
\end{eqnarray}
This choice of variables is therefore instrumental to obtain the most simple structure. However,
it will sometimes not naturally appear in the physical differential equations, requesting associated variable 
transformations in general. 

This becomes more and more involved in higher hypergeometric cases, which is already illustrated in the case of
the generalized hypergeometric function $_3F_2$. Its differential equation (\ref{eq:D3F2}) implies the recurrence
for $f[n]$ (\ref{eq:R3F2}) with $f[n]=1$, which has the solution
\begin{eqnarray}
f[n] = \frac{\prod_{i_1=1}^n 
[-C
+B_1 \big(
        1-i_1\big)
-B_2 \big(
        2-i_1
\big)
{ \big(1-i_1\big) }
-\big(
        3-i_1
\big)
\big(2-i_1
\big)
\big(1-i_1\big)]}
{n!~\prod_{i_1=1}^n
[A_1
-A_2 \big(
        1-i_1\big)
+\big(
        2-i_1
\big)
\big(1-i_1\big)]}.
\label{eq:3F2a}
\end{eqnarray}
Eq.~(\ref{eq:3F2a}) can be rewritten in terms of radicals by
\begin{eqnarray}
f[n] = \frac{
(\alpha_1)_n
(\alpha_2)_n
(\alpha_3)_n}{ {n!} \big(
        -\frac{1}{2}
        +\frac{A_2 }{2}
        -\frac{z_5}{2}
\big)_n \big(
        -\frac{1}{2}
        +\frac{A_2 }{2}
        +\frac{z_5}{2}
\big)_n},
\end{eqnarray}
with
\begin{eqnarray}
\alpha_1 &=& 
        1
        -\frac{z_4}{3}
        +\frac{\sqrt[3]{z_1
        +z_2}
        }{6 \sqrt[3]{2}}
        -\frac{i \sqrt[3]{z_1
        +z_2
        }}{2 \sqrt[3]{2} \sqrt{3}}
        +\frac{\sqrt[3]{-2} z_3}{3 \sqrt[3]{z_1
        +z_2
        }}
\\
\alpha_2 &=&
        1
        -\frac{z_4}{3}
        -\frac{\sqrt[3]{z_1
        +z_2
        }}{3 \sqrt[3]{2}}
        -\frac{\sqrt[3]{2} z_3}{3 \sqrt[3]{z_1
        +z_2
        }}
\\
\alpha_3 &=& 
        1
        -\frac{z_4}{3}
        +\frac{\sqrt[3]{z_1
        +z_2
        }}{6 \sqrt[3]{2}}
        +\frac{i \sqrt[3]{z_1
        +z_2
        }}{2 \sqrt[3]{2} \sqrt{3}}
        -\frac{(-1)^{2/3} \sqrt[3]{2} z_3}{3 \sqrt[3]{z_1
        +z_2
        }}
\\
z_1 &=&  27 C + (3 + B_2) (9 B_1 + B_2 (3 + 2 B_2))
\\
z_2 &=&  \sqrt{-4 (3 + 3 B_1 + 
    B_2 (3 + B_2))^3 + (27 C + (3 + B_2) (9 B_1 + 
      B_2 (3 + 2 B_2)))^2}
\\
z_3 &=&  3 + 3 B_1 + B_2 (3 + B_2)
\\
z_4 &=&  6 + B_2
\\
z_5 &=&  \sqrt{-4 A_1 + (A_2-1)^2}.
\end{eqnarray}

After performing the replacements
\begin{eqnarray} A_{{2}} &\rightarrow& b_1+b_2 +1,~~ B_{{2}} \rightarrow -(3 + a_1 + a_2 +a_3),~~A_{{1}} \rightarrow b_1 b_2, 
\nonumber\\ 
B_{{1}} &\rightarrow& -(a_2 a_1 + a_3 a_1 + a_2 a_3 + a_1 +
a_2 +a_3 +1),~~C \rightarrow - a_1 a_2 a_3
\label{eq:VIETA}
\end{eqnarray}
one obtains
\begin{eqnarray} 
f[n] = \frac{(a_1)_n (a_2)_n (a_3)_n}{(b_1)_n (b_2)_n n!}.
\end{eqnarray}
One observes that the substitutions (\ref{eq:VIETA}) are related to the root--relations by Vieta's theorem 
\cite{VIETA} for the 
roots $r_i$ of the algebraic equation
\begin{eqnarray} 
x^n + \sum_{k = 1}^{n} a_{n-k} x^{n-k} = 0,
\end{eqnarray}
which obey
\begin{eqnarray} 
-a_{n-1} &=& r_1 + ... + r_n
\nonumber\\
 a_{n-2} &=& r_1 (r_2 + ... + r_n) + r_2(r_3+ ... r_n) + ... r_{n-1} r_n 
\nonumber\\
&\vdots& 
\nonumber\\
(-1)^n a_0 &=& r_1 ... r_n.
\end{eqnarray}
In all cases, which can be solved by a single recurrence at the time the above procedures are applied.
Considering the generalized hypergeometric function $_{p+1}F_p$ the product--solution for the expansion coefficient
reads
\begin{eqnarray} 
{
f[n] = 
\prod_{i=1}^n \frac{\frac{1}{(i-p-2)!} - \sum_{k=1}^p \frac{B_k}{(i-k-1)!} - \frac{C}{(i-1)!}}
{\frac{i}{(i-p-1)!} + \sum_{k=1}^p \frac{A_k i}{(i-k)!} }
}
\end{eqnarray}
and one may factorize the corresponding product as in the above examples. However, the corresponding roots
can in general only be obtained numerically. Still one may work with the corresponding symbolic expressions.
This is in particular useful w.r.t.\ their expansion in the dimensional parameter $\ep$, which is contained 
in the quantities $A_n, B_n$ and $C$ in polynomial form.

We turn now to the multivariate case. Here we have explicit formal solutions which apply to 
all concrete cases listed in the appendix, resp.\ in the files attached, but may cover even more cases.
To find this out for concrete parameter settings one is advised to check 
whether these particular solutions obey the 
corresponding difference equations.

For the Horn--type functions one obtains {from Eqs. \eqref{APP1}, \eqref{APP2}} the product solution for $f[m,n]$
\begin{eqnarray}
f_{\rm H}[m,n] &=& 
\Biggl[
        \prod_{i_1=1}^m \frac{-a
        +b
        -2 e
        -f n
        +h n
        +j n
        -j n^2
        -b i_1
        +3 e i_1
        -h n i_1
        -e i_1^2
        }{\big(
                c
                -d
                +g n
                +d i_1
        \big) i_1}\Biggr]
\nonumber\\&& \times        
        \Biggl[\prod_{i_1=1}^n \frac{-a_1
+b_1
-2 e_1
-b_1 i_1
+3 e_1 i_1
-e_1 i_1^2
}{\big(
        c_1
        -d_1
        +d_1 i_1
\big) i_1}\Biggr]
\label{eq:HORN2}
\end{eqnarray}
and for the functions $S_1$ and $S_2$ one has
\begin{eqnarray}
f_{\rm S_1}[m,n] &=& 
\Biggl[
        \prod_{i_1=1}^n \frac{- a_1 
        + b_1 
        -2  e_1 
        - b_1  i_1
        +3  e_1  i_1
        - e_1  i_1^2
        }{\big(
                 c_1 
                - d_1 
                + d_1  i_1
        \big) i_1}\Biggr] 
\nonumber\\&& \times    
\Biggl[   
        \prod_{i_1=1}^m 
\bigg(        
        \frac{1}{\big(
        c
        -d
        +2 l
        +g n
        -n q
        +d i_1
        -3 l i_1
        +n q i_1
        +l i_1^2
\big) i_1}	(-a
+b
\nonumber\\&&
-2 e
-f n
+h n
+j n
-j n^2
+6 p
-2 n r
-n s
+n^2 s
-b i_1
\nonumber\\&&
+3 e i_1
-h n i_1
-11 p i_1
+3 n r i_1
+n s i_1
-n^2 s i_1
-e i_1^2
+6 p i_1^2
-n r i_1^2
\nonumber\\&&
-p i_1^3
)
\bigg)
\bigg]
\\
f_{\rm S_2}[m,n] &=& 
\bigg(
        \prod_{i_1=1}^n \frac{- a_1 
        + b_1 
        -2  e_1 
        - b_1  i_1
        +3  e_1  i_1
        - e_1  i_1^2
        }{\big(
                 c_1 
                -d_1
                +2  p_1 
                +d_1 i_1
                -3  p_1  i_1
                + p_1  i_1^2
        \big) i_1}\bigg)
\nonumber\\&& \times     
\bigg[   
         \prod_{i_1=1}^m 
\bigg(         
         \frac{1}{\big(
        c
        -d
        +2 l
        +g n
        -n q
        +d i_1
        -3 l i_1
        +n q i_1
        +l i_1^2
\big) i_1}
(-a
+b
-2 e
-f n
\nonumber\\&&
+h n
+j n
-j n^2
+6 p
-2 n r
-n s
+n^2 s
-b i_1
+3 e i_1
-h n i_1
-11 p i_1
+3 n r i_1
\nonumber\\&&
+n s i_1
-n^2 s i_1
-e i_1^2
+6 p i_1^2
-n r i_1^2
-p i_1^3
)
\bigg)\Bigr].
\end{eqnarray}
Eq.~(\ref{eq:HORN2}) can be rewritten as
\begin{eqnarray} 
f[n,m] &=& \frac{\displaystyle (-1)^{m+n}} 
{\displaystyle m! n! \Biggl(
        \frac{c_1}{d_1}\Biggr)_n \Biggl(
        \frac{c}{d}
        +\frac{g n}{d}
\Biggr)_m}
\Biggl(
        \frac{e}{d}\Biggr)^m \Biggl(
        \frac{e_1}{d_1}\Biggr)^n \Biggl(
        -\frac{1}{2}
        +\frac{b_1}{2 e_1}
        -\frac{r_1}{2 e_1}
\Biggr)_n 
\nonumber\\ && \times
\Biggl(
        -\frac{1}{2}
        +\frac{b_1}{2 e_1}
        +\frac{r_1}{2 e_1}
\Biggr)_n \Biggl(
        -\frac{1}{2}
        +\frac{b}{2 e}
        +\frac{h n}{2 e}
        -\frac{r_2}{2 e}
\Biggr)_m \Biggl(
        -\frac{1}{2}
        +\frac{b}{2 e}
        +\frac{h n}{2 e}
        +\frac{r_2}{2 e}
\Biggr)_m,
\nonumber\\
\label{eq:FMN}
\end{eqnarray}
with
\begin{eqnarray}\label{Equ:AlgebraicExt}
r_1 &=& \sqrt{(b_1 - e_1)^2 - 4 a_1 e_1}
\\
r_2 &=& 
{
\sqrt{(b - 3 e + h n)^2 - 4 e (a - b + 2 e + f n - h n - j n + j n^2)}.
}
\end{eqnarray}
The Pochhammer-form of (\ref{eq:FMN}) directly allows the $\ep$--expansion, if the free parameters in the 
Pochhammer symbols are replaced accordingly by expressions that contain also the $\ep$-parameter. Further simplifications are obtained using the 
replacement rules given in Appendix~\ref{sec:A}. E.g. for the Appell function $F_1$ one obtains
\begin{eqnarray}
f[n,m] = \frac{(\alpha )_{m+n} (\beta )_m (\beta')_n}{m! n! (\gamma )_{m+n}},
\end{eqnarray}
by using the replacements given in (\ref{eq:F1a}), Appendix~\ref{sec:B}.

In the tri--variate cases one obtains
\begin{eqnarray}
	f[m,n,p] &=&
\bigg(
        \prod_{i_1=1}^m \frac{-A
        + B_1
        -2  E_1
        - B_1 i_1
        +3  E_1 i_1
        - E_1 i_1^2
        }{\big(
                B_0 
                -E_0 
                +E_0  i_1
        \big) i_1}
\bigg)
\nonumber\\&& \times
\bigg[\prod_{i_1=1}^n 
\bigg(
\frac{1}{\big(
                 C'_0 
                - F'_0 
                + H'_2  m
                + F'_0  i_1
        \big) i_1}
		(- A' 
        + C'_1 
        -2  F'_1 
        - B'_1  m
\nonumber\\&&        
        + E'_1  m
        + H'_1  m
        - E'_1  m^2
        - C'_1  i_1
        +3  F'_1  i_1
        - H'_1  m i_1
        - F'_1  i_1^2
        )  
\bigg)              
        \bigg]
\nonumber\\&& \times   
\bigg[     
         \prod_{i_1=1}^p 
\bigg(         
         \frac{1}{\big(
        D''_0 
        - G''_0 
        + L''_2  m
        +n  S''_2 
        + G''_0  i_1
\big) i_1}
(-A'' 
+ D''_1 
-2  G''_1 
- B''_1  m
+ E''_1  m
\nonumber\\&&
+ L''_1  m
- E''_1  m^2
- C''_1  n
+ F''_1  n
- H''_1  m n
- F''_1  n^2
+n  S''_1 
- D''_1  i_1
+3  G''_1  i_1
- L''_1  m i_1
\nonumber\\&&
-n  S''_1  i_1
- G''_1  i_1^2
)
\bigg)
\bigg].
\end{eqnarray}

Finally, in the four--variable case the product solution reads 
\begin{eqnarray}
f[m,n,p,q] &=&	\bigg(
        \prod_{i_1=1}^m 
        \frac{-A
        +B_1
        -2 F_1
        -B_1 i_1
        +3 F_1 i_1
        -F_1 i_1^2
        }{\big(
                B_0
                -F_0
                +F_0 i_1
        \big) i_1}
\bigg)
\nonumber\\&&\times
\bigg[
\prod_{i_1=1}^n
\frac{1}{\big(
                C'_0
                -G'_0
                +m M'_2
                +G'_0 i_1
        \big) i_1}
        (-A'
        +C'_1
        -2 G'_1
        -B'_1 m
        +F'_1 m
\nonumber\\&&        
        -F'_1 m^2
        +m M'_1
        -C'_1 i_1
        +3 G'_1 i_1
        -m M'_1 i_1
        -G'_1 i_1^2
        )
\bigg]
\nonumber\\&&\times
\bigg[
\prod_{i_1=1}^p 
\frac{1}{\big(
                D''_0
                -H''_0
                +m N''_2
                +n Q''_2
                +H''_0 i_1
        \big) i_1}
(-A''
        +D''_1
        -2 H''_1
        -B''_1 m
        +F''_1 m
\nonumber\\&&        
        -F''_1 m^2
        -C''_1 n
        +G''_1 n
        -m M''_1 n
        -G''_1 n^2
        +m N''_1
        +n Q''_1
        -D''_1 i_1
        +3 H''_1 i_1
\nonumber\\&&        
        -m N''_1 i_1
        -n Q''_1 i_1
        -H''_1 i_1^2
        )        
        \bigg]
\nonumber\\&&\times
\bigg[
         \prod_{i_1=1}^q \frac{1}{\big(
        E'''_0
        -L'''_0
        +m P'''_2
        +n R'''_2
        +p S'''_2
        +L'''_0 i_1
\big) i_1}
(-A'''
+E'''_1
-2 L'''_1
-B'''_1 m
\nonumber\\&& 
+F'''_1 m
-F'''_1 m^2
-C'''_1 n
+G'''_1 n
-m M'''_1 n
-G'''_1 n^2
-D'''_1 p
+H'''_1 p
-m N'''_1 p
\nonumber\\&& 
-H'''_1 p^2
+m P'''_1
-n p Q'''_1
+n R'''_1
+p S'''_1
-E'''_1 i_1
+3 L'''_1 i_1
-m P'''_1 i_1
-n R'''_1 i_1
\nonumber\\&& 
-p S'''_1 i_1
-L'''_1 i_1^2
)
\bigg].
\end{eqnarray}
Pochhammer solutions are of advantage, since the $\ep$--expansion can be derived more easily compared to
the case of the product solutions.
The contributing powers in $i_1$ in the above products determine the degree, {\sf d}, of the algebraic 
equations to switch to the associated Pochhammer form, which is in many cases {\sf d = 2} and {\sf d = 3} for 
the $S_{1,(2)}$ functions, for the functions considered in the present paper. In the $_pF_q$ case it 
can be even of higher order. Here complex solutions will appear in general for the Pochhammer symbols.
The first index of the Pochhammer symbol will imply a new constant
in the ground field to be used in the summation problem. 
However, still the solutions remain real. If the corresponding algebraic equations can be solved in closed form 
the special conditions discussed in Appendix~\ref{sec:B} need not to be obeyed.

In any case, if the degree {\sf d} of the algebraic 
equations is too high, in particular, if the algebraic extensions get too complicated, one can use the 
general tools developed in Section~\ref{Sec:EpExpansion} to derive the $\ep$--expansion by introducing 
generalized 
versions of harmonic sums and Hurwitz type sums where the summands have denominators which do not factor linearly. 

\section{Computing the expansion in \boldmath $\ep$} 
\label{sec:epExpansion}

\vspace*{1mm} 
\noindent 
Performing the expansion in the dimensional parameter $\ep$ on the basis of series representation 
around $x_i$ in the vicinity of zero, the convergence region of the respective series has to be known 
in general. For the one--parameter series we consider the $_pF_q$ functions, which converge for $|x| < 
1$ for $p \leq q+1$, \cite{SLATER1}, which we are going to consider. In the two--variable case one has 
\cite{SLATER1} 
\begin{eqnarray}
F_1,  
F_3  
&:&  |x|, |y| < 1,
\\
F_2 &:&  |x| + |y| < 1.
\\
F_4 &:& \sqrt{|x|} + \sqrt{|y|} < 1  
\end{eqnarray}
In the attachment {\tt converg.m} we present the corresponding convergence conditions for all functions 
up to three variables, as have been given in \cite{SRIKARL}, in computer-readable form, for the convenience 
of the user. These conditions are partly very involved.
An example is $f_{26f}$
\begin{eqnarray}
f_{26f} &:& |z| < \frac{1}{4}, |x| < \frac{1}{1+2\sqrt{|z|}}, |y|< 
  \sqrt{1+|x|(1+2\sqrt{|z|})}  
- \sqrt{|x|(1+2\sqrt{|z|})},  
\end{eqnarray}
\cite{SRIKARL}.
More involved conditions are obtained in the four-variable case. They may be derived using d'Alemberts
ratio test \cite{WW,DAL} to these cases.

In general, multi--sums appear with complicated hypergeometric products and one may try to apply, e.g., the 
package \texttt{EvaluateMultiSums}~\cite{Ablinger:2010pb,Blumlein:2012hg,Schneider:2013zna,Schneider:19} (utilizing the difference ring algorithms~\cite{DR,TermAlgebra,LinearSolver} available in {\tt Sigma}) to represent these sums to indefinite nested sums. In general, this seems not possible. But we will show how this goal can be accomplished for various interesting cases with our computer algebra tools. In particular, if the products depend on the dimensional parameter $\ep$ 
and one is interested in its $\ep$--expansion, the best tactic is to 
perform 
the $\ep$--expansion of the innermost summand, given in terms of hypergeometric products, and to apply 
afterwards the summation quantifiers to the coefficients of the expansion; here one has to take care that the interchange of infinite summation quantifiers and the differential operator w.r.t.\ $\ep$ is possible. 
To accomplish this task, we will first explain how such products can be expanded in full generality. 
Afterwards we will focus on the task to carry out the summations on top of the $\ep$--expansion.

\subsection{The \boldmath $\ep$--expansion of the summand}\label{Sec:EpExpansion}

In general the summand is built by a product of the form 
\begin{equation}\label{Equ:HypProd}
\prod_{i=\ell}^n h(\ep,i),
\end{equation}
where $h(\ep,x)\in\KK(\ep,x)$ is a rational function in the variables $\ep$ and $x$, or by a linear combination of power products of such products; for concrete examples see~\eqref{eq:HORN2} and below. For simplicity we suppress further summation variables that may arise in $h$ and move them to the ground field (e.g., for the variables $m,n$ in~\eqref{eq:HORN2} we take the rational function field $\KK=K(m,n)$ over a field $K$ of characteristic $0$).

Before expanding in the dimensional parameter $\ep$ one may map to Pochhammer symbols. 
In such a representation $\ep$ occurs usually in the form
\begin{eqnarray}\label{Equ:PochhammerForm}
(a + r \ep)_n,~~~~\text{with}~~r \in \mathbb{Q}. 
\end{eqnarray}
The series expansion is then given in terms of harmonic sums \cite{Vermaseren:1998uu,Blumlein:1998if} 
at argument $a$ and $a+n$, with $a \in \mathbb{C} \backslash \mathbb{Z}_-$,
\begin{eqnarray}
(a + r \ep)_n &=& \frac{\Gamma(n+a)}{\Gamma(a)} =
(a)_n \Biggl\{ 1
+ r \ep \Biggl[
        \frac{n}{a (a
        +n
        )}
        -S_1(a)
        +S_1(a+n)
\Biggr]
+ r^2 \ep^2 \Biggl[
        -\frac{n}{a (a
        +n
        )^2}
\nonumber\\ &&
        +\Biggl(
                -\frac{n}{a (a
                +n)}
                -S_1(a+n)
        \Biggr) S_1(a)
        +\frac{1}{2} S_1^2(a)
        +\frac{n S_1(a+n)}{a (a
        +n
        )}
        +\frac{1}{2} \Biggl(S_1^2(a+n)
\nonumber\\ &&
        + S_2(a)
        - S_2(a+n) \Biggr)
\Biggr]
+r^3 \ep^3 \Biggl[
        \frac{n}{a (a
        +n
        )^3}
        +\Biggl(
                \frac{n}{a (a
                +n
                )^2}
                -\frac{n S_1(a+n)}{a (a
                +n
                )}
\nonumber\\ &&                
                -\frac{1}{2} \Biggl(S_1^2(a+n)
                + S_2(a)
- S_2(a+n)\Biggr)
        \Biggr) S_1(a)
        +\Biggl(
                \frac{n}{2 a (a
                +n
                )}
                +\frac{1}{2} S_1(a+n)
        \Biggr) S_1^2(a)
\nonumber\\ &&
        -\frac{1}{6} S_1^3(a)
        +\Biggl(
                -\frac{n}{a (a
                +n
                )^2}
                +\frac{1}{2} S_2(a)
                -\frac{1}{2} S_2(a+n)
        \Biggr) S_1(a+n)
        +\frac{n S_1^2(a+n)}{2 a (a
        +n
        )}
\nonumber\\ &&
        +\frac{1}{6} S_1^3(a+n)
        +\frac{n S_2(a)}{2 a (a
        +n
        )}
        -\frac{n S_2(a+n)}{2 a (a
        +n
        )}
        -\frac{1}{3} S_3(a)
        +\frac{1}{3} S_3(a+n)
\Biggr]
\Biggr\} + O(\ep^4).
\nonumber\\ \label{Equ:PochhammerExpansion}
\end{eqnarray}
Here the harmonic sums \cite{Vermaseren:1998uu,Blumlein:1998if}
are defined by
\begin{eqnarray}
S_{b,\vec{a}}(N) = \sum_{k=1}^N \frac{({\rm sign}(b))^k}{k^{|b|}} 
S_{\vec{a}}(k),~~~S_\emptyset = 1,~~ a_i,b \in \mathbb{N} \backslash 
\{0\}. 
\end{eqnarray}

Analogous expressions are obtained in the case that the Pochhammer symbols depend on $\ep$ polynomially.
The harmonic sums $S_{\vec{c}}(a;n)$
will be called {\it Hurwitz harmonic sums}, since they converge to the Hurwitz 
$\zeta$-values \cite{HURWITZ} in the limit $n \rightarrow \infty$. 

These sums are defined by 
\begin{eqnarray}
S_1(a;n) &=& \sum_{k=1}^n \frac{1}{a+k}
\\
S_{c,\vec{b}}(a;n) &=& \sum_{k=1}^n \frac{({\rm sign}(c))^k}{(a+k)^{|c|}} S_{\vec{b}}(a;k).
\end{eqnarray}
Single Hurwitz harmonic sums are given by
\begin{eqnarray}
S_l(a;n) \equiv S_l(a+n) -  S_l(a),~~a \in \mathbb{C}.
\end{eqnarray}
Here the harmonic sums are understood as derived from their Mellin transformation,
cf.~Ref.~\cite{Blumlein:1998if}. More involved relations of this kind hold also for
nested sums.
In course of further summations in the multivariate case also the Hurwitz generalizations of the sum having been dealt with 
in Refs.~\cite{Ablinger:2011te,
Ablinger:2013cf,Ablinger:2014bra,Ablinger:2021fnc} can occur.

If the multiplicands $h(i,\ep)$ of the arsing products~\eqref{Equ:HypProd} 
do not factorize linearly over the given field, one has to introduce algebraic extensions, such as given in~\eqref{Equ:AlgebraicExt}, in order to obtain the product representations as given in~\eqref{Equ:PochhammerForm}. 
In the case that one wants to avoid such non--trivial field extensions 
(which 
are often hard to handle with symbolical tools), we propose the following general and rather flexible method.

Let $\ell$ be an integer and suppose that $f_i(\ep)$ are functions in $\ep$ which are nonzero and complex differentiable around $0$ (and thus infinitely many times complex differentiable) for all integers $i\geq\ell$. By the product rule (and the quotient rule for the second identity) it follows that 
\begin{align}
\partial_{\ep}\prod_{i=\ell}^n f_i(\ep)&=\left(\prod_{i=\ell}^n f_i(\ep)\right)\sum_{i=\ell}^n\frac{\partial_{\ep}f_i(\ep)}{f_i(\ep)}\label{Equ:StandardRule},\\
\partial_{\ep}\prod_{i=\ell}^n \frac1{f_i(\ep)}&=-\left(\prod_{i=\ell}^n \frac1{f_i(\ep)}\right)\sum_{i=\ell}^n \frac{\partial_{\ep}f_i(\ep)}{f_i(\ep)}\label{Equ:InverseRule}
\end{align}
holds for all $n\geq\ell$. 

Since $f_i(\ep)$ for $i\geq\ell$ is infinitely many times differentiable around $0$, also the summand
$\frac{\partial_{\ep}f_i(\ep)}{f_i(\ep)}$ and thus the finite sums in~\eqref{Equ:StandardRule} and~\eqref{Equ:InverseRule} are infinitely many times differentiable around $0$. E.g., we get 
$$\partial_{\ep}\sum_{i=\ell}^n\frac{\partial_{\ep}f_i(\ep)}{f_i(\ep)}=
\sum_{i=\ell}^n\frac{f_i(\ep)\partial_{\ep}^2f_i(\ep)-f_i(\ep)\partial_{\ep}f_i(\ep)}{f_i(\ep)^2}.$$
As a consequence, we can apply $D^r_{\ep}$ iteratively on $\prod_{i=\ell}^n f_i(\ep)$ and obtain an explicit expression $F_r(\ep)$ given by the product $\prod_{i=\ell}^n f_i(\ep)$ itself times a polynomial expression in terms of sums where the denominator is of the form $f_i(n)^r$ and the numerator is built by a linear combination of power products of the form $f_i(\ep)^{e_0}(\partial f_i(\ep))^{e_1}\dots(\partial^r f_i(\ep))^{e_r}$ with $e_1+\dots+e_r=r$ and $e_1<r$. 

To calculate $\ep$--expansions for such products, we assume from now on in addition that $f_i(0)\neq0$ 
holds for all $i\geq\ell$. Then 
$$F_r(0)=\frac{\partial^r_{\ep}f_i(\ep)}{f_i(\ep)}\Bigg|_{\ep=0}$$
is well defined and by Taylor's formula we get the power series expansion
$$\prod_{i=\ell}^n f_i(\ep)=F_u(0)\ep^u+F_{u+1}(0)\ep^{u+1}+F_2(0)\ep^{u+2}+\dots$$
with order $u\geq0$ where $F_u(0)\neq0$.

Within the package \texttt{EvaluateMultiSums} we specialized this general mechanism to the product case~\eqref{Equ:HypProd}, i.e., we assume that $f_i(\ep)=h(\ep,x)$ where $h(\ep,x)\in\KK(\ep,i)$ is a rational function in the variables $\ep$ and $x$. If $h(0,i)$ is zero for some $i\geq\ell$, we take $\ell'\geq\ell$ as the minimal value such that this is not the case and extract the critical part with
$$\prod_{i=\ell}^n f_i(\ep)=r(\ep)\prod_{i=\ell'}^n h(\ep,i)$$
where $r(\ep)=\prod_{i=\ell}^{\ell'-1}h(\ep,i)\in\KK(\ep)$ is a rational 
function in $\ep$. We may assume that $r(\ep)=\ep^s\frac{p(\ep)}{q(\ep)}$ for an integer $s$ where $p,q$ are coprime polynomials in $\ep$ with $p(0)q(0)\neq0$.
Applying now the above machinery to $\prod_{i=\ell'}^n h(\ep,i)$ leads to a power series expansion of order $u\geq0$ as stated above. In addition, we can compute a Laurent series expansion of $r$ in $\ep$ of order $s$. Thus by the Cauchy product we end up at the Laurent series expansion
$$\prod_{i=\ell}^n h(\ep,i)=H_{t}\ep^t+H_{t+1}\ep^{t+1}+H_{t+2}\ep^{t+2}+\dots$$
of order $t=u+s$. Here each coefficient $H_r$ is given by $\prod_{i=\ell'}^n h(0,i)$ times a polynomial expression in terms of single sums $\sum_{i=\ell'}^n \frac{a(i)}{h(0,i)^r}$ where $a(i)$ is built by a linear combinations of power products of the form $h(\ep,i)^{e_0}(\partial h(\ep,i))^{e_1}\dots(\partial^r h(\ep,i))^{e_r}|_{\ep=0}$ with $e_1+\dots+e_r=r$ and $e_1<r$.

In order to obtain a nicer output, we factorize the input multiplicand 
$h(\ep,i)$ over the fixed field $\KK$ and pull over the product sign to each irreducible factor. In addition, we replace any product $\big(\prod_{i=\ell'}^n p(\ep,i)\big)^{-z}$ with $z>0$ to $\big(\prod_{i=\ell'}^n\frac1{p(\ep,i)}\big)^{z}$ with positive exponents $z$ and use~\eqref{Equ:InverseRule} instead of~\eqref{Equ:StandardRule}. Carrying out the expansions for each product and combining them by the Cauchy product yield an expression in terms of the input product $\prod_{i=\ell}^n h(0,i)$ (or a power product built by the irreducible parts  $\big(\prod_{i=\ell'}^n p(0,i)\big)^{-z}$) times a polynomial expressions of sums of the form $\sum_{i=\ell'}^n \frac{a(i)}{p(0,i)^r}$ where $a(x)\in\KK[x]$ is a polynomial of degree smaller than the degree of the polynomial $p(0,x)^r$.

The following remarks should be stated. First, applying this general method to $(a + r 
\ep)_n=\prod_{i=1}^np(\ep,i)$ with $p(\ep,x)=(-1 + a + \ep r + x)$ we rediscover precisely the 
$\ep$--expansion given in~\eqref{Equ:PochhammerExpansion}. Second, the polynomial $p(0,i)$ may be reducible and thus the denominators in the sums can be split further. In this case the routines of package \texttt{EvaluateMultiSums} (using the summation tools of \texttt{Sigma}) split the sums automatically further. E.g., calling the command \texttt{SeriesForProduct[SigmaProduct[}$2 \ep + 2 i + \ep i + 3 i^2 + 6 \ep i^2 + i^3 + \ep i^3$\texttt{,\{i,1,n\}],\{$\ep$,0,2\},\{n\}]} of \texttt{EvaluateMultiSums} yields the $\ep$--expansion
\begin{align*}
\prod_{i=1}^n h(\ep,i)=&n!^3\Big\{\frac{\ep^0}{2} (1+n)^2 (2+n)\\
&\hspace*{0.3cm}+\frac{\ep^1}{2}(1+n) \big(
n \big(
-6-3 n+n^2\big)
+3 (1+n) (2+n) S_1({n})
\big)\\
&\hspace*{0.3cm}+\frac{\ep^2}{4} \Big(
n \big(
498+597 n+185 n^2-9 n^3+n^4\big)
+6 (1+n) \big(
-2-9 n-4 n^2+n^3\big) S_1({n})\\
&\quad\quad+9 (1+n)^2 (2+n) S_1({n})^2
-101 (1+n)^2 (2+n) S_2({n})
\Big) \Big\}+O(\ep^3)
\end{align*}
for the irreducible polynomial $h(\ep,x)=2 \ep + 2 x + \ep x + 3 x^2 + 6 \ep x^2 + x^3 + \ep x^3\in\mathbb Q[\ep,x]$ which factorizes linearly to  $h(0,x)=x(x+1)(x+2)$ when it is evaluated at $\ep=0$. 

However, for a generic irreducible polynomial $p(\ep,x)$, also $p(0,x)$ is irreducible. For instance, consider the product expression
\begin{equation}\label{Equ:NonTrivialSummand}
f[\ep,n]=\frac{\prod_{i=1}^n \big(
	2-\ep
	+B_1
	-C
	-3 i
	-B_1 i
	+i^2
	\big)}{n! (A_1-4\ep)_n},
\end{equation}
where $f[0,n]$ equals to the product given in~\eqref{eq:T1}. Then applying the command \texttt{SeriesForProduct} to this expression gives
\begin{equation}\label{Equ:ExpandPrductExpr}
\begin{split}
f[\ep,n]=&\frac{\prod_{i=1}^n \big(
	2
	+B_1
	-C
	-3 i
	-B_1 i
	+i^2
	\big)}{n! (A_1)_n}\times\\
&\times\Big\{\ep^0+\ep^1 \Big(4\big(
\frac{n}{A (A
	+n
	)}
-S_1({A})
+S_1({A+n})
+
\sum_{i=1}^n \tfrac{1}{2
	+B
	-C
	-(3+B) i
	+i^2
}\Big)
\Big\}+O(\ep^2).
\end{split}
\end{equation}
If necessary or appropriate, depending on the application, the found sum 
solutions (and the products) can be factorized further (within an 
appropriate algebraic field extension).

\subsection{Symbolic summation}

We consider now multi--sums over such products (e.g., a single sum over the discrete parameter $n$, or sums 
over further discrete parameters that appear in other products or even inside of products). Applying the package \texttt{EvaluateMultiSum} one can now try to work from the inner sum towards the outermost sum and to transform the definite sums stepwise to indefinite nested versions. 
Internally, one computes stepwise recurrences and tries to solve these recurrences within the class of indefinite nested sums; for details see~\cite{SIG2}. 
In the case that a sum has an infinite upper bound, one first considers a 
truncated version with the upper bound $N$, applies the symbolic summation tools to this version and performs afterwards the limit $N \rightarrow \infty$ using
procedures available in the package {\tt HarmonicSums} 
\cite{HARMSU,Vermaseren:1998uu,Blumlein:1998if,Remiddi:1999ew,Ablinger:2011te,
	Ablinger:2013cf,Ablinger:2014bra,Ablinger:2021fnc}. 
Since in our application also the formal parameters $x_i$ are involved, it may
turn out in course of the summation that the summation problem cannot be solved for certain classes of cases, while it is 
possible in others. In particular, if the $\ep$ parameter appears in the innermost summand, it is of great 
advantage to first expand in $\ep$  and to apply afterwards the summation tools to the coefficients of the expansion which are free of $\ep$. To carry out the infinite sums after the $\ep$--expansion, the infinite power series have to be considered
in their convergence region around zero to perform the infinite sums, see {\tt converg.m} for the cases up to three variables.

For instance, we take the summand in~\eqref{Equ:NonTrivialSummand} and specialize it further to $A\to3, B\to-2, C\to-1$. Then with the 
expansion~\eqref{Equ:ExpandPrductExpr} we obtain
$$\sum_{n=0}^{\infty}\frac{\prod_{i=1}^n(\ep+1-i+i^2)}{n!(3-4\ep)_n}=G_0+\ep\,G_1+O(\ep^2),$$
with
\begin{align*}
G_0&=\sum_{n=0}^{\infty}\frac{\prod_{i=1}^n(1-i+i^2)}{n!(3)_n}\\
G_1&=\sum_{n=0}^{\infty}\frac{\prod_{i=1}^n(1-i+i^2)}{n!(3)_n}\Big(-6
+\frac{4}{1+n}
+\frac{4}{2+n}
+4 S_1({n})+\sum_{i=1}^n \frac{1}{1-i+i^2}\Big).
\end{align*}
Given this expansion, we apply our summation tools to the $\ep$-free sums in the second step. For $G_0$ we consider first the truncated version and get the simplification
$$\sum_{n=0}^{N}\frac{\prod_{i=1}^n(1-i+i^2)}{n!(3)_n}=\frac{(3+N) \big(
	1+N+N^2\big)}{3}\frac{\prod_{i=1}^N \big(
	1-i+i^2\big)}{N! (3)_N}.$$
Finally, we perform $N\to\infty$ yielding
$$G_0=\frac{2{\rm cosh}(\tfrac {\sqrt {3}\pi} {2})}{3\pi}.$$
The sum $G_1$ is more complicated and the command \texttt{EvaluateMultiSum[$G_1$]} produces the output
$$G_1= -\frac{8}{3}
+\frac{8 C}{3}
-\frac{20 \cosh \big(
	\frac{\sqrt{3} \pi }{2}\big)}{9 \pi }
+\frac{2 \cosh \big(
	\frac{\sqrt{3} \pi }{2}\big)}{3 \pi}\sum_{i=1}^{\infty} 
\frac{1}{1-i+i^2},$$
with the extra constant 
\begin{eqnarray}
C = \sum_{k=1}^\infty \left(\frac{1}{\pi k} {\rm cosh}\left[\frac{\sqrt{3} \pi}{2}\right] - \frac{1}{(k!)^2}
\prod_{l=1}^k (1 - l + l^2)\right).
\label{eq:C1}
\end{eqnarray}
As will be shown by non--trivial considerations in Appendix~\ref{sec:E} 
this convergent sum can be simplified to 
\begin{eqnarray}
C &=& 1 + \frac{2 \cosh \left[\frac{\sqrt{3} \pi }{2}\right]
	\left\{
	\Re\left[\psi
	\left(\frac{1}{2}+\frac{i
		\sqrt{3}}{2}\right)\right] + \gamma_E \right\}}{\pi },
\label{eq:C3}
\end{eqnarray}
with $\gamma_E$ the Euler--Mascheroni constant and $\psi(x)$ the digamma 
function. 

If this transformation works successfully (in particular, 
if the recurrences arising within the course of the transformation can
be fully solved within the class of indefinite nested sums defined over
hypergeometric products), 
one obtains finally an expression in terms of special functions $f(x_1, ..., x_n)$, which are the results 
of the $\ep$--expansion 
of the respective higher transcendental function. In this process one tries to keep the parameters symbolically and one finally 
inserts the respective function of the parameters of the original differential equations. This will in general 
lead to representations in radicals. For numerical representations this is not problematic, while the analytic 
representations are involved. Calculating the respective amplitudes for off--shell invariants one may use 
these quantities
in principle in higher loop diagrams by observing the respective kinematics. Whether this will be a practical method compared
to the direct calculation of the higher loop diagrams has to be seen in the respective cases.

Summarizing, the $\ep$--expansion leads to (multiple) infinite sums which can be simplified further by 
symbolic summation in many non--trivial applications. These are functions of the corresponding
set of variables, either in terms of functions which also appear in other quantum-fieldtheoretic calculations 
\cite{Remiddi:1999ew,Ablinger:2011te,Ablinger:2013cf,Ablinger:2014bra} or higher transcendental functions. Frequently the 
different letters appear within root--valued expressions.  

In the examples that we will present in Section~\ref{sec:fullmachinery} below or in the {\tt Mathematica} notebooks attached one obtains e.g.\ the following sums
\begin{eqnarray}
s_1 &=& \sum_{i=1}^{\infty} \frac{y^{i} \big(
	\frac{3}{2}\big)_{i}}{i! i}
\\
s_2 &=& 
\sum_{i=1}^{\infty } \frac{x^{2 i} \big(\frac{3}{2}\big)_{i}}{i!}
	\displaystyle\sum_{j=1}^{i} \frac{y^{-j} j!}{\big(
		\frac{3}{2}\big)_{j}}
\end{eqnarray}
and much more complicated structures and variables, as shown in the attachment in several examples. 
The above sums evaluate to
\begin{eqnarray}
s_1 &=& 
-2
+2 \frac{1}{\sqrt{1-y}}
-2 \ln \left[
\frac{1}{2} \big(
1+\sqrt{1
	-y
}\big)\right]
\\
s_2 &=& 
-\frac{y}{(1-x^2) (x^2-y)}
-\frac{1}{(1-x^2)^{3/2}}
-\frac{y}{(1-x^2)^{3/2} \sqrt{1-y}}
\Biggl[{\rm arctanh}\left(
\frac{1}{\sqrt{1-y}}\right) 
\nonumber\\ &&
+ {\rm arctanh}\left(
\frac{\sqrt{1-x^2}}{\sqrt{1-y}}\right) \Biggr].
\end{eqnarray}
In general one has to introduce integral representations successively as has been described in Ref.~\cite{Ablinger:2014bra}
in detail.

\section{The full machinery}
\label{sec:fullmachinery}

In the following we consider two--variable examples starting from its partial differential equation down to 
its infinite sum representation and $\ep$--expansion to illustrate the principle formalism.
\subsection{Example 1}

\vspace*{1mm}
\noindent
Consider for example the system of equations
\begin{eqnarray}
	\Bigg[ (x-1) y \partial_{x,y}^2+ \Big[x \Big(2 \varepsilon +\frac{7}{2}\Big)-\varepsilon +1\Big]\partial_x + (x-1) x \partial_x^2
&\nonumber\\
	+y (2 \varepsilon +1) \partial_y +\frac{3}{2} (2 \varepsilon +1) \Bigg] f(x,y) &= 0, \\
	\Bigg[ x (y-1) \partial_{x,y}^2 +x (4-\varepsilon ) \partial_x+ \Big[y \Big(\frac{13}{2}-\varepsilon \Big)-\varepsilon +1\Big]\partial_y 
	&\nonumber\\
	+(y-1) y \partial_y^2+\frac{3 (4-\varepsilon )}{2}
	\Bigg]f(x,y) &= 0,
\end{eqnarray}
for which we search for a solution of the form~\eqref{eq:hyp-series} with $r=2$ where $x_1=x$ and $x_2=y$. 
Computing a first--order recurrence system of $A(n_1,n_2)=A(m,n)$ and solving it by the method presented in Section~\ref{sec:solveProd} provides the solution
\begin{equation}
	f(x,y) = \sum_{m,n= 0}^\infty A(m,n) = \sum_{m,n= 0}^\infty \frac{x^m y^n \big(
        \frac{3}{2}\big)_{m
+n
} (4-\varepsilon )_n (1+2 \varepsilon )_m}{m! n! (-1+\varepsilon )_{m
+n
}}.
\label{eq:example-summand}
\end{equation}
A series expansion of the summand $A(m,n)$ in \eqref{eq:example-summand} up to $\mathcal O(\varepsilon^0)$ gives
\begin{eqnarray}
A(m,n)&=&	-\frac{1}{6} \frac{x^m y^n (3+n)! \big(
        \frac{3}{2}\big)_{m
+n
}}{n! (-2
+m
+n
)! \varepsilon}
+
\frac{1}{36} \bigg[
        -\frac{1}{(1+n) (2+n) (3+n) (m
        +n
        ) (-1
        +m
        +n
        )} 
\nonumber\\&& \times        
        \big(
                -36
                -30 n
                +17 n^2
                +97 n^3
                +79 n^4
                +17 n^5
                +m^2 \big(
                        36+115 n+84 n^2+17 n^3\big)
\nonumber\\&&                
                +m \big(
                        36+89 n
                        +218 n^2
                        +163 n^3+34 n^4\big)
        \big)
        -12 S_1({m})
        +6 S_1({n})
        +6 S_1({m+n})
\bigg] 
\nonumber\\&& \times
\frac{x^m y^n (3+n)! \big(
        \frac{3}{2}\big)_{m
+n
}}{n! (-2
+m
+n
)!}
+\mathcal O(\varepsilon).
\end{eqnarray}
A series expansion of \eqref{eq:example-summand} in the region $0<x<\sqrt{y}$, $0<y<\frac{1}{2}$,
\begin{equation}
	f(x,y) = \frac{1}{\varepsilon} f_{-1}(x,y) + f_0(x,y) +\mathcal O(\varepsilon)
\end{equation}
 is possible using {\tt EvaluateMultiSum} and results in an expression involving the sums
\begin{eqnarray}
	R_0 &=& \sum_{i=1}^{\infty } \frac{x^i \big(\frac{3}{2}\big)_i}{i!} = -1+\frac{1}{(1-x)^{3/2}}
\\
	R_1 &=& \sum_{i=1}^{\infty } \frac{y^i \big(\frac{3}{2}\big)_i}{i!} = -1+\frac{1}{(1-y)^{3/2}}
\end{eqnarray}
at $\mathcal O(\varepsilon^{-1})$. The function $f_{-1}(x,y)$ reads
\begin{eqnarray}
	f_{-1}(x,y) &=& -\frac{15 x^6 }{4 (x-y)^4(1-x)^{7/2}}
-\frac{15 y^3}{64 (x-y)^4 (1-y)^{13/2}} \big[
        y^3 \big(
                160+80 y-10 y^2+y^3\big)
\nonumber\\&&                                
        -x y^2 \big(
                576+176 y
                -64 y^2+5 y^3\big)
        +x^3 \big(
                -320+120 y-36 y^2+5 y^3\big)
\nonumber\\&&
        +3 x^2 y \big(
                240+8 y-22 y^2+5 y^3\big)
\big] .
\end{eqnarray}
In addition, one encounters at $\mathcal O(\varepsilon^0)$ the sums 
\begin{eqnarray}
R_2 &=& \sum_{i=1}^{\infty } \frac{x^i \big(
        \frac{3}{2}\big)_i}{(1+2 i)^2 i!} = -1 + \frac{1}{\sqrt{x}} \arcsin\big( \sqrt{x}\big)
\\ 
R_3 &=& \sum_{i=1}^{\infty } \frac{y^i \big(
        \frac{3}{2}\big)_i}{i i!} = -2
+2 \ln(2)
+2 \frac{1}{\sqrt{1-y}}
-2 H_{-1}\big(
        \sqrt{1-y}\big)
\\ 
R_4 &=& \sum_{i_1=1}^{\infty } \frac{x^{i_1} \big(\frac{3}{2}\big)_{i_1}
}{i_1!} 
        \sum_{i_2=1}^{i_1} \frac{1}{1+2 i_2}
= \frac{1}{2} \frac{H_1(x)}{(1-x)^{3/2}}
\\ 
R_5 &=& \sum_{i_1=1}^{\infty } \frac{x^{i_1} 
\big(\frac{3}{2}\big)_{i_1}}{i_1!} 
        \sum_{i_2=1}^{i_1} \frac{y^{-i_2} i_2!}{\big(
                \frac{3}{2}\big)_{i_2}}
\nonumber\\&=&
\frac{y}{(1-x) (y-x)}
-\frac{1}{(1-x)^{3/2}}
+\frac{y}{2(1-x)^{3/2} \sqrt{1-y}} \Big[
        i \pi 
        -H_0\big( \sqrt{1-y} -\sqrt{1-x} \big)
\nonumber\\&&
        -2 H_{-1}\big( \sqrt{1-y}\big)
        +H_0(y)
        +H_0\big( \sqrt{1-y}+\sqrt{1-x} \big)
\Big] 
\\ 
R_6 &=& \sum_{i_1=1}^{\infty } \frac{x^{i_1} y^{i_1}
\big(\big(
                \frac{3}{2}\big)_{i_1}\big)^2}{\big(
        i_1!\big)^2}  
        \sum_{i_2=1}^{i_1} \frac{y^{-i_2} i_2!}{\big(
                \frac{3}{2}\big)_{i_2}}
\nonumber\\&=&
\frac{1}{2}\int_0^1 dt \bigg[
\frac{-1+t}{\pi  (-1
+t
+y
)} \Big[
        \frac{1}{(1
        -(1-t) x
        )^2}
         \big[
                4 E(x -t x)
                -2 [1
                -(1-t) x
                ] K(x-t x)
        \big]
\nonumber\\&&
        -\frac{4 E(x y)
        +2 (-1
        +x y
        ) K(x y)
        }{(-1
        +x y
        )^2}
\Big] \frac{1}{\sqrt{t}}
\bigg]
\\ 
R_7 &=& \sum_{i_1=1}^{\infty } \frac{x^{i_1} y^{i_1} 
\big(\big(
                \frac{3}{2}\big)_{i_1}\big)^2}{\big(
        i_1!\big)^2 \big(
        1+2 i_1\big)^2} 
        \sum_{i_2=1}^{i_1} \frac{y^{-i_2} i_2!}{\big(
                \frac{3}{2}\big)_{i_2}}
=
{ \frac{1}{\pi} } \int_0^1 dt \frac{t-1}{\sqrt{t}(t+y-1) } \big[ K(x(1-t))
-K(x y) \big]
\\ 
R_8 &=& \sum_{i_1=1}^{\infty } \frac{y^{i_1} 
\big(\frac{3}{2}\big)_{i_1}}{i_1! \big(
        1+2 i_1\big)} 
        \sum_{i_2=1}^{i_1} \frac{x^{i_2} y^{-i_2}}{i_2}
=
-2 \frac{H_0\big(
        \sqrt{1-x}+\sqrt{1-y}\big)}{\sqrt{1-y}}
+2 \frac{H_{-1}\big(
        \sqrt{1-y}\big)}{\sqrt{1-y}}
\\ 
R_9 &=& \sum_{i_1=1}^{\infty } \frac{x^{i_1} 
\big(\frac{3}{2}\big)_{i_1}}{i_1!} \big(
        \sum_{i_2=1}^{i_1} \frac{y^{-i_2} i_2!}{\big(
                \frac{3}{2}\big)_{i_2}}
\big)
\big(
        \sum_{i_2=1}^{i_1} \frac{y^{i_2} \big(
                \frac{3}{2}\big)_{i_2}}{i_2!}
\big)
\nonumber\\&=&
\frac{1}{(1-y)^{5/2}} \Big[ \left(\frac{1}{(1-x)^{3/2}}-1\right) \left((1-y)^{3/2}-1\right) 
\biggl(\sqrt{1-y} y
   \Big(H_{-1}\big(\sqrt{1-y}\big)
\nonumber\\&&   
   -\frac{H_0(y)}{2}+\frac{i \pi }{2}\Big)-y+1\biggr) \Big]
+\sum_{i_ 1=1}^{\infty } \bigg\{
        \frac{1}{\pi  
(1-y)^2 \sqrt{1- y} \Gamma \big(
                1+i_ 1\big) \Gamma \big(
                2+i_ 1\big)} 
\nonumber\\&& \times
                \bigg[ 4 x^{i_ 1} y^{1+i_ 1} \Big[
                1
                -y
                +\sqrt{1-y} y \Big(
                        \frac{i \pi }{2}
                        -\frac{1}{2} H_ 0(y)
                        +H_ {-1}\big(
                                \sqrt{1-y}\big)
                \Big)
        \Big] \Gamma \big(
                \frac{3}{2}+i_ 1\big) 
\nonumber\\&& \times                
                \Gamma \big(
                \frac{5}{2}+i_ 1\big)
                 \,_ 2F_ 1\Big(
                -\frac{1}{2},1+i_ 1;2+i_ 1;y\Big) \bigg]
        -\frac{1}{(-1+y) 
\sqrt{\pi -\pi  y} \Gamma \big(
                1+i_ 1\big)}
\nonumber\\&& \times                                
                 \Big[2 x^{i_ 1} \Gamma \big(
                \frac{3}{2}+i_ 1\big) 
                \,_ 2F_ 1\Big(
                -\frac{1}{2},1+i_ 1;2+i_ 1;y\Big) 
                \,_ 2F_ 1\Big(
                1,2+i_ 1;\frac{5}{2}+i_ 1;\frac{1}{y}\Big) \Big]
\nonumber\\&&
        -\frac{1}{(-1+y) \sqrt{\pi -\pi  y} \big(
                3+2 i_ 1\big)}           
                \Big[ 2 \sqrt{\pi } x^{i_ 1} y^{-1-i_ 1} \big(
                -1+(1
                -y
                )^{3/2}\big)
\nonumber\\&& \times                
                 \,_ 2F_ 1\Big(
                1,2+i
                _ 1;\frac{5}{2}+i_ 1;\frac{1}{y}
        \Big)
\big(1+i_ 1\big) \Big] 
\bigg\}
\\ 
R_{10} &=& \sum_{i_1=1}^{\infty } \frac{y^{i_1} \big(
        \frac{3}{2}\big)_{i_1} S_1\big({i_1}\big) i_1}{i_1!} 
=
-3 \ln(2) y \frac{1}{(1-y)^{5/2}}
+\frac{3}{2} y \frac{H_1(y)}{(1-y)^{5/2}}
\nonumber\\&&
+3 y \frac{H_{-1}\big(
        \sqrt{1-y}\big)}{(1-y)^{5/2}}
+\Big[
        1
        +y \big(
                3-2 \sqrt{1
                -y
                }\big)
        -\sqrt{1-y}
\Big] \frac{1}{(1-y)^{5/2}},
\end{eqnarray}
as well as the combination
\begin{eqnarray}
R_{11} &=&\big( 1-\big(1-x\big)^{3/2} \big)
        \sum_{i_1=1}^{\infty } \frac{y^{i_1} 
\big(\frac{3}{2}\big)_{i_1}}{i_1! \big(
                1+2 i_1\big)}
                \sum_{i_2=1}^{i_1} \frac{y^{-i_2} i_2!}{\big(
                        \frac{3}{2}\big)_{i_2}}
\nonumber\\&&
        -\big(
                1-x\big)^{3/2} 
        \sum_{i_1=1}^{\infty } \frac{y^{i_1} 
\big(\frac{3}{2}\big)_{i_1}}{i_1! \big(
                1+2 i_1\big)}\Big(
                \sum_{i_2=1}^{i_1} \frac{y^{-i_2} i_2!}{\big(
                        \frac{3}{2}\big)_{i_2}}
        \Big)
\Big(
                \sum_{i_2=1}^{i_1} \frac{x^{i_2} \big(
                        \frac{3}{2}\big)_{i_2}}{i_2!}
        \Big)
\nonumber\\&=&
\frac{1}{4} \bigg\{
        \Big[
                2
                -2 \sqrt{1-x}
                +2 x \sqrt{1-x}
                -3 x F_1\Big(
                        \frac{5}{2};\frac{1}{2},1;2;x y,x
                \Big)
\big(1-x\big)^{3/2}
        \Big]
\big[i \pi  y
\nonumber\\&&
                +2 \sqrt{1-y}                
                -y H_0(y)
                +2 y H_{-1} \big(\sqrt{1-y}\big)
        \big]\bigg\} \frac{1}{\sqrt{1-y}}
\nonumber\\&&
-
\sum_{i_1=1}^{\infty } \Big[ \frac{x^{1+ i_1} \big(
        1-x\big)^{3/2} \Gamma \big(
        \frac{1}{2}+i_1\big)  }{\sqrt{\pi } y \Gamma \big(
        1+i_1\big)}     
        \,_2F_1\Big(
        1,2+i_1;\frac{5}{2}+i_1;\frac{1}{y}\Big)
\nonumber\\&& \times
        \,_2F_1\Big(
        1,\frac{5}{2}+i_1;2+i_1;x\Big) \Big].
\end{eqnarray}
The harmonic polylogarithms \cite{Remiddi:1999ew} are defined by
\begin{eqnarray}
H_{b,\vec{a}}(x) &=& \int_0^x dy f_b(y) H_{\vec{a}}(y),~~~H_\emptyset = 
1,~~~b, a_i \in \{0,-1,1\},
\end{eqnarray}
with
\begin{eqnarray}
f_0(x) = \frac{1}{x},~~
f_{-1}(x) = \frac{1}{1+x},~~
f_1(x) = \frac{1}{1-x}.
\end{eqnarray}

One can further employ the relations
\begin{eqnarray}
_2F_1\Big(\frac{3}{2},\frac{3}{2};1,z\Big) &=& \frac{2(z-1)K(z)+4E(z)}{\pi(z-1)^2}
\\
K(z) &=& \int_0^1 \frac{1}{\sqrt{(1-t^2)(1-z t^2)}} dt = \frac{\pi}{2} \,_2F_1\Bigr(\frac{1}{2},\frac{1}{2};1;z  \Bigl)
\\
E(z) &=& \int_0^1 \frac{\sqrt{1-z t^2}}{\sqrt{1-t^2}} dt = \frac{\pi}{2} \,_2F_1\Bigl(-\frac{1}{2},\frac{1}{2},1,z  \Bigr).
\end{eqnarray}
The function $f_0(x,y)$ reads
{
\begin{eqnarray}
f_0(x,y) &=& 
\frac{5 R_{10} y^2}{16 (1-y)^4 (x
-y
)^4} \Bigl[
        y^3 \big(
                160+80 y-10 y^2+y^3\big)
        -x y^2 \big(
                576+176 y-64 y^2+5 y^3\big) 
\nonumber\\&&
        +x^3 \big(
                -320+120 y-36 y^2+5 y^3\big)
        +3 x^2 y \big(
                240+8 y-22 y^2+5 y^3\big)
\Bigr]
\nonumber\\&&
-\frac{15 R_8 y^3}{32 (1-y)^6 (x
-y
)^4} \Bigl[
        y^3 \big(
                160+80 y-10 y^2+y^3\big)
        -x y^2 \big(
                576+176 y-64 y^2+5 y^3\big)
\nonumber\\&&
        +x^3 \big(
                -320+120 y-36 y^2+5 y^3\big)
        +3 x^2 y \big(
                240+8 y-22 y^2+5 y^3\big)
\Bigr]
\nonumber\\&&
+\frac{1}{128 (1-x)^2 (1-y)^6 y (x
-y
)^4} \big(
        -960 x^7 (-1+y)^7
        +y^5 \big(
                128-1344 y+1536 y^2
\nonumber\\&&                
                -4240 y^3+4110 y^4-223 y^5+33 
y^6\big)
        +8 x^6 y \big(
                4-24 y+60 y^2-4880 y^3+1860 y^4
\nonumber\\&&                
                -564 y^5+79 y^6\big)
        +x^5 y \big(
                -832+2752 y-1920 y^2+76160 y^3+69280 y^4-8772 
y^5
\nonumber\\&&
+2427 y^6-495 y^7\big)
        -x y^4 \big(
                512-4544 y+3264 y^2-34480 y^3+4240 y^4+3363 y^5
\nonumber\\&&                
                -141 
y^6+66 y^7\big)
        +x^3 y^2 \big(
                -512+384 y+11008 y^2+83680 y^3+169980 y^4+6287 
y^5
\nonumber\\&&
+5931 y^6+747 y^7-305 y^8\big)
        +x^2 y^3 \big(
                768-4736 y-2272 y^2-84288 y^3-56570 y^4
\nonumber\\&&                
                +11627 
y^5-3549 y^6+387 y^7+33 y^8\big)
        +x^4 y \big(
                128+1984 y-9792 y^2-29440 y^3
\nonumber\\&&                
                -180320 y^4-63768 
y^5+10536 y^6-8583 y^7+2055 y^8\big)
\big)
\nonumber\\&&
+R_1
 \biggl[
        -
        \frac{1}{128 (1-x)^2 (1-y)^5 y (x
        -y
        )^5} \Bigl[
                960 x^7 (-1+y)^6
                -y^6 \big(
                        128+4800 y-4640 y^2
\nonumber\\&&                        
                        +4480 y^3-270 y^4+33 
y^5\big)
                +2 x y^5 \big(
                        320+10944 y-4016 y^2+4664 y^3+1749 y^4
\nonumber\\&&                        
                        -101 
y^5+33 y^6\big)
                +x^6 y \big(
                        -448+1920 y-12800 y^2+5440 y^3+1920 y^4-628 
y^5+65 y^6\big)
\nonumber\\&&
                -x^2 y^4 \big(
                        1280+38080 y+20128 y^2-6304 y^3+18546 
y^4-4204 y^5+406 y^6+33 y^7\big)
\nonumber\\&&
                +2 x^3 y^3 \big(
                        640+15040 y+32080 y^2-9896 y^3+11559 y^4-3791 
y^5-491 y^6+169 y^7\big)
\nonumber\\&&
                -5 x^4 y^2 \big(
                        128+1856 y+11712 y^2+1856 y^3-2076 y^4+1727 
y^5-2058 y^6+448 y^7\big)
\nonumber\\&&
                +2 x^5 y \big(
                        64+320 y+8000 y^2+15360 y^3-12080 y^4+5164 
y^5-4170 y^6+935 y^7\big)
        \Bigr]
\nonumber\\&&
        -\frac{15 R_5 x^6 (1-y)^2}{2 (1-x)^2 y (x
        -y
        )^4}
\biggr]
+R_0 \biggl[
        \frac{1}{16 (1-x)^2 (1-y)^6 y (x
        -y
        )^5} \Bigl[
                -120 x^8 (-1+y)^7
\nonumber\\&&
                +y^6 \big(
                        -32+384 y+2988 y^2+140 y^3-15 y^4\big)
                +5 x y^5 \big(
                        32-352 y-2676 y^2-1772 y^3
\nonumber\\&&                        
                        -95 y^4+12 y^5\big)
                +5 x^2 y^4 \big(
                        -64+608 y+4688 y^2+7516 y^3+1723 y^4+100 
y^5-18 y^6\big)
\nonumber\\&&
                +x^6 y \big(
                        20-624 y+3324 y^2+8040 y^3+18380 y^4-6720 
y^5+2164 y^6-329 y^7\big)
\nonumber\\&&
                +5 x^3 y^3 \big(
                        64-448 y-4056 y^2-12396 y^3-6809 y^4-612 
y^5-10 y^6+12 y^7\big)
\nonumber\\&&
                -5 x^4 y^2 \big(
                        32-64 y-1888 y^2-9120 y^3-11664 y^4-1436 
y^5-158 y^6+40 y^7+3 y^8\big)
\nonumber\\&&
                +x^5 y \big(
                        32+416 y-2848 y^2-12000 y^3-46800 y^4-11880 
y^5+430 y^6-200 y^7+85 y^8\big)
\nonumber\\&&
                +x^7 \big(
                        -60+304 y-444 y^2-360 y^3-2780 y^4-1080 
y^5+1536 y^6-701 y^7+120 y^8\big)
        \Bigr]
\nonumber\\&&
        +\frac{15 R_3
         x^6}{4 (1-x)^2 (x
        -y
        )^4}
        -\frac{15 R_1 x^6 (1-y)}{2 (1-x)^2 y (x
        -y
        )^4}
\biggr]
+\frac{15 R_9 x^6 (1-y)^2}{2 (1-x)^2 y (x
-y
)^4}
\nonumber\\&&
+\frac{15 R_6 x^6 (1-y) (-1
+(2+x) y
)}{4 (1-x)^2 y (x
-y
)^4}
+\frac{15 R_3 x^6}{4 (1-x)^2 (x
-y
)^4}
+\frac{15 R_2 x^6 (1-y)}{4 (1-x)^2 y (x
-y
)^4}
\nonumber\\&&
+\frac{15 R_4 x^6 (1-y)}{2 (1-x)^2 y (x
-y
)^4}
-\frac{15 R_7 x^6 (1-y)}{4 (1-x)^2 y (x
-y
)^4}
-\frac{15 R_{11} x^6 (1-y) }{2 y (x
-y)^4 (1-x)^{7/2}}.
\end{eqnarray}

}

\subsection{Example 2}
Consider for example the system of equations
\begin{eqnarray}
	1+\varepsilon +(2-x+\varepsilon ) \partial_x+2 x (1+x) 
\partial_x^2=0\\
	2-\varepsilon +(1-2 y+2 \varepsilon ) \partial_y+y (3+y) \partial_y^2=0
\end{eqnarray}
We can write its solution as
\begin{equation}
	\mathcal F(x,y) = \sum_{x,y\ge 0}A(m,n) x^m y^n
\end{equation}
with
\begin{equation}
	A(m,n) = \bigg(
        \prod_{i_1=1}^m \frac{-6
        -\varepsilon 
        +7 i_1
        -2 i_1^2
        }{\big(
                \varepsilon 
                +2 i_1
        \big) i_1}\bigg) \prod_{i_1=1}^n \frac{-6
+\varepsilon 
+5 i_1
-i_1^2
}{\big(
        -2
        +2 \varepsilon 
        +3 i_1
\big) i_1}.
\end{equation}
The quantity $A(m,n)$ can also be expressed as
\begin{eqnarray}
	A(m,n) &=& \frac{(-1)^m  \big(
        -\frac{3}{4}-\frac{1}{4} \sqrt{1
        -8 \varepsilon 
        }\big)_m \big(
        \frac{1}{4} \big(
                -3+\sqrt{1
                -8 \varepsilon 
                }\big)\big)_m }{\big(
        1+\frac{\varepsilon }{2}\big)_m  \Gamma (1+m) }
\nonumber\\&&\times
\frac{ (-1)^n 3^{-n} \big(
        -\frac{3}{2}-\frac{1}{2} \sqrt{1
        +4 \varepsilon 
        }\big)_n \big(
        \frac{1}{2} \big(
                -3+\sqrt{1
                +4 \varepsilon 
                }\big)\big)_n}{\big(
        \frac{1}{3}+\frac{2 \varepsilon }{3}\big)_n \Gamma (1+n)}
\end{eqnarray}
and $\mathcal F(x,y)$ can be rewritten as 
\begin{eqnarray}
	\mathcal F(x,y) &=& \Big( \sum_{m\ge 0} x^m f_1(m,\varepsilon) \Big) \Big( \sum_{n\ge 0}y^n f_2(n,\varepsilon) \Big).
	\\
	&=& F_1(x,\varepsilon) F_2(y,\varepsilon).
\end{eqnarray}
Expanding $F_1$ and $F_2$ in a series in $\varepsilon$ using {\tt EvaluateMultiSums}, one can write an expression containing infinite (nested) sums. These are rewritten as iterated integrals following \cite{Ablinger:2014bra}. Two of the sums are written in semi-analytic form as definite integrals by writing part of the summand as the Mellin transform of a function. For example, we encounter the sum
\begin{equation}
	s_1=\sum_{i=1}^{\infty } \frac{(-1)^i x^i \big(
        -\frac{3}{2}+i\big)! \big(
        \sum_{j=1}^i \frac{1}{1+2 j}\big) S_1({i})}{i i!}.
\end{equation}
By isolating the term $i=1$ and applying the Legendre duplication formula
\begin{equation}
	\Gamma\Big(z+\frac{1}{2}\Big) = \sqrt{\pi} \frac{\Gamma(2z)}{2^{2z-1} \Gamma(z)}
\end{equation}
and the identity 
\begin{equation}
	\Gamma(2z) = \frac{1}{2} \binom{2z}{z} \Gamma(z) \Gamma(z+1)
\end{equation}
we write
\begin{eqnarray}
	s_1 &=& -\frac{1}{3} x \sqrt{\pi }
+
\sum_{i=1}^{\infty } \frac{(-1)^{1+i} 2^{-2 i} \sqrt{\pi } x^{1+i} 
\binom{2 i}{i}}{(1+i)^3 (3+2 i)}
+
\sum_{i=1}^{\infty } \frac{(-1)^{1+i} 2^{-2 i} \sqrt{\pi } x^{1+i} 
\binom{2 i}{i} 
\sum_{j=1}^i \frac{1}{1+2 j}}{(1+i)^3}
\nonumber\\&&
+
\sum_{i=1}^{\infty } \frac{(-1)^{1+i} 2^{-2 i} \sqrt{\pi } x^{1+i} 
\binom{2 i}{i} S_1({i})}{(1+i)^2 (3+2 i)}
\nonumber\\&&
+
\sum_{i=1}^{\infty } \frac{(-1)^{1+i} 2^{-2 i} \sqrt{\pi } x^{1+i} 
\binom{2 i}{i} \big(
        \sum_{j=1}^i \frac{1}{1+2 j}\big) S_1({i})}{(1+i)^2}.
\end{eqnarray}
The first three sums are treated following \cite{Ablinger:2014bra}. The fourth sum can be written as
\begin{eqnarray}
	t_1 &=& \sum_{i=1}^{\infty } \frac{(-1)^{1+i} 2^{-2 i} \sqrt{\pi } x^{1+i} 
\binom{2 i}{i} \big(
        \sum_{j=1}^i \frac{1}{1+2 j}\big) S_1({i})}{(1+i)^2}
\nonumber\\&=&
        \sum_{i=1}^\infty \frac{(-1)^{1+i}2^{-2i}\sqrt{\pi} {x^{1+i}} \binom{2i}{i}}{(1+i)^2} \int_0^1 dz \bigg\{ (z^i-1) \frac{1}{2 (-1+z)} \bigg[
        -2
        +2 z
        +\big(
                1+\sqrt{z}\big) G\Big(
                \frac{\sqrt{\tau }}{1-\tau };z\Big)
\nonumber\\&&                
        +2 \sqrt{z} \big(1-\ln (2)\big)
\bigg] \bigg\}
\nonumber\\&=&
\int_0^1 dz \bigg\{ { \frac{1}{2(z-1)} } \bigg[
        -2
        +2 z
        +\big(
                1+\sqrt{z}\big) G\Big(
                \frac{\sqrt{\tau }}{1-\tau };z\Big)    
        +2 \sqrt{z} \big(1-\ln (2)\big)
\bigg]
\nonumber\\&& \times
\sum_{i=1}^\infty \frac{\big(
        -1+z^i\big) (-1)^{1+i} 2^{-2 i} \sqrt{\pi } x^{1+i} \binom{2 i}{i}}{(1+i)^2}
\bigg\}
\nonumber\\&=&
\int_0^1 dz \bigg\{ { \frac{1}{2(z-1)} }
\bigg[
        -2
        +2 z
        +\big(
                1+\sqrt{z}\big) G\Big(
                \frac{\sqrt{\tau }}{1-\tau };z\Big)    
        +2 \sqrt{z} \big(1-\ln (2)\big)
\bigg]
\bigg[
-\frac{4}{z} \Big[
        -1
        +z
\nonumber\\&&        
        +\sqrt{1
        +x z
        }
        -z \sqrt{1+x}
        +z H_0\Big(
                \frac{1}{2} \big(
                        1+\sqrt{1+x}\big)\Big)
        -H_0\big(
                \frac{1}{2} \Big(
                        1+\sqrt{1+x z}\big)\Big)
\Big] \sqrt{\pi }
\bigg]
\bigg\}.
\end{eqnarray}
The $\varepsilon$ expansion of $F_1(x,\varepsilon)$ and $F_2(x,\varepsilon)$ then can be written by
\begin{eqnarray}
	F_1(x,\varepsilon) &=& 1
-
\frac{x}{2}
+\varepsilon  \biggl\{
        -1
        +\sqrt{1+x}
        +\frac{1}{4} x \big(
                -9+4 \sqrt{1
                +x
                }\big)
        +\frac{1}{2} (-2+x) H_0(x)
\nonumber\\&&        
        +\frac{1}{2} (2-x)  G_{3}(x)
\biggr\}
+\varepsilon ^2 \Biggl\{
        \frac{1}{8} \Bigl[
                20 \big(
                        -1+\sqrt{1
                        +x
                        }\big)
                +x \big(
                        -33
                        +4 x
                        +20 \sqrt{1+x}
                \big)
        \Bigr]
\nonumber\\&&
        +(-2+x) \Bigl(
                -\frac{1}{4} H_0(x)^2
                +\frac{1}{4} H_{0,-1}(x)
                +\frac{ G_{11}(x)}{4}
                -\frac{ G_{12}(x)}{4}
                +\frac{ G_{5}(x)^2}{8}
        \Bigr)
\nonumber\\&&        
        +\frac{1}{2} (2-x) \bigl( G_{8}(x)
        + G_{9}(x)
        \bigr)
        +\frac{1}{4} (-4+13 x) H_0(x)
        +\frac{1}{2} (1+x)^{3/2} H_{-1}(x)
\nonumber\\&&        
        +\biggl[
                1
                -\frac{13 x}{4}
                +\frac{1}{2} (-2+x) H_0(x)
        \biggr]  G_{3}(x)
        +\biggl[
                \frac{1}{2} (1+x)^{3/2}
                +\frac{1}{4} (2-x) H_0(x)
        \biggr]  G_{5}(x)
\Biggr\}
\nonumber\\&&
+\varepsilon ^3 \Biggl\{
        \frac{ G_{5}(x)^2}{16} \Bigl[
                x \big(
                        17-4 \sqrt{1
                        +x
                        }\big)
                -4 \big(
                        3+\sqrt{1
                        +x
                        }\big)
        \Bigr]
        + G_{11}(x) \Bigl[
                \frac{1}{8} \big(
                        x \big(
                                13-6 \sqrt{1
                                +x
                                }\big)
\nonumber\\&&                                
                        -6 \sqrt{1+x}
                \big)
                +\frac{1}{8} (-2+x) H_0(x)
        \Bigr]
        + G_{12}(x) \biggl[
                \frac{1}{8} \Bigl[
                        2 \big(
                                6+\sqrt{1
                                +x
                                }\big)
                        +x \big(
                                -17+2 \sqrt{1
                                +x
                                }\big)
                \Bigr]
\nonumber\\&&                
                +\frac{3}{8} (-2+x) H_0(x)
        \biggr]
        + G_{3}(x) \biggl[
                \frac{9}{2}
                -\frac{105 x}{8}
                +\frac{1}{4} (-14+17 x) H_0(x)
                +\frac{1}{4} (2-x) H_0(x)^2
        \biggr]
\nonumber\\&&        
        + G_{5}(x) \Biggl[
                (-2+x) \Bigl(
                        -\frac{1}{16} H_0(x)^2
                        +\frac{1}{8} H_{0,-1}(x)
                        +\frac{3  G_{11}(x)}{8}
                        -\frac{ G_{12}(x)}{8}
                \Bigr)
\nonumber\\&&                
                +\frac{1}{4} \biggl[
                        -6 x
                        -2 x^2
                        +(1+x) (5+16 x) \sqrt{1+x}
                \biggr]
                +\frac{1}{8} \biggl[
                        x \big(
                                -13+6 \sqrt{1
                                +x
                                }\big)
\nonumber\\&&                                
                        +6 \sqrt{1+x}
                \biggr] H_0(x)
                -\frac{1}{4} (1+x)^{3/2} H_{-1}(x)
        \Biggr]
        +(-2+x) \big(
                \frac{1}{12} H_0(x)^3
                -\frac{3}{8} H_{0,0,-1}(x)
\nonumber\\&&                
                -\frac{1}{8} H_{0,-1,-1}(x)
                -\frac{ G_{18}(x)}{8}
                -\frac{5  G_{19}(x)}{8}
                -\frac{3  G_{20}(x)}{8}
                +\frac{ G_{21}(x)}{8}
                -\frac{3  G_{22}(x)}{4}
                +\frac{ G_{23}(x)}{4}
\nonumber\\&&                
                -\frac{1}{24}  G_{5}(x)^3
        \big)
        +\frac{1}{2} (2-x) ( G_{14}(x)
        + G_{15}(x)
        + G_{16}(x)
        + G_{17}(x)
        )
\nonumber\\&&        
        +\frac{1}{240} \Biggl\{
                -60 (-2+x) 
                        \int_0^1 dz \biggl\{
                                -\frac{1}{(-1+z) z}2 \sqrt{\pi } \Bigl[
                                        -2
                                        +2 z
                                        +\big(
                                                1+\sqrt{z}\big) G_1(z)
\nonumber\\&&
                                        -2 \sqrt{z} \bigl(-1+\ln (2)\bigr)
                                \Bigr]
\biggl[-1
                                        +z
                                        -\sqrt{1+x} z
                                        +\sqrt{1
                                        +x z
                                        }
                                        +z H_0\Big(
                                                \frac{1}{2} \big(
                                                        1+\sqrt{1+x}\big)\Big)
\nonumber\\&&                                                        
                                        -H_0\Big(
                                                \frac{1}{2} \big(
                                                        1+\sqrt{1+x 
z}\big)\Big)
                                \biggr] \biggr\}
                +180 x 
                        \int_0^1 dz \biggl\{
                                -
                                \frac{1}{4 \sqrt{1
                                +x z
                                }}\sqrt{\pi } x \big(
                                        -1+\sqrt{1
                                        +x z
                                        }
                                \big)
\biggl[4
\nonumber\\&&
                                        +2 \sqrt{z} \Bigl[
                                                -8
                                                +2 z^{3/2}
                                                -3 \sqrt{z} \big(
                                                        2
                                                        +\zeta_2
                                                \big)
                                                +z \big(6-4 \ln (2)\big)
                                                +8 \ln (2)
                                        \Bigr]
                                        +4 H_0(z)
\nonumber\\&&
                                        +4 H_1(z)
                                        +2 \Bigl[
                                                -3
                                                -4 \sqrt{z}
                                                +3 z
                                                +2 z^{3/2}
                                                +(-1+z) H_0(z)
                                                +(-1+z) H_1(z)
                                                +2 \ln (2)
\nonumber\\&&
                                                -2 z \ln (2)
                                        \Bigr]  G_{1}(z)
                                        +(-1+z)  G_{1}(z)^2
                                        -2 (-1+z)  G_{6}(z)
                                        -2 (-1+z)  G_{7}(z)
                                        +6 \zeta_2
                                \biggr]\biggr\}
\nonumber\\&&                                
                +\biggl[
                        x \Bigl[
                                -2215
                                +4 x \big(
                                        435
                                        +96 x
                                        -500 \sqrt{1+x}
                                \big)
                                +60 \sqrt{1+x}
                        \Bigr]
\nonumber\\&&                        
                        +2060 \big(
                                -1+\sqrt{1
                                +x
                                }\big)
                \biggr] \sqrt{\pi }
        \Biggr\} \frac{1}{\sqrt{\pi }}
        +\frac{3}{8} (-12+35 x) H_0(x)
        +\frac{1}{8} (14-17 x) H_0(x)^2
\nonumber\\&&        
        +\frac{1}{4} (11-16 x) (1+x)^{3/2} H_{-1}(x)
        +\frac{5}{8} (1+x)^{3/2} H_{-1}(x)^2
        +\frac{1}{8} \Bigl[
                x \big(
                        17-2 \sqrt{1
                        +x
                        }\big)
\nonumber\\&&                        
                -2 \big(
                        6+\sqrt{1
                        +x
                        }\big)
        \Bigr] H_{0,-1}(x)
        +8 \sqrt{x}  G_{10}(x)
        +8 \sqrt{x}  G_{13}(x)
        +\frac{ G_{2}(x)}{2}
\nonumber\\&&        
        +\Bigl[
                2 (17-4 x) \sqrt{x}
                -8 \sqrt{x}  G_{5}(x)
        \Bigr]  G_{4}(x)
        +2 (-9+4 x) \sqrt{x}  G_{4}(x)
        +\biggl[
                \frac{7}{2}
                -\frac{17 x}{4}
\nonumber\\&&                
                +\frac{1}{2} (-2+x) H_0(x)
        \biggr]  G_{8}(x)
        +\biggl[
                -\frac{7}{2} (-1+x)
                +\frac{1}{2} (-2+x) H_0(x)
        \biggr]  G_{9}(x)
\Biggr\}
\nonumber\\&&
+\mathcal{O}(\varepsilon^4),
\end{eqnarray}
\begin{eqnarray}
	F_2(y,\varepsilon) &=& 1
-2 y
-
\frac{1}{4} (-20+y) y \varepsilon 
+\varepsilon ^2 \biggl\{
        -\frac{1}{48} y \big(
                480-765 y-56 y^2+64 y^3+12 y^4\big)
\nonumber\\&&
        +\frac{1}{4} (-9+4 y) y^{2/3} (3+y)^{4/3}  G_{26}(y)
        +(1-2 y)  G_{30}(y)
\biggr\}
+\varepsilon ^3 \Biggl\{
        \frac{1}{192} y \big(
                3840-21453 y
\nonumber\\&&
-1672 y^2+1638 y^3+280 y^4-6 y^5\big)
        + G_{26}(y) \biggl[
                \frac{1}{16} \big(
                        243-108 y+2 y^2\big) y^{2/3} (3+y)^{4/3}
\nonumber\\&&
                +\frac{1}{6} (9-4 y) y^{2/3} (3+y)^{4/3} H_0(y)
                +\frac{1}{6} (-9+4 y) y^{2/3} (3+y)^{4/3} H_{-3}(y)
\nonumber\\&&
                +\frac{1}{270} \big(
                        -1215+108 y+4 y^2\big) y^{2/3}  G_{24}(y)
                +\frac{7}{5} (-1+2 y)  G_{25}(y)
                +\frac{2}{3} (-1+2 y)  G_{28}(y)
\nonumber\\&&
                -\frac{2}{3} (-1+2 y)  G_{29}(y)
        \biggr]
        +\big(
                -1215+108 y+4 y^2
        \big)
\biggl[-\frac{1}{270} y^{2/3}  G_{27}(y)
                -\frac{1}{270} y^{2/3}  G_{33}(y)
        \biggr]
\nonumber\\&&
        +(-1+2 y) \Bigl(
                -\frac{7  G_{34}(y)}{5}
                +\frac{2  G_{35}(y)}{3}
        \Bigr)
        +\biggl[
                \big(
                        \frac{2}{3}-\frac{4 y}{3}\big)  G_{31}(y)
                +\frac{2}{3} (-1+2 y)  G_{32}(y)
        \biggr]  G_{25}(y)
\nonumber\\&&
        +\frac{1}{20} \big(
                -52+204 y-5 y^2\big)  G_{30}(y)
        +\frac{1}{6} (-9+4 y) y^{2/3}
         (3+y)^{4/3}  G_{31}(y)
\nonumber\\&&
        -\frac{1}{6} (3+y) (-9+4 y) y^{2/3} \sqrt[3]{3+y}  G_{32}(y)
        -\frac{2}{3} (-1+2 y)  G_{36}(y)
\Biggr\}+\mathcal{O}(\varepsilon^4).
\end{eqnarray}
By multiplying $F_1(x)$ and $F_2(y)$ one obtains the series expansion of $\mathcal F(x,y)$, with $0<x<1, 0<y<1$.
The functions $G_i$ are the iterated integrals of the type 
\cite{Ablinger:2014bra}
\begin{eqnarray}
G(f_1(\tau), ... f_n(\tau); x)  
= \int_0^x dy f_1(y) G(f_2(\tau), ..., f_n(\tau);y),
\end{eqnarray}
with
\begin{eqnarray}
         G_{1} (z) &=&  G \Big(
                \frac{\sqrt{\tau }}{1-\tau };z\Big)=-2 \sqrt{z}
+H_1\big(
        \sqrt{z}\big)
+H_{-1}\big(
        \sqrt{z}\big)
,
\\
         G_{2} (x) &=&  G \Big(
                \sqrt{1+\tau };x\Big)=-\frac{2}{3}
+\frac{2 \sqrt{1+x}}{3}
+\frac{2}{3} x \sqrt{1+x}
,
\\
         G_{3} (x) &=&  G \Big(
                \frac{\sqrt{1+\tau }}{\tau };x\Big)=-2
+2 \ln(2)
-i \pi 
+2 \sqrt{1+x}
-H_1\big(
        \sqrt{1+x}\big)
\nonumber\\&&        
-H_{-1}\big(
        \sqrt{1+x}\big)
,
\\
         G_{4} (x) &=&  G \big(
                \sqrt{\tau } \sqrt{1+\tau };x\big)=\frac{1}{4} 
\sqrt{x (1+x)}
+\frac{1}{2} x \sqrt{x (1+x)}
-\frac{1}{4} \ln \big(
        \sqrt{x}
        +\sqrt{1+x}
\big)
,
\\
         G_{5} (x) &=&  G \Big(
                \frac{1-\sqrt{1
                +\tau 
                }}{\tau };x\Big)=2
-2 \ln(2)
-2 \sqrt{1+x}
+2 H_{-1}\big(
        \sqrt{1+x}\big)
,
\\
         G_{6} (z) &=&  G \Big(
                \frac{\sqrt{\tau }}{1-\tau },\frac{1}{1-\tau };z\Big)=-4 \sqrt{z}
-2 \sqrt{z} H_1(z)
+H_1\big(
        \sqrt{z}\big) H_1(z)
+H_{-1}\big(
        \sqrt{z}\big) H_1(z)
\nonumber\\&&        
+2 H_1\big(
        \sqrt{z}\big)
-H_{-1}\big(
        \sqrt{z}\big) H_1\big(
        \sqrt{z}\big)
-\frac{1}{2} H_1^2\big(\sqrt{z}\big)
+2 H_{-1}\big(\sqrt{z}\big)
+
\frac{1}{2} H_{-1}^2\big(\sqrt{z}\big)
\nonumber\\&&        
+2 H_{-1,1}\big(
        \sqrt{z}\big)
,
\\
         G_{7} (z) &=&  G \Big(
                \frac{\sqrt{\tau }}{1-\tau },\frac{1}{\tau };z\Big)=4 \sqrt{z}
-2 \sqrt{z} H_0(z)
+H_{-1}\big(
        \sqrt{z}\big) H_0(z)
+H_0(z) H_1\big(
        \sqrt{z}\big)
\nonumber\\&&        
-2 H_{0,1}\big(
        \sqrt{z}\big)
-2 H_{0,-1}\big(
        \sqrt{z}\big)
,
\\
         G_{8} (x) &=&  G \Big(
                \frac{\sqrt{1+\tau }}{\tau },\frac{1}{\tau 
};x\Big)=4
+2 i \pi 
-\frac{2 \pi ^2}{3}
-4 \sqrt{1+x}
-4 \ln(2)
-2 i \pi  \ln(2)
\nonumber\\&&
+2 \ln^2(2)
+2 \sqrt{1+x} H_0(x)
-H_{-1}\big(
        \sqrt{1+x}\big) H_0(x)
+2 H_1\big(
        \sqrt{1+x}\big)
\nonumber\\&&        
-H_0(x) H_1\big(
        \sqrt{1+x}\big)
-H_{-1}\big(
        \sqrt{1+x}\big) H_1\big(
        \sqrt{1+x}\big)
-\frac{1}{2} H_1^2\big(
        \sqrt{1+x}\big)
\nonumber\\&&        
+2 H_{-1}\big(
        \sqrt{1+x}\big)
+\frac{1}{2} H_{-1}^2\big(
        \sqrt{1+x}\big)
+2 H_{-1,1}\big(
        \sqrt{1+x}\big)
,
\\
         G_{9} (x) &=&  G \Big(
                \frac{\sqrt{1+\tau }}{\tau },\frac{1}{1+\tau };x\Big)=4
-\frac{\pi ^2}{2}
-4 \sqrt{1+x}
-H_{-1}(x) H_1\big(
        \sqrt{1+x}\big)
+\frac{2 H_{-1}(x)}{\sqrt{1+x}}
\nonumber\\&&
+\frac{2 x H_{-1}
(x)}{\sqrt{1+x}}
-H_{-1}(x) H_{-1}\big(
        \sqrt{1+x}\big)
-2 H_{0,1}\big(
        -\sqrt{1+x}\big)
+2 H_{0,1}\big(
        \sqrt{1+x}\big)
,
\\
         G_{10} (x) &=&  G \Big(
                \sqrt{\tau } \sqrt{1+\tau },
                \frac{1}{1+\tau };x\Big)=
        \frac{1}{48} \bigg\{
        6 \sqrt{x} \sqrt{1+x}
        -12 x^{3/2} \sqrt{1+x}
\nonumber\\&&        
        -6 H_0\big(
                \sqrt{x}+\sqrt{1+x}\big)
        -12 H_{-1}(x) H_0\big(
                \sqrt{x}+\sqrt{1+x}\big)
\nonumber\\&&                
        +24 H_{-1}\Big(
                \big(
                        \sqrt{x}+\sqrt{1+x}\big)^2\Big) H_0\big(
                \sqrt{x}+\sqrt{1+x}\big)
        -12 H_0^2\big(
                \sqrt{x}+\sqrt{1+x}\big)
\nonumber\\&&                
        +12 \sqrt{x} \sqrt{1+x} H_{-1}(x)
        +24 x^{3/2} \sqrt{1+x} H_{-1}(x)
        -12 H_{0,-1}\Big(
                \big(
                        \sqrt{x}+\sqrt{1+x}\big)^2\Big)
\nonumber\\&&                        
        +6 \zeta_2
\bigg\},
\\
         G_{11} (x) &=&  G \Big(
                \frac{1-\sqrt{1
                +\tau 
                }}{\tau },\frac{1}{\tau };x\Big)=-4
-2 i \pi 
+\frac{\pi ^2}{6}
+4 \sqrt{1+x}
+4 \ln(2)
+2 i \pi  \ln(2)
\nonumber\\&&
-2 \ln^2(2)
-2 \sqrt{1+x} H_0(x)
+H_0(-x) H_0(x)
+H_{-1}\big(
        \sqrt{1+x}\big) H_0(x)
-\frac{1}{2} H_0^2(-x)
\nonumber\\&&
-2 H_1\big(
        \sqrt{1+x}\big)
+H_0(x) H_1\big(
        \sqrt{1+x}\big)
+H_{-1}\big(
        \sqrt{1+x}\big) H_1\big(
        \sqrt{1+x}\big)
\nonumber\\&&        
+\frac{1}{2} H_1^2\big(
        \sqrt{1+x}\big)
-2 H_{-1}\big(
        \sqrt{1+x}\big)
-\frac{1}{2} H_{-1}^2\big(
        \sqrt{1+x}
\big)
-2 H_{-1,1}\big(
        \sqrt{1+x}\big)
,
\\
         G_{12} (x) &=&  G \Big(
                \frac{1-\sqrt{1
                +\tau 
                }}{\tau },\frac{1}{1+\tau };x\Big)=-4
+\frac{\pi ^2}{3}
+4 \sqrt{1+x}
+H_{-1}(x) H_0(1)
\nonumber\\&&
+H_{-1}(x) H_0(-x)
+H_0(-x) H_1(-x)
+H_{-1}(x) H_1\big(
        \sqrt{1+x}\big)
-2 \sqrt{1+x} H_{-1}(x)
\nonumber\\&&
+H_{-1}(x) H_{-1}\big(
        \sqrt{1+x}\big)
+\zeta_2
-H_{0,1}(-x)
-2 H_{0,1}\big(
        \sqrt{1+x}\big)
\nonumber\\&&        
-2 H_{0,-1}\big(
        \sqrt{1+x}\big)
,
\\
         G_{13} (x) &=&  G \Big(
                \frac{1-\sqrt{1
                +\tau 
                }}{\tau },\sqrt{\tau } \sqrt{1+\tau 
};x\Big)=\frac{1}{40} \bigg[
        -40 \sqrt{x}
        -20 x^{3/2}
        -8 x^{5/2}
        +15 \sqrt{x (1+x)}
\nonumber\\&&        
        +10 x \sqrt{x (1+x)}
\bigg]
+\bigg[
        \frac{1}{8} \big(
                1+4 \sqrt{1
                +x
                }\big)
        -\frac{1}{2} H_1\big(
                \sqrt{x}+\sqrt{1+x}\big)
\nonumber\\&&                
        +\frac{1}{2} H_1\Big(
                \big(
                        \sqrt{x}+\sqrt{1+x}\big)^2\Big)
        -\frac{1}{2} H_{-1}\big(
                \sqrt{x}+\sqrt{1+x}\big)
\bigg] H_0\big(
        \sqrt{x}+\sqrt{1+x}\big)
\nonumber\\&&        
+\frac{1}{4} H_0^2\big(
        \sqrt{x}+\sqrt{1+x}\big)
-\frac{1}{2} \zeta_2
+H_{0,-1}\big(
        \sqrt{x}+\sqrt{1+x}\big),
\\
         G_{14} (x) &=&  G \Big(
                \frac{\sqrt{1+\tau }}{\tau },\frac{1}{\tau 
},\frac{1}{\tau };x\Big)
=
-8
+
\frac{4 \ln^3(2)}{3}
-4 i \pi 
+\frac{4 \pi ^2}{3}
+\frac{5 i \pi ^3}{6}
-2 \ln^2(2) (2+i \pi )
\nonumber\\&&
+\frac{1}{3} \ln(2) \big(
        24+12 i \pi -\pi ^2\big)
+8 \sqrt{1+x}
+\Big[
        -i \ln(2) \pi 
        +\frac{3 \pi ^2}{2}
        -4 \sqrt{1+x}
\nonumber\\&&        
        -2 (-1+i \pi ) H_1\big(
                \sqrt{1+x}\big)
        -\frac{1}{2} H_1^2\big(
                \sqrt{1+x}\big)
        +2 H_{-1,1}\big(
                \sqrt{1+x}\big)
\Big] H_0(x)
\nonumber\\&&
+\Big[
        -i \pi 
        +\sqrt{1+x}
        -\frac{1}{2} H_1\big(
                \sqrt{1+x}\big)
\Big] H_0^2(x)
+\Big[
        -4
        -i \ln(2) \pi 
        +\frac{3 \pi ^2}{2}
\nonumber\\&&        
        +2 H_{-1,1}\big(
                \sqrt{1+x}\big)
\Big] H_1\big(
        \sqrt{1+x}\big)
+(1-i \pi ) H_1^2\big(
        \sqrt{1+x}\big)
-\frac{1}{6} H_1^3\big(
        \sqrt{1+x}\big)
\nonumber\\&&        
+\bigg[
        -4
        +i \ln(2) \pi 
        -\frac{3 \pi ^2}{2}
        +\Big[
                2
                +2 i \pi 
                -H_1\big(
                        \sqrt{1+x}\big)
        \Big] H_0(x)
        -\frac{1}{2} H_0^2(x)
\nonumber\\&&        
        +2 (1+i \pi ) H_1\big(
                \sqrt{1+x}\big)
        -\frac{1}{2} H_1^2\big(
                \sqrt{1+x}\big)
\bigg] H_{-1}\big(
        \sqrt{1+x}\big)
+\Big[
        -1
        -i \pi 
        +\frac{1}{2} H_0(x)
\nonumber\\&&        
        +\frac{1}{2} H_1\big(
                \sqrt{1+x}\big)
\Big] H_{-1}^2\big(
        \sqrt{1+x}\big)
-\frac{1}{6} H_{-1}^3\big(
        \sqrt{1+x}\big)
-4 H_{-1,1}
\big(
        \sqrt{1+x}\big)
\nonumber\\&&        
-2 H_{-1,1,1}\big(
        \sqrt{1+x}\big)
-2 H_{-1,-1,1}\big(
        \sqrt{1+x}\big)
+2 \zeta_3
,
\\
         G_{15} (x) &=&  G \Big(
                \frac{\sqrt{1+\tau }}{\tau },\frac{1}{\tau 
},\frac{1}{1+\tau };x\Big)
=
8 \big(
        -1+\sqrt{1
        +x
        }\big)
-H_{-1}^2\big(
        \sqrt{1+x}\big) H_0\big(
        \sqrt{1+x}\big)
\nonumber\\&&        
-8 \sqrt{1+x} H_{-1}\big(
        -1+\sqrt{1+x}\big)
+\Big[
        4 \big(
                1+\sqrt{1
                +x
                }\big) H_0\big(
                \sqrt{1+x}\big)
        +2 H_{0,-1}\big(
                \sqrt{1+x}\big)
\nonumber\\&&                
        -\zeta_2
\Big] H_{-1}\big(
        \sqrt{1+x}\big)
+4 \big(
        -1+\sqrt{1
        +x
        }\big) H_{0,-1}\big(
        -1+\sqrt{1+x}\big)
\nonumber\\&&        
-4 \big(
        1+\sqrt{1
        +x
        }\big) H_{0,-1}\big(
        \sqrt{1+x}\big)
+2 H_{0,0,-1}\big(
        -1+\sqrt{1+x}\big)
-2 H_{0,-1,-1}\big(
        \sqrt{1+x}\big)
\nonumber\\&&        
+2 H_{0,-2,-1}\big(
        -1+\sqrt{1+x}\big)
-2 H_{-2,0,-1}\big(
        -1+\sqrt{1+x}\big)
+2 \big(
        1+\sqrt{1
        +x
        }\big) \zeta_2
\nonumber\\&&        
+\frac{1}{4} \zeta_3
,
\\
         G_{16} (x) &=&  G \Big(
                \frac{\sqrt{1+\tau }}{\tau },\frac{1}{1+\tau 
},\frac{1}{\tau };x\Big)
=
8 \big(
        -1+\sqrt{1
        +x
        }\big)
-4 \ln(2) \big(
        -1+\sqrt{1
        +x
        }\big)
\nonumber\\&&        
+\Big[
        -4 \big(
                -1+\sqrt{1
                +x
                }\big)
        +2 H_{0,-1}\big(
                -1+\sqrt{1+x}\big)
\Big] H_0\big(
        -1+\sqrt{1+x}\big)
\nonumber\\&&
+\Big[
        4 \ln(2) \sqrt{1+x}
        +4 \sqrt{1+x} H_0\big(
                -1+\sqrt{1+x}\big)
\Big] H_{-1}\big(
        -1+\sqrt{1+x}\big)
\nonumber\\&&        
-4 \big(
        1+\sqrt{1
        +x
        }\big) H_{-2}\big(
        -1+\sqrt{1+x}\big)
+\Big[
        2 \ln(2)
        -4 \sqrt{1+x}
\Big] H_{0,-1}\big(
        -1+\sqrt{1+x}\big)
\nonumber\\&&        
+4 \sqrt{1+x} H_{-1,-2}\big(
        -1+\sqrt{1+x}\big)
-2 \ln(2) H_{-2,-1}\big(
        -1+\sqrt{1+x}\big)
\nonumber\\&&        
-4 H_{0,0,-1}\big(
        -1+\sqrt{1+x}\big)
+2 H_{0,-1,-2}\big(
        -1+\sqrt{1+x}\big)
\nonumber\\&&        
-2 H_{-2,-1,0}\big(
        -1+\sqrt{1+x}\big)
-2 H_{-2,-1,-2}\big(
        -1+\sqrt{1+x}\big)
,
\\
         G_{17} (x) &=&  G \Big(
                \frac{\sqrt{1+\tau }}{\tau },\frac{1}{1+\tau 
},\frac{1}{1+\tau };x\Big)
=
8 \big(
        -1+\sqrt{1
        +x
        }\big)
-8 \sqrt{1+x} H_{-1}\big(
        -1+\sqrt{1+x}\big)
\nonumber\\&&        
+4 \sqrt{1+x} H_{-1}^2\big(
        -1+\sqrt{1+x}\big)
+4 H_{0,-1,-1}\big(
        -1+\sqrt{1+x}\big)
\nonumber\\&&        
-4 H_{-2,-1,-1}\big(
        -1+\sqrt{1+x}\big)
,
\\
         G_{18} (x) &=&  G \Big(
                \frac{1-\sqrt{1
                +\tau 
                }}{\tau },\frac{1}{\tau },\frac{1}{\tau 
};x\Big)
=
8
-
\frac{4 \ln^3(2)}{3}
+2 \ln^2(2) (2+i \pi )
+4 i \pi 
-\frac{5 i \pi ^3}{6}
\nonumber\\&&
+\ln(2) \big(
        -8
        -4 i \pi 
        +2 \zeta_2
\big)
-8 \sqrt{1+x}
+\Big[
        i \ln(2) \pi 
        +4 \sqrt{1+x}
        +2 (-1+i \pi ) H_1\big(
                \sqrt{1+x}\big)
\nonumber\\&&
        +\frac{1}{2} H_1^2\big(
                \sqrt{1+x}\big)             
        -2 H_{-1,1}\big(
                \sqrt{1+x}\big)
        -9 \zeta_2
\Big] H_0(x)
+\Big[
        \frac{i \pi }{2}
        -\sqrt{1+x}
        +\frac{1}{2} H_0(-x)
\nonumber\\&&        
        +\frac{1}{2} H_1\big(
                \sqrt{1+x}\big)
\Big] H_0^2(x)
-\frac{1}{3} H_0^3(x)
+\Big[
        4
        +i \ln(2) \pi 
        -2 H_{-1,1}\big(
                \sqrt{1+x}\big)
\nonumber\\&&                
        -9 \zeta_2
\Big] H_1\big(
        \sqrt{1+x}\big)
+ (-1+i \pi ) H_1^2\big(
        \sqrt{1+x}\big)
+\frac{1}{6} H_1^3\big(
        \sqrt{1+x}\big)
+\bigg[
        4
        -i \ln(2) \pi 
\nonumber\\&&        
        +\frac{1}{2} H_0^2(x)
        -2 (1+i\pi ) H_1\big(
                \sqrt{1+x}\big)
        +\frac{1}{2} H_1^2\big(
                \sqrt{1+x}\big)
        +9 \zeta_2
        +H_0(x) \Big[
                -2-2 i \pi 
\nonumber\\&&                
                +H_1\big(
                        \sqrt{1+x}\big)\Big]
\bigg] H_{-1}\big(
        \sqrt{1+x}\big)
+\Big[
        1
        +i \pi 
        -\frac{1}{2} H_0(x)
\nonumber\\&&        
        -\frac{1}{2} H_1\big(
                \sqrt{1+x}\big)
\Big] H_{-1}^2\big(
        \sqrt{1+x}\big)
+\frac{1}{6} H_{-1}^3\big(
        \sqrt{1+x}\big)
+4 H_{-1,1}\big(
        \sqrt{1+x}
\big)
\nonumber\\&&
+2 H_{-1,1,1}\big(
        \sqrt{1+x}\big)
+2 H_{-1,-1,1}\big(
        \sqrt{1+x}\big)
-8 \zeta_2
-2 \zeta_3
,
\\
         G_{19} (x) &=&  G \Big(
                \frac{1-\sqrt{1
                +\tau 
                }}{\tau },\frac{1}{\tau },\frac{1}{1+\tau 
};x\Big)
=
-8 \big(
        -1+\sqrt{1
        +x
        }\big)
+8 \sqrt{1+x} H_{-1}\big(
        -1+\sqrt{1+x}\big)
\nonumber\\&&        
-4 \big(
        -1+\sqrt{1
        +x
        }\big) H_{0,-1}\big(
        -1+\sqrt{1+x}\big)
-4 \big(
        1+\sqrt{1
        +x
        }\big) H_{-2,-1}\big(
        -1+\sqrt{1+x}\big)
\nonumber\\&&        
-2 H_{0,0,-1}\big(
        -1+\sqrt{1+x}\big)
-2 H_{0,-2,-1}\big(
        -1+\sqrt{1+x}\big)
+2 H_{-2,0,-1}\big(
        -1+\sqrt{1+x}\big)
\nonumber\\&&        
+2 H_{-2,-2,-1}\big(
        -1+\sqrt{1+x}\big)
+H_{0,0,-1}(x)
,
\\
         G_{20} (x) &=&  G \Big(
                \frac{1-\sqrt{1
                +\tau 
                }}{\tau },\frac{1}{1+\tau },\frac{1}{\tau 
};x\Big)
=
4 (-2+\ln(2)) \big(
        -1+\sqrt{1
        +x
        }\big)
\nonumber\\&&        
+\Big[
        4 \big(
                -1+\sqrt{1
                +x
                }\big)
        -2 H_{0,-1}\big(
                -1+\sqrt{1+x}\big)
\Big] H_0\big(
        -1+\sqrt{1+x}\big)
\nonumber\\&&        
+\Big[
        -4 \ln(2) \sqrt{1+x}
        -4 \sqrt{1+x} H_0\big(
                -1+\sqrt{1+x}\big)
\Big] H_{-1}\big(
        -1+\sqrt{1+x}\big)
\nonumber\\&&        
+4 \big(
        1+\sqrt{1
        +x
        }\big) H_{-2}\big(
        -1+\sqrt{1+x}\big)
+H_0(x) H_{0,-1}(x)
\nonumber\\&&
+\Big[
        -2 \ln(2)
        +4 \sqrt{1+x}
\Big] H_{0,-1}\big(
        -1+\sqrt{1+x}\big)
-4 \sqrt{1+x} H_{-1,-2}\big(
        -1+\sqrt{1+x}\big)
\nonumber\\&&        
+2 \ln(2) H_{-2,-1}\big(
        -1+\sqrt{1+x}\big)
-2 H_{0,0,-1}(x)
+4 H_{0,0,-1}\big(
        -1+\sqrt{1+x}\big)
\nonumber\\&&        
-2 H_{0,-1,-2}\big(
        -1+\sqrt{1+x}\big)
+2 H_{-2,-1,0}\big(
        -1+\sqrt{1+x}\big)
\nonumber\\&&        
+2 H_{-2,-1,-2}\big(
        -1+\sqrt{1+x}\big)
,
\\
         G_{21} (x) &=&  G \Big(
                \frac{1-\sqrt{1
                +\tau 
                }}{\tau },\frac{1}{1+\tau },\frac{1}{1+\tau 
};x\Big)
=
-8 \big(
        -1+\sqrt{1
        +x
        }\big)
+\Big[
        4 \sqrt{1+x}
\nonumber\\&&        
        -2 H_{0,1}\big(
                \sqrt{1+x}\big)
        -2 H_{0,-1}\big(
                \sqrt{1+x}\big)
\Big] H_{-1}(x)
+\Big[
        -\sqrt{1+x}
\nonumber\\&&        
        +\frac{1}{2} H_1\big(
                \sqrt{1+x}\big)
        +\frac{1}{2} H_{-1}\big(
                \sqrt{1+x}\big)
\Big] H_{-1}(x)^2
+4 H_{0,0,1}\big(
        \sqrt{1+x}\big)
\nonumber\\&&        
+4 H_{0,0,-1}\big(
        \sqrt{1+x}\big)        
-7 \zeta_3
+H_{0,-1,-1}(x),
\\
G_{22} (x) &=&  G \Big(
                \frac{1-\sqrt{1
                +\tau 
                }}{\tau },\frac{1-\sqrt{1
                +\tau 
                }}{\tau },\frac{1}{\tau };x\Big)
=
-16
+
\frac{8 \ln^3(2)}{3}
-7 i \pi 
-6 x
+4 \ln(2) \Big[
        4
        +2 i \pi 
\nonumber\\&&        
        -2 \sqrt{1+x}
        -i \pi  \sqrt{1+x}
        +7 \zeta_2
\Big]
+16 \sqrt{1+x}
+4 i \pi  \sqrt{1+x}
+\ln^2(2) \Big[
        -8+3 i \pi 
\nonumber\\&&        
        +4 \sqrt{1
        +x
        }\Big]
+\bigg[
        -3
        +5 \ln^2(2)
        +\ln(2) (4-5 i \pi )
        +2 x
        +\Big[
                4-15 i \pi 
\nonumber\\&&                
                -2 \sqrt{1
                +x
                }\Big] H_1\big(
                \sqrt{1+x}\big)
        -\frac{11}{2} H_1^2\big(
                \sqrt{1+x}\big)
        +42 \zeta_2
\bigg] H_0(x)
+\Big[
        \frac{1}{2} \big(
                4-15 i \pi -2 \sqrt{1
                +x
                }\big)
\nonumber\\&&                
        -\frac{11}{2} H_1\big(
                \sqrt{1+x}\big)
\Big] H_0^2(x)
-\frac{11}{6} H_0^3(x)
+\Big[
        -7
        +5 \ln^2(2)
        +\ln(2) (4-5 i \pi )
\nonumber\\&&        
        +4 \sqrt{1+x}
        +42 \zeta_2
\Big] H_1\big(
        \sqrt{1+x}\big)
+\frac{1}{2} \Big(
        4-15 i \pi -2 \sqrt{1
        +x
        }\Big) H_1^2\big(
        \sqrt{1+x}\big)
\nonumber\\&&        
-\frac{11}{6} H_1^3\big(
        \sqrt{1+x}\big)
+\bigg[
        -9
        -9 \ln^2(2)
        +\ln(2) (4+9 i \pi )
        -4 i \pi 
        +4 \sqrt{1+x}
        +\Big[
                -4
\nonumber\\&&                
                +15 i \pi 
                -2 \sqrt{1+x}
                +13 H_1\big(
                        \sqrt{1+x}\big)
        \Big] H_0(x)
        +\frac{13}{2} H_0^2(x)
        +\Big(
                -4+15 i \pi 
\nonumber\\&&                
                -2 \sqrt{1
                +x
                }\Big) H_1\big(
                \sqrt{1+x}\big)
        +\frac{13}{2} H_1^2\big(
                \sqrt{1+x}
        \big)
        -34 \zeta_2
\bigg] H_{-1}\big(
        \sqrt{1+x}\big)
\nonumber\\&&        
+\Big[
        \frac{1}{2} \big(
                -15 i \pi +2 \sqrt{1
                +x
                }\big)
        -\frac{11}{2} H_0(x)
        -\frac{11}{2} H_1\big(
                \sqrt{1+x}\big)
\Big] H_{-1}^2\big(
        \sqrt{1+x}\big)
\nonumber\\&&        
+\frac{3}{2} H_{-1}^3\big(
        \sqrt{1+x}\big)
+4 \big(
        -1+\sqrt{1
        +x
        }\big) H_{-1,1}\big(
        \sqrt{1+x}\big)
-4 H_{-1,-1,1}\big(
        \sqrt{1+x}\big)
\nonumber\\&&        
+14 \zeta_2
+8 i \pi  \zeta_2
-8 \sqrt{1+x} \zeta_2
+\frac{1}{2} \zeta_3                
,
\\
G_{23} (x) &=&  G \Big(
                \frac{1-\sqrt{1
                +\tau 
                }}{\tau },\frac{1-\sqrt{1
                +\tau 
                }}{\tau },\frac{1}{1+\tau };x\Big)
=
-16
+16 \ln(2)
-4 i \pi 
-6 x
+16 \sqrt{1+x}
\nonumber\\&&
+2 \big(
        -2
        +\zeta_2
\big) H_0(x)
-4 H_{-1}^2\big(
        \sqrt{1+x}\big) H_0\big(
        -\sqrt{1+x}\big)
\nonumber\\&&        
+2 \big(
        -2
        +\zeta_2
\big) H_1\big(
        \sqrt{1+x}\big)
+\Big[
        2 (1+x)
        -4 \sqrt{1+x} H_{-1}\big(
                \sqrt{1+x}\big)
\nonumber\\&&                
        +2 H_{-1}^2\big(
                \sqrt{1+x}\big)
\Big] H_{-1}(x)
+\Big[
        2 \big(
                -6
                +5 \zeta_2
        \big)
\nonumber\\&&        
        +8 \sqrt{1+x} H_0\big(
                -\sqrt{1+x}\big)
\Big] H_{-1}\big(
        \sqrt{1+x}\big)        
+8 \sqrt{1+x} H_{0,1}\big(
        1+\sqrt{1+x}\big)
\nonumber\\&&        
-8 H_{0,0,1}\big(
        1+\sqrt{1+x}\big)
+2 i \pi  \zeta_2
-12 \sqrt{1+x} \zeta_2
+7 \zeta_3
\\
G_{24} (y) &=&  G\Big(
        \frac{(3+\tau )^{1/3}}{\tau };y\Big)
=
\frac{1}{6} \bigg\{
        3 \bigg[
                -6
                +3 \ln (3)
                -2 (-1)^{1/3} \Big[
                        -\ln (3)
\nonumber\\&&                        
                        +\ln \Big(
                                3
                                -(-3)^{2/3} (3+y)^{1/3}
                        \Big)\Big]
                +2 (-1)^{2/3} \ln \Big[
                        1+3^{-1/3}(-1)^{1/3}(3+y)^{1/3}
                        \Big]
\nonumber\\&&                        
                +2 \ln \big(
                        -3+3^{2/3} (3+y)^{1/3}\big)
        \bigg] 3^{1/3}
        +\pi  3^{5/6}
        +18 (3+y)^{1/3}
\bigg\}
,
\\        
G_{25} (y) &=&  G\Big(
        \tau ^{2/3} (3+\tau )^{1/3};y\Big)
=
\frac{1}{2} \bigg\{
        (1+y) (3+y)^{1/3}
        -3^{1/3} \,_2F_1\Big[
                \frac{2}{3},\frac{2}{3};\frac{5}{3};-\frac{y}{3}\Big]
\bigg\} y^{2/3}
,
\\        
G_{26} (y) &=&  G\Big(
        \tau ^{1/3} (3+\tau )^{2/3};y\Big)
=
\frac{1}{2} \bigg\{
        (2+y) (3+y)^{2/3}
        -2 \times 3^{2/3} \,_2F_1\Big[
                \frac{1}{3},\frac{1}{3};\frac{4}{3};-\frac{y}{3}\Big]
\bigg\} y^{1/3}
, 
\\        
G_{27} (y) &=&  G\Big(
        \frac{(3+\tau )^{1/3}}{\tau },\tau ^{1/3} (3+\tau )^{2/3};y\Big), 
\\        
G_{28} (y) &=&  G\Big(
        \tau ^{2/3} (3+\tau )^{1/3},\frac{1}{\tau };y\Big), 
\\        
G_{29} (y) &=&  G\big(
        \tau ^{2/3} (3+\tau )^{1/3},\frac{1}{3+\tau };y\Big), 
\\        
G_{30} (y) &=&  G\Big(
        \tau ^{2/3} (3+\tau )^{1/3},\tau ^{1/3} (3+\tau )^{2/3};y\Big), 
\\        
        G_{31} (y) &=&  G\Big(
        \tau ^{1/3} (3+\tau )^{2/3},\frac{1}{\tau };y\Big), 
\\        
G_{32} (y) &=&  G\Big(
        \tau ^{1/3} (3+\tau )^{2/3},\frac{1}{3+\tau };y\Big), 
\\        
G_{33} (y) &=&  G\Big(
        \tau ^{1/3} (3+\tau )^{2/3},\frac{(3+\tau )^{1/3}}{\tau 
};y\Big), 
\\
G_{34} (y) &=&  G\Big(
        \tau ^{1/3} (3+\tau )^{2/3},\tau ^{2/3} (3+\tau )^{1/3};y\Big), 
\\        
G_{35} (y) &=&  G\Big(
        \tau ^{1/3} (3+\tau )^{2/3},\frac{1}{\tau },\tau ^{2/3} (3+\tau )^{1/3};y\Big), 
\\        
G_{36} (y) &=&  G\Big(
        \tau ^{1/3}
         (3+\tau )^{2/3},\frac{1}{3+\tau },\tau ^{2/3} 
(3+\tau )^{1/3};y\Big).
\end{eqnarray}

\subsection{Example 3}

Consider as an example the system of two differential equations 
in two variables implied by the following differential operators,
\begin{eqnarray}
&&	-\alpha  \beta -x^6 \partial_x^6-\beta  x^5 \partial_x^5-5 x^5 y \partial_x^5 \partial_y-15 x^5 \partial_x^5-10 \beta  x^4 \partial_x^4-10 x^4 y^2 \partial_x^4 \partial_y^2-5 \beta  x^4 y \partial_x^4 \partial_y
\nonumber\\&&	
	-60 x^4 y \partial_x^4 \partial_y-66 x^4 \partial_x^4-26 \beta  x^3 \partial_x^3-10 x^3 y^3 \partial_x^3 \partial_y^3-10 \beta  x^3 y^2 \partial_x^3 \partial_y^2-90 x^3 y^2 \partial_x^3 \partial_y^2
\nonumber\\&&	
	-40 \beta x^3 y \partial_x^3 \partial_y-198 x^3 y \partial_x^3
   \partial_y-96 x^3 \partial_x^3-18 \beta  x^2 \partial_x^2-5 x^2 y^4 \partial_x^2 \partial_y^4-10 \beta  x^2 y^3 \partial_x^2 \partial_y^3
\nonumber\\&&   
   -60 x^2 y^3 \partial_x^2 \partial_y^3-60 \beta  x^2 y^2 \partial_x^2 \partial_y^2-198 x^2 y^2
   \partial_x^2 \partial_y^2-78 \beta  x^2 y \partial_x^2 \partial_y-192 x^2 y \partial_x^2 \partial_y
\nonumber\\&&   
   -38 x^2 \partial_x^2-\alpha  x \partial_x-2 \beta  x \partial_x+\gamma  \partial_x-x y^5 \partial_x \partial_y^5-5 \beta  x y^4 \partial_x
   \partial_y^4-15 x y^4 \partial_x \partial_y^4
\nonumber\\&&   
   -40 \beta  x y^3 \partial_x \partial_y^3-66 x y^3 \partial_x \partial_y^3-78 \beta  x y^2 \partial_x \partial_y^2-96 x y^2 \partial_x \partial_y^2-36 \beta  x y \partial_{x y}-38 x y \partial_{x y}
\nonumber\\&&   
   +y
   \partial_{x y}+x \partial_x^2-2 x \partial_x-\beta  y^5 \partial_y^5-10 \beta  y^4 \partial_y^4-26 \beta  y^3 \partial_y^3-18 \beta  y^2 \partial_y^2-2 \beta  y \partial_y,
\\  [1cm]
&& 
-\alpha  \beta_1 -\beta_1  x^5 \partial_x^5-x^5 y \partial_x^5 \partial_y-10 \beta_1  x^4 \partial_x^4-5 x^4 y^2 \partial_x^4 \partial_y^2-5 \beta_1  x^4 y \partial_x^4
   \partial_y-15 x^4 y \partial_x^4 \partial_y
\nonumber\\&&   
   -26 \beta_1  x^3 \partial_x^3-10 x^3 y^3 \partial_x^3 \partial_y^3-10 \beta_1  x^3 y^2 \partial_x^3 \partial_y^2-60 x^3 y^2 \partial_x^3 \partial_y^2-40 {\beta_1} x^3 y \partial_x^3 \partial_y
\nonumber\\&&   
   -66 x^3 y \partial_x^3 \partial_y-18 \beta_1  x^2 \partial_x^2-10 x^2 y^4 \partial_x^2 \partial_y^4-10 \beta_1  x^2 y^3 \partial_x^2 \partial_y^3-90 x^2 y^3 \partial_x^2
   \partial_y^3
\nonumber\\&&   
   -60 \beta_1  x^2 y^2 \partial_x^2 \partial_y^2-198 x^2 y^2 \partial_x^2 \partial_y^2-78 \beta_1  x^2 y \partial_x^2 \partial_y-96 x^2 y \partial_x^2 \partial_y-2 \beta_1  x \partial_x
\nonumber\\&&   
   -5 x
   y^5 \partial_x \partial_y^5-5 \beta_1  x y^4 \partial_x \partial_y^4-60 x y^4 \partial_x \partial_y^4-40 \beta_1  x y^3 \partial_x \partial_y^3-198 x y^3 \partial_x \partial_y^3
\nonumber\\&&   
   -78 \beta_1  x y^2 \partial_x
   \partial_y^2-192 x y^2 \partial_x \partial_y^2-36 \beta_1  x y \partial_{x y}+x \partial_{x y}-38 x y \partial_{x y}-y^6 \partial_y^6-\beta_1  y^5 \partial_y^5
\nonumber\\&&   
   -15 y^5 \partial_y^5-10 \beta_1 
   y^4 \partial_y^4-66 y^4 \partial_y^4-26 \beta_1  y^3 \partial_y^3-96 y^3 \partial_y^3-18 \beta_1  y^2 \partial_y^2-38 y^2 \partial_y^2-\alpha  y \partial_y
\nonumber\\&&   
   -2 \beta_1  y \partial_y+\gamma
    \partial_y+y \partial_y^2-2 y \partial_y.
\end{eqnarray}

Assuming a hypergeometric solution
\begin{equation}
	\mathcal F(x,y) = \sum_{m,n\ge 0} f(m,n) x^m y^n,
\end{equation}
the coefficients $f(m,n)$ must obey
\begin{eqnarray}
	(m+1) (\gamma +m+n) f(m+1,n)-(\beta +m) \left(\alpha +(m+n)^5+(m+n)^3\right) f(m,n)=0,\\
	(n+1) (\gamma +m+n) f(m,n+1)-(\beta_1 +n) \left(\alpha +(m+n)^5+(m+n)^3\right) f(m,n)=0.
\end{eqnarray}
Solving these two equations with the help of {\tt Sigma}, one obtains
\begin{eqnarray}
	f(m,n) &=& \bigg(
        \prod_{i_1=1}^n \frac{\big(
                -1
                +{\beta_1}
                +i_1
        \big)
\big(-2
                +\alpha 
                +8 i_1
                -13 i_1^2
                +11 i_1^3
                -5 i_1^4
                +i_1^5
        \big)}{\big(
                -1
                +\gamma 
                +i_1
        \big) i_1}\bigg) 
\nonumber\\&& \times        
        \prod_{i_1=1}^m \frac{\big(
        -1
        +\beta 
        +i_1
\big)
}{\big(
        -1
        +n
        +\gamma 
        +i_1
\big) i_1}
\big(-2
        +8 n
        -13 n^2
        +11 n^3
        -5 n^4
        +n^5
        +\alpha 
        +8 i_1
        -26 n i_1
\nonumber\\&&        
        +33 n^2 i_1
        -20 n^3 i_1
        +5 n^4 i_1
        -13 i_1^2
        +33 n i_1^2
        -30 n^2 i_1^2
        +10 n^3 i_1^2
        +11 i_1^3
        -20 n i_1^3
\nonumber\\&&        
        +10 n^2 i_1^3
        -5 i_1^4
        +5 n i_1^4
        +i_1^5
\big)
\end{eqnarray}
This quantity cannot be analytically expressed as a product of Pochhammer symbols due to the high degree of the polynomials appearing.

\section{Partial difference equations with rational coefficients}
\label{sec:PLDEsolver}

\vspace*{1mm}
\noindent
In a series of problems also partial linear difference equations in various variables with polynomial coefficients  occur,  with  the  target  solution  space  being  that  of  rational  functions in several variables, possibly also including harmonic sums or Pochhammer symbols in the numerator. In various interesting situations, see Section~\ref{sec:fullmachinery}, one can derive solutions iteratively by solving 
first--order linear recurrences. More generally, one may solve higher-order linear recurrences using 
difference ring algorithms~\cite{abramov71,LinearSolver} implemented in \texttt{Sigma}.
However, in the general case of multivariate linear difference equations, there are only very few algorithms available to find the solution if compared to the case of the univariate difference equations.  To support possible future challenges in applications, we developed a {\tt Mathematica} implementation  of  the  algorithms  of  Refs.  \cite{kauers10,kauers11},  which  are  a  multivariate generalization of the algorithm described in Ref. \cite{abramov71}. In addition, we enhanced these methods by further heuristic techniques that may be useful for the calculation of Feynman integrals.
The basic idea of these algorithms is to constrain the denominator of the solution.  From this,  finding the numerator of the solution using an ansatz becomes easier.  In particular, it only requires the solution of a linear system of equations.
In the following, we give a survey on how to constrain the denominator of 
the solution of a partial linear difference equation (PLDE). 
In Section~\ref{sec:definitions}, we describe the notation used and in 
Section~\ref{sec:denBounds} we describe the concepts of aperiodic and 
periodic denominator bounds given in the literature, and in Section \ref{sec:num} we discuss the determination of the numerator of the solution. In particular, we explain how one can deal with a hypergeometric prefactor in the solution in Section \ref{sec:hypergeometric-prefactor} and how one can search in addition for solutions in terms nested sums in Section~\ref{sec:findingNestedSums}. After commenting on the problem to combine the solutions using initial values in Section~\ref{sec:initVal} we turn in Section~\ref{sec:expansion} to tools to obtain a Laurent expansion in the dimensional parameter $\varepsilon$ efficiently by successively solving a set of difference equations where the parameter no longer appears in the coefficients. This section is supplemented by Section \ref{sec:D} where we describe the commands available in our {\tt Mathematica} implementation \texttt{solvePartialLDE}.

\subsection{The basic problem description}
\label{sec:definitions}

\vspace*{1mm}
\noindent
With $y(n_1,\ldots,n_r)\in\KK(n_1,\dots,n_r)$ a rational function in $r$ variables, we define the \emph{shift operators} $N_\mathbf s$ with respect to the \emph{shift} $\mathbf s=(s_1,\ldots,s_r)\in\mathbb Z^r$ as
\begin{equation}
N_\mathbf s y = y(n_1+s_1,\ldots,n_r+s_r).
\end{equation}
Partial linear difference equations are equations of the type
\begin{equation}
\sum_{\mathbf s \in S} a_\mathbf s N_\mathbf s y = f,
\label{eq:PLDE}
\end{equation}
where $S$ is a finite subset of $\mathbb Z^r$, $a_\mathbf s$ and $f$ are polynomials in the variables $n_1,\ldots,n_r$, and $y$ is an unknown rational function to be determined; the set $S$ of all shifts appearing in the equation is called the \emph{shift set} or \emph{structure set}. Because the equation is linear, the general solution is the sum of a particular solution of \eqref{eq:PLDE} and of the homogeneous equation with $f=0$.

An example of the type of equation under consideration is:
\begin{equation}
-(1 + k + n^2) y(n, k) + (4 + k + 2 n + n^2) y(1 + n, 2 + k) = 0.
\end{equation}
It has the shift set $S=\{(0,0), (1,2)\}$ and its coefficients are
\begin{eqnarray}
a_{(0,0)} &=& -(1+k+n^2),\\
a_{(1,2)} &=& (4 + k + 2 n + n^2).
\end{eqnarray}

A distinction used in the literature~\cite{DR,LinearSolver,kauers10} is the one between \emph{periodic} and \emph{aperiodic} polynomials. A polynomial $p$ is periodic if there exist infinitely many shifts,  mapping $p$ into $p'$, such that $\mathrm{gcd}(p,p')\neq 1$. A polynomial is called aperiodic if it is not periodic. 
For example, the polynomial $(n+k+2)$ is periodic, and the polynomial $(n^2+k+6)$ is aperiodic.
An important fact is that any polynomial can be factorized into a periodic and an aperiodic part.

Given a partial linear difference equation, algorithms exist to constrain what denominators may appear in the solution. These algorithms target separately the periodic and the aperiodic part of the denominator of the solution of \eqref{eq:PLDE}. In our package we have implemented and enhanced the algorithms described in \cite{kauers10,kauers11} and we describe our implementation choices and their rationale in the following.

\subsection{Denominator bounds}
\label{sec:denBounds}

\vspace*{1mm}
\noindent
Let us first review the reason why the calculation of a denominator bound for the solution of a PLDE is valuable. One naive way, which one aims to avoid, to search for solutions of \eqref{eq:PLDE} among the space of rational functions would be to start with an ansatz; for example, one can naively write a generic rational function in the variables $n_1,\ldots,n_r$ with undetermined coefficients $c_{\mathbf k}$ and $c_{\mathbf k'}$,
\begin{equation}
y(n_1,\ldots,n_r)=\frac{\sum\limits_{\mathbf k} c_\mathbf k \prod\limits_i n_i^{k_i} }{\sum\limits_{\mathbf k'} c_{\mathbf k'} \prod\limits_i n_i^{k'_i}}.
\label{eq:rational-ansatz}
\end{equation}
By plugging the ansatz \eqref{eq:rational-ansatz} into \eqref{eq:PLDE}, 
one 
obtains equations for the unknown coefficients $c_\mathbf k$ and 
$c_{\mathbf k'}$ by imposing the equality of every monomial in the 
variables $n_i$ on both sides of the equation, and one finds in this way, 
if they exist, the solutions having numerator and denominator of degree lower or equal to the degree chosen 
for the ansatz. However, the equations obtained in this way will be, in general, non--linear, and therefore difficult to solve.

As observed in the univariate case~\cite{abramov71} the situation improves if we are able to find a \emph{denominator bound} for the solutions. A denominator bound $d$ is a polynomial such that for any solution $y=\frac{n}{p}$ of \eqref{eq:PLDE}  it must be $p|d$: the denominator of the solution is a divisor of the denominator bound.

If we were able to calculate $d$ algorithmically, then only an ansatz for the numerator of the solution would be required, and the equations for the unknown coefficients $c_\mathbf k$ would be linear, and therefore easier to solve. It is possible to formulate an ansatz for the numerator which also includes terms involving harmonic sums \cite{Vermaseren:1998uu,Blumlein:1998if}, satisfying a wide class of recurrence equations.

If we write the solution to a partial linear difference equation as $y=\frac{n}{uv}$ with $u$ aperiodic and $v$ periodic, it is always possible to calculate a bound $d_a$ for the aperiodic part $u$ of the denominator. We refer to \cite{kauers10} for a description of how the aperiodic denominator bound is calculated. 

For the periodic part $v$ it is not always possible to obtain a complete denominator bound for a PLDE. This is illustrated for example by the equation
\begin{equation}
y(n+1,k) - y(n,k+1) = 0,
\label{eq:noDenBound}
\end{equation}
which is satisfied by $\frac{1}{(n+k)^\alpha}$ for any $\alpha \in \mathbb N$. Clearly, no polynomial can be a denominator bound for equation \eqref{eq:noDenBound}.

In other words, one cannot expect to obtain a complete denominator bound 
(due to the intrinsic problem that periodic factors might arise with 
arbitrary powers). Nevertheless, it is often possible to calculate a partial bound, and to identify what shape the factors of $v$, that cannot be predicted, 
must have. (A partial bound is a bound for some, but not all, the periodic factors).
The algorithm in \cite{kauers11} works by successively examining the 
periodic factors $u$ of the coefficients $a_\mathbf p$ when $\mathbf p$ is 
a ``corner point", see \cite{kauers11} for a definition. Applying all the tactics described in this article, one obtains an explicitly given polynomial $d_{p}$, a finite set of polynomials $P$ and a set of generators that spans a lattice $V$ in $\mathbb Z^r$ such for any solution of the given PLDE one can predict the aperiodic denominator part $v$ as follows:
$$v\mid d_p \cdot v_{\text{semi-known}} \cdot v_{\text{unknown}}$$
where $v_{\text{semi-known}}$ is a polynomial whose factors are take from the set $\{N_\mathbf s p^m\mid \mathbf 
s\in\mathbb Z^r, m\in\mathbb Z\}$ and $v_{\text{unknown}}$ is a polynomial such that\footnote{The spread of a polynomial $u$ is closely related to Abramov's definition of the dispersion~\cite{abramov71} and is defined by $\mathrm{spread}(u)=\{\mathbf s\in\mathbb Z^r\mid \gcd(u,N_\mathbf s u)\neq1\}$.} $\mathrm{spread}(v_{\text{unknown}})=V$.  

If $P\neq\{\}$ or $V\neq\{\}$, the implementation will print out the 
corresponding data in order to support the user to guess the missing parts 
$v_{\text{semi-known}}$ and/or $v_{\text{unknown}}$. Summarizing, the user 
will obtain a guidance in formulating an ansatz $d_{user}$ for the missing 
factors in the denominator of the solution. To force their inclusion in 
the search when looking for the numerator of the solution, one can use the 
option {\tt InsertDenFactor $\rightarrow$} $d_{\text{user}}$ of our 
package, cf.\ Section~\ref{sec:num} and Appendix~\ref{sec:D}.

\subsection{Determination of the numerator}
\label{sec:num}

\vspace*{1mm}
\noindent

Once the aperiodic and periodic denominator bounds $d_a,d_p$ are calculated, and possibly an ansatz $d_{user}$ for missing factors in the denominator has been set, one can search for the numerator contribution. In general, it has been shown in~\cite{AP:12} based on~\cite{Hilbert10} that this problem is unsolvable: given a homogeneous PLDE with polynomial coefficients, there does not exist an algorithm that can determine all polynomial solutions. Nevertheless, one can search for the desired polynomial solutions by taking as ansatz a general polynomial $\text{num}(c_i)$ with undetermined coefficients $c_i$ where the polynomial degree is set sufficiently high. Then one may substitute the rational function
\begin{equation}\label{Equ:PLDEAnsatz}
y=\frac{\text{num}(c_i)}{d_a d_p d_{\text{user}}}
\end{equation}
into the equation \eqref{eq:PLDE}. Then finding non--trivial solutions of the the underlying linear system 
allows one to specialize the $c_i$ such that $y$ is a solution of the given PLDE. 

In many cases, it is the determination of the $c_i$ which requires the largest computation time, whereas the denominator bounds can be computed quite quickly. For this reason we propose the following strategy to reduce the computation time.

In the cases where the PLDE does not contain any symbolic parameters (such 
as 
the dimensional regulator $\varepsilon$, or ratios of invariants) other than the shift variables, one may obtain constraints on the undetermined $c_i$ simply by plugging, sufficiently many times, random numerical values for the shift variables. Then one quickly obtains a linear system for the $c_i$.

If there are symbols present, instead, one may consider performing a first 
pass with the symbols replaced by random numbers, with the purpose of 
identifying and removing redundant constraints. Then, after removing the redundant equations for the $c_i$, the system can be solved in a stepwise manner, i.e. considering one at a time the constraints produced by one monomial, and plugging the result in the rest of the equation. This is what our package does when the function {\tt SolvePLDE} is called, cf. 
Appendix~\ref{sec:D}.

It is certainly possible that the use of random numbers to generate constraints can cause the system to generate two equations for the $c_i$ which are not independent. The probability of such an occurrence can be made arbitrarily small by choosing a sufficiently large range over which the random numbers are chosen. In any event, the consequence of an unfortunate draw of random numbers can only cause the software to output more functions misidentified as solutions when in fact they are not; it cannot cause the software to miss any solutions. By explicitly checking the result, one can guard against this remote possibility, at the expense of additional computation time.

In the following we elaborate further enhancements in order to extend the solution space from the rational function case to more general classes of functions. Besides the examples below, further examples for each aspect can be found in the {\tt Mathematica} notebook auxiliary to this paper.

\subsubsection{Treatment of a hypergeometric prefactor}
\label{sec:hypergeometric-prefactor}

\vspace*{1mm}
\noindent
Given a partial linear difference equation \eqref{eq:PLDE}, 
\begin{equation}
\sum_{\mathbf s \in S} a_\mathbf s N_\mathbf s y = f,
\label{eq:PLDE_2}
\end{equation}
it is possible to derive another difference equation 
\begin{equation}
\sum_{\mathbf s \in S} a'_\mathbf s N_\mathbf s y' = f',
\label{eq:PLDE1}
\end{equation}
whose solution $y'$ is related to $y$ by
\begin{equation}
y' = r y
\end{equation}
with $r=r(n_i)$ a hypergeometric function of its arguments, i.e. a function such that the ratio
\begin{equation}
\frac{N_{\mathbf e_i} r}{r} = \frac{r(n_1,\ldots,n_i+1,\ldots n_r)}{r(n_1,\ldots,n_i,\ldots,n_r)} 
\end{equation}
is for all $i$ a rational function of the variables $n_i$. Examples of 
hypergeometric functions are Pochhammer symbols, factorials, 
$\Gamma$-functions, binomial symbols, and obviously rational functions and polynomials.

The transformation from \eqref{eq:PLDE_2} to \eqref{eq:PLDE1} is useful whenever it is possible to formulate an ansatz for $r$. Once some specific form can be postulated for $r$, the equation \eqref{eq:PLDE1} is obtained by substitution and by exploiting the hypergeometric property.

Consider for example the equation
\begin{eqnarray}
&&
(1 + k) (\varepsilon + k) (1 + k + n^2) \, y(n, k) - 2 k (2 + k + n^2) \, y(n, 1 + k) 
\nonumber\\ &&
+ (1 + k) (\varepsilon + k) (2 + k + 2 n + n^2) \, y(1 + n, k) = 0.
\label{eq:example_5}
\end{eqnarray}
We assume that its solution is 
\begin{equation}
y(n,k) = (\varepsilon)_k \, y'(n,k)
\end{equation}
with $y'$ a rational function of $n$ and $k$ and $(\varepsilon)_k$ the Pochhammer symbol
\begin{equation}
(\varepsilon)_k = \varepsilon(\varepsilon+1)\cdots(\varepsilon+k-1).
\end{equation}
Then one derives a difference equation for $y'$, namely
\begin{eqnarray}
&& (\varepsilon + k) \big[(1 + k^2 + n^2 + k (2 + n^2)) y'(n, k) - 
2 k (2 + k + n^2) y'(n, 1 + k)
\nonumber\\&&
+ (2 + k^2 + 2 n + n^2 + k (3 + 2 n + n^2)) y'(1 + n, k)\big] = 0 .
\end{eqnarray}
We can now solve the new equation, obtaining
\begin{equation}
y'(n,k)=\frac{k}{1 + k + n^2}.
\end{equation}
From this we conclude that the solution of \eqref{eq:example_5} is 
\begin{equation}
y(n,k) = (\varepsilon)_k \, \frac{k}{1 + k + n^2} C,
\end{equation}
for some constant $C\in\KK(\varepsilon)$.

\subsubsection{Finding solutions in terms of nested sums}\label{sec:findingNestedSums}

Solutions connected to Feynman integrals involve often also indefinite nested sums, such as  (cyclotomic) harmonic sums \cite{Vermaseren:1998uu,Blumlein:1998if,Ablinger:2011te} or generalized versions, like Hurwitz
harmonic sums. A straightforward modification of the ansatz~\eqref{Equ:PLDEAnsatz} is to search for a numerator $\text{num}(c_i)$ that is built not only by a polynomial in $\KK[n_1,\dots,n_r]$ with the unknown coefficients $c_i$ but to search for polynomial expressions of a finite set of nested sums, i.e., one takes a linear combination of power products in terms of the given nested sums whose coefficients are polynomials in  $\KK[n_1,\dots,n_r]$ with unknown coefficients. In practice, it is important for this list of nested sums to be shift-stable, meaning that a shift in any of the variables must not introduce new harmonic sums not already included in the list, and they also should be linearly independent. To guarantee this property, one can use quasi-shuffle algebras or difference ring methods~\cite{AlgebraicRelations,TermAlgebra}. The nested sums at shifted arguments can then be rewritten through the repeated application of identities of the type
\begin{equation}\label{Equ:SumShiftRules}
S_1(n+i) = \frac{1}{n+i} + S_1(n+i-1)
\end{equation}
and similarly for all other nested sums, until only unshifted nested sums appear. After clearing denominators, the PLDE implies a set of linear constraints on the undetermined parameters $c_i$, obtained by coefficient comparison in all the power products which appear when $y$ is plugged back into the equation~\eqref{eq:PLDE}.
Note that the number of unknowns $c_i$ increases strongly: one tries to determine not only one numerator polynomial but numerous polynomials for each power product. In this regard, the homomorphic image techniques described in the beginning of Section~\ref{sec:num} are instrumental to perform these calculations in reasonable time.

\medskip

This heuristic method provides in many cases the desired solution. For instance, consider the equation
\begin{eqnarray}
	&&(-k-1) \left(k+n^2+2 n+1\right) f(n,k)
	+k \left(k+n^2+2 n+2\right) f(n,k+1)
\nonumber\\&&		
	+2 (k+1) \left(k+n^2+4 n+4\right) f(n+1,k)
	-2 k \left(k+n^2+4 n+5\right) f(n+1,k+1)
\nonumber\\&&	
	-(k+1) \left(k+n^2+6 n+9\right) f(n+2,k)
	+k \left(k+n^2+6 n+10\right) f(n+2,k+1)=0,
\nonumber\\
\end{eqnarray}
Looking for solutions of the form described, with a numerator of degree up to 2, containing the harmonic sums $S_1(n),S_1(k),S_{2,1}(n)$ the algorithm finds the denominator
\begin{equation}
	d_p=1 + k + 2 n + n^2
\end{equation}
and the corresponding numerators of the solutions of the homogeneous equation:
\begin{eqnarray}
	1, k, k^2, n, k n, S_1(k), k S_1(k), n S_1(k), S_1(k)^2,k S_1(n), k S_{2,1}(n).
\end{eqnarray}

\medskip
\noindent\textit{Remark.} A more advanced (and also less heuristic) tactic is to apply a recursive strategy as worked out in~\cite{LinearSolver}: one defines an order of the nested sums $(s_1,s_2,\dots,s_e)$ where a sum $s_i$ does not arise inside of any of the sums $s_1,\dots,s_{i-1}$.
Then one makes an ansatz for the solution $y=p_0+p_1\,s_e\dots+p_d s_e^d$ where $p_0,\dots,p_d$ are polynomial expressions in terms of the remaining nested sums $s_0,\dots,s_{e-1}$ with coefficients from the ground field $\KK(n_1,\dots,n_r)$. Here one has to set up $d$ sufficiently high in order to guarantee that the desired solution can be derived. Then one plugs $y$ into~\eqref{eq:PLDE}, applies the shift rules such as~\eqref{Equ:SumShiftRules}, clears denominators and compares the coefficients of the highest term $s_e^d$. This yields a new PLDE in terms of the unknown $p_d$. Now we compute by recursion all the solutions of this new PLDE in terms of the remaining sums $s_1,\dots,s_{e-1}$, plug the solutions into $p_d$ of the original system and obtain an updated PLDE of~\eqref{eq:PLDE} where $s_e$ occurs only up to degree $d-1$. Now we proceed by degree reduction to compute the remaining coefficients $p_0,\dots,p_{d-1}$ in order to obtain the final solution $y$. We remark that in the base cases, i.e., when all sums are removed within the recursion one ends up to solve several PLDEs purely in the ground field $\KK(n_1,\dots,n_r)$, i.e., the machinery described in the beginning of Section~\ref{sec:PLDEsolver} is applied. It is our plan in the near future to implement this more advanced machinery within the formal setting  of $R\Pi\Sigma$-difference ring extensions~\cite{DR} based on the reduction strategy given in~\cite{LinearSolver}.

\subsubsection{Matching the solution to initial values}
\label{sec:initVal}

\vspace*{1mm}
\noindent
If initial values are provided, it is possible to look for a general solution that conforms to them.
This general solution is found by building a linear combination with 
undetermined coefficients of the solutions of the homogeneous equation, 
plus a particular solution of the equation. Next, the initial values are 
plugged in, and a system of equations is obtained. In the case that the 
system contains symbolic parameters other than the shift variables, the 
undetermined coefficients to be searched for are not just numbers. In that case, the coefficients of the 
linear combination are taken to be general rational functions in the parameters up to some chosen degree. 
The combination of the solutions will be of particular importance for the next subsection.
\subsubsection{Finding the solution in a series expansion}
\label{sec:expansion}

\vspace*{1mm}
\noindent
In many applications it is desirable to obtain the Laurent series expansion of the solution of a difference 
equation. This may be easier to achieve than the derivation of a complete solution, because, at each order in the expansion, it is possible to derive a difference equation where the expansion parameter is absent, therefore the linear system to find the coefficients $c_i$ can potentially be solved much more quickly. The procedure, described in the following, generalizes the univariate case given in~\cite{Blumlein:2010zv}. It assumes that the initial values of the solution in its $\ep$--expansion are known. 

Consider for instance of \eqref{eq:PLDE}, possibly containing a parameter 
$\varepsilon$ in the coefficients:
\begin{equation}
\sum_{\mathbf s \in S} a_\mathbf s(n_i,\varepsilon) N_\mathbf s y(n_i) = f(n_i,\varepsilon) ,
\label{eq:PLDE-eps}
\end{equation}
where the coefficients $a_{\mathbf s}(n_i,\varepsilon)$ are polynomials in the shift variables and in the parameter $\varepsilon$. Assume that the solution of \eqref{eq:PLDE-eps} has, around $\varepsilon=0$, a Laurent expansion starting from the power $\varepsilon^{-\ell}$ of the parameter, with $\ell$ known,
\begin{equation}
	y_\varepsilon(n_i) = \varepsilon^{-\ell} y_{-\ell}(n_i) + \cdots + y_0(n_i) + \varepsilon y_1(n_i) + \cdots + \varepsilon^c y_c(n_i) ,
\label{eq:y_eps}	
\end{equation}
and that the right-hand side of the equation can be expanded in a series in $\varepsilon$ as
\begin{equation}
	f =  \varepsilon^{-\ell} f_{-\ell}(n_i) + \varepsilon^{-\ell+1} f_{-\ell+1} (n_i) + \cdots .
\label{eq:f_eps}	
\end{equation}
Assume also that the $a_{\mathbf s}(n_i,\varepsilon=0)$ are not all zero, so that an overall power of $\varepsilon$, if present in the equation, has been factored out.
Then, one may proceed by inserting \eqref{eq:y_eps} and \eqref{eq:f_eps} into \eqref{eq:PLDE-eps} and doing a coefficient comparison of the $\varepsilon^{-\ell}$ terms, obtaining
\begin{equation}
	\sum_{\mathbf s \in S} a_\mathbf s(n_i,\varepsilon=0) N_\mathbf s y_{-\ell}(n_i) = f_{-\ell}(n_i).
\label{eq:PLDE_eps-l}
\end{equation}
Equation \eqref{eq:PLDE_eps-l} is now free of $\varepsilon$, which facilitates the task of finding a solution and reduces the computational time required. If \eqref{eq:PLDE_eps-l} can be uniquely solved for $y_{-\ell}$ and the solution matched to initial values, one can move to the next higher power in $\varepsilon$ by plugging the solution into \eqref{eq:y_eps}. In this new equation one does a coefficient comparison of the next power in $\varepsilon$ and solves for $y_{\ell+1}$. The process is repeated as many times as needed until all the terms of interest in the Laurent expansion are obtained.

For instance, consider the equation
\begin{eqnarray}
&&\big[3 (k+1) (n+1)+4 (n+1)+1\big] (4 k n^2 \varepsilon ^3+5 n \varepsilon +6 \varepsilon ^2+1) f(n+1,k+1)
\nonumber\\&&
-(3 k n+4 n+1) \big[4 (k+1) (n+1)^2 \varepsilon ^3+5 (n+1)
   \varepsilon +6 \varepsilon ^2+1\big] f(n {,k} ) = 0 .
\label{eq:example-series}   
\end{eqnarray}
Together with a list of 25 initial values, our procedure to compute the expansion encounters at order 
$\varepsilon^{-2},\varepsilon^{-1},\varepsilon^0$ the equations
\begin{eqnarray}
	(-3 k n-4 n-1) f(n,k)+(3 k n+3 k+7 n+8) f(n+1,k+1) &=& \tau , 
\end{eqnarray}
with $\tau = 0, 5, 0$, respectively,
which are free of $\varepsilon$. The series solution of \eqref{eq:example-series} is found to be
\begin{equation}
	f(n,k) = \frac{1}{\varepsilon ^2 (3 k n+4 n+1)}+\frac{5 n}{\varepsilon  (3 k n+4 n+1)}+\frac{6}{3 k n+4 n+1} + \mathcal O(\varepsilon).
\end{equation}

\section{Conclusions}
\label{sec:conclusion}

\vspace*{1mm}
\noindent
We reviewed some techniques, algorithms and implementation choices for the solution of 
partial linear differential equations in form of multivariate power series representations. Here we extract the underlying partial linear difference equations of the power series coefficients (see, e.g., Section~\ref{sec:reclist}) and try to solve them in terms of special functions. For this task we presented an algorithm that can solve frequently arsing hypergeometric systems in terms of hypergeometric products (see Section~\ref{sec:recsol}) and elaborated heuristic methods to find such solutions (also in terms of nested sums) for the general higher-order case (see Section~\ref{sec:PLDEsolver}). Special care has been put on the 
$\ep$--expansion of such solutions (see Sections~\ref{Sec:EpExpansion} and~\ref{sec:expansion}) where in 
addition, e.g., Hurwitz harmonic sums and generalized versions may therefore arise. Finally, we utilize the 
available summation tools in the setting of difference rings to simplify the found sum solutions in terms of indefinite nested sums over hypergeometric products. In particular, various concrete examples of this computer algebra machinery has been elaborated (see Section~\ref{sec:fullmachinery}).

\vspace{5mm}\noindent
{\bf Acknowledgment.}~
We thank J.~Ablinger, D.~Broadhurst, and  P.~Marquard for discussions. This project has received 
funding from the European Union's Horizon 2020 research and innovation programme under the Marie 
Sk\l{}odowska--Curie grant agreement No. 764850, SAGEX and from the Austrian Science Fund (FWF) 
grant SFB F50 (F5009-N15) and P33530.

\appendix
\section{The multiple series representation} 
\label{sec:A}

\vspace*{1mm} 
\noindent 
In the following we summarize the different multiple series representations that can be found in the existing literature. We note that starting with their partial linear differential representation our methods described above can also provide the the presented sum representations. The expansion coefficients are given
as rational functions of Pochhammer symbols, which requires a corresponding re-parameterization of the coefficients 
of the foregoing differential and difference equations.

One of the simplest function of these classes is Gau\ss{}' hypergeometric 
function
\begin{eqnarray}
\pFq{2}{1}{a_1,a_2}{b_1}{z} =
\sum_{l=0}^\infty \frac{(a_1)_l (a_2)_l}{(b_1)_l} \frac{z^l}{l!}.
\end{eqnarray}
Its generalizations, the generalized hypergeometric functions, are given by
\begin{eqnarray}
\pFq{p}{q}{a_1,...,a_p}{b_1,...b_q}{z} =
\sum_{l=0}^\infty \frac{(a_1)_l ... (a_p)_l}{(b_1)_l ... (b_q)_l} \frac{z^l}{l!}.
\end{eqnarray}
The bi--variate Appell series \cite{APPELL1,APPELL2} have the representation
\begin{eqnarray}
F_1(a;b,b';c;x,y) &=& \sum_{m=0}^\infty \sum_{n=0}^\infty 
\frac{(a)_{m+n} (b)_m (b')_n}{m! n! (c)_{m+n}} x^m y^n
\\
F_2(a;b,b';c,c';x,y) &=& \sum_{m=0}^\infty \sum_{n=0}^\infty 
\frac{(a)_{m+n} (b)_m (b')_n}{m! n! (c)_m (c')_n} x^m y^n
\\
F_3(a,a';b,b';c;x,y) &=& \sum_{m=0}^\infty \sum_{n=0}^\infty 
\frac{(a)_m  (a')_n (b)_m (b')_n}{m! n! (c)_{m+n}}x^m y^n
\\
F_4(a;b;c,c';x,y) &=& \sum_{m=0}^\infty \sum_{n=0}^\infty 
\frac{(a)_{m+n} (b)_{m+n}}{m! n! (c)_m (c')_n}x^m y^n.
\end{eqnarray}
This set is extended to the Horn-type bi--variate series \cite{HORN}
\begin{eqnarray}
G_1(a;b,b';x,y) &=& \sum_{m=0}^\infty \sum_{n=0}^\infty 
(a)_{m+n} (b)_{n-m} (b')_{m-n} \frac{x^m y^n}{m! n!}
\\
G_2(a,a';b,b';x,y) &=& \sum_{m=0}^\infty \sum_{n=0}^\infty 
(a)_m  (a')_n (b)_{n-m} (b')_{m-n} \frac{x^m y^n}{m! n!}
\\
G_3(a,a';x,y) &=& \sum_{m=0}^\infty \sum_{n=0}^\infty 
(a)_{2n-m}  (a')_{2m-n} \frac{x^m y^n}{m! n!}
\\
H_1(a;b;c;d;x,y) &=& \sum_{m=0}^\infty \sum_{n=0}^\infty 
\frac{(a)_{m-n}  (b)_{m+n}  (c)_n}{(d)_m} \frac{x^m y^n}{m! n!}
\\
{H_2(a;b;c;d;e;x,y)} &=& \sum_{m=0}^\infty \sum_{n=0}^\infty 
{
\frac{(a)_{m-n}  (b)_m  (c)_n (d)_n }{(e)_m} \frac{x^m y^n}{m! n!}
}
\\
H_3(a;b;c;x,y) &=& \sum_{m=0}^\infty \sum_{n=0}^\infty 
\frac{(a)_{2m+n}  (b)_{n}}{(c)_{m+n}}\frac{x^m y^n}{m! n!}
\\
H_4(a;b;c;d;x,y) &=& \sum_{m=0}^\infty \sum_{n=0}^\infty 
\frac{(a)_{2m+n}  (b)_{n}}{(c)_{m} (d)_n}\frac{x^m y^n}{m! n!}
\\
H_5(a;b;c;x,y) &=& \sum_{m=0}^\infty \sum_{n=0}^\infty 
\frac{(a)_{2m+n}  (b)_{n-m}}{(c)_{n}}\frac{x^m y^n}{m! n!}
\\
H_6(a;b;c;x,y) &=& \sum_{m=0}^\infty \sum_{n=0}^\infty 
(a)_{2m-n}  (b)_{n-m} (c)_{n} \frac{x^m y^n}{m! n!}
\\
H_7(a;b;c;d;x,y) &=& \sum_{m=0}^\infty \sum_{n=0}^\infty 
\frac{(a)_{2m-n}  (b)_{n} (c)_{n}}{(d)_m} \frac{x^m y^n}{m! n!}.
\end{eqnarray}
As in the case of the generalized hypergeometric functions there are confluent forms of the bi--variate
functions.

There are also the further bi--variate series \cite{Anastasiou:1999ui,Anastasiou:1999cx}
\begin{eqnarray}
S_1(a,a',b;c,d;x,y) &=& \sum_{m=0}^\infty \sum_{n=0}^\infty
\frac{(a)_{m+n}  (a')_{m+n} (b)_{m}}{(c)_{m+n} (d)_m} \frac{x^m y^n}{m! n!}.
\\
S_2(a,a',b, b';c;x,y) &=& \sum_{m=0}^\infty \sum_{n=0}^\infty
\frac{(a)_{m-n}  (a')_{m-n} (b)_{n} (b')_m}{(c)_{m-n}} \frac{x^m y^n}{m! n!}.
\end{eqnarray}

The generalization of the bi--variate hypergeometric functions is the Kamp\'e de F\'eriet function \cite{KAMPE2}
\begin{eqnarray}
F_{C;D;D'}^{A;B:B'}\biggl[\genfrac..{0pt}{}{a,b,b'}{c,d,d'};x,y\biggr] = \sum_{m,n = 0}^\infty
\frac{
(a_1)_{m+n} ... (a_A)_{m+n} 
(b_1)_{m} ... (b_B)_{m} 
(b'_1)_{n} ... (b'_{B'})_{n}}
{(c_1)_{m+n} ... (c_C)_{m+n}
(d_1)_{m} ... (d_D)_{m}  
(d'_1)_{n} ... (d'_{D'})_{n}} \frac{x^m y^n}{m! n!}. 
\end{eqnarray}

Triple hypergeometric functions of the Lauricella--Saran type 
\cite{Lauricella:1893,Saran:1954,Saran:1955} are
\begin{eqnarray}
\hat{F}_4 &=& F_E(a_1,a_1,a_1,b_1,b_2;c_1,c_2,c_3;x,y,z) = 
\sum_{m=0,n=0,p=0}^\infty 
\frac{(a_1)_{m+n+p} (b_1)_m (b_2)_{n+p}}{(c_1)_m (c_1)_n (c_1)_p} \frac{x^m y^n z^p}{m! n! p!}
\nonumber\\
\\
\hat{F}_{14} &=& F_F(a_1,a_1,a_1,b_1,b_2,b_1;c_1,c_2,c_2;x,y,z) = 
\sum_{m=0,n=0,p=0}^\infty 
\frac{(a_1)_{m+n+p} (b_1)_{m+p} (b_2)_{n}}{(c_1)_m (c_2)_{n+p}} \frac{x^m y^n z^p}{m! n! p!}
\nonumber\\
\\
\hat{F}_{8} &=& F_G(a_1,a_1,a_1,b_1,b_2,b_3;c_1,c_2,c_2;x,y,z) = 
\sum_{m=0,n=0,p=0}^\infty
\frac{(a_1)_{m+n+p} (b_1)_{m} (b_2)_{n} (b_3)_{p}}{(c_1)_m (c_2)_{n+p}} \frac{x^m y^n z^p}{m! n! p!}
\nonumber\\
\\
\hat{F}_{3} &=& F_K(a_1,a_2,a_2,b_1,b_2,b_1;c_1,c_2,c_3;x,y,z) = 
\sum_{m=0,n=0,p=0}^\infty
\frac{(a_1)_{m} (a_2)_{n+p}(b_1)_{m+p} (b_2)_{n}}{(c_1)_m (c_2)_{n} (c_3)_p} \frac{x^m y^n z^p}{m! n! p!}
\nonumber\\
\\
\hat{F}_{11} &=& F_M(a_1,a_2,a_1,b_1,b_2,b_1;c_1,c_2,c_2;x,y,z) = 
\sum_{m=0,n=0,p=0}^\infty
\frac{(a_1)_{m} (a_2)_{n+p}(b_1)_{m+p} (b_2)_{n}}{(c_1)_m (c_2)_{n+p}} \frac{x^m y^n z^p}{m! n! p!}
\nonumber\\
\\
\hat{F}_{6} &=& F_N(a_1,a_2,a_3,b_1,b_2,b_1;c_1,c_2,c_2;x,y,z) = 
\sum_{m=0,n=0,p=0}^\infty
\frac{(a_1)_{m} (a_2)_{n} (a_3)_{p} (b_1)_{m+p} (b_2)_{n}}{(c_1)_m (c_2)_{n+p}} \frac{x^m y^n z^p}{m! n! p!}
\nonumber\\
\\
\hat{F}_{12} &=& F_P(a_1,a_2,a_1,b_1,b_1,b_2;c_1,c_2,c_2;x,y,z) = 
\sum_{m=0,n=0,p=0}^\infty
\frac{(a_1)_{m+p} (a_2)_{n} (b_1)_{m+n} (b_2)_{p}}{(c_1)_m (c_2)_{n+p}} \frac{x^m y^n z^p}{m! n! p!}
\nonumber\\
\\
\hat{F}_{10} &=& F_P(a_1,a_2,a_1,b_1,b_2,b_1;c_1,c_2,c_2;x,y,z) = 
\sum_{m=0,n=0,p=0}^\infty
\frac{(a_1)_{m+p} (a_2)_{n} (b_1)_{m+p} (b_2)_{n}}{(c_1)_m (c_2)_{n+p}} \frac{x^m y^n z^p}{m! n! p!}
\nonumber\\
\\
\hat{F}_{7} &=& F_P(a_1,a_2,a_2,b_1,b_2,b_3;c_1,c_1,c_1;x,y,z) = 
\sum_{m=0,n=0,p=0}^\infty
\frac{(a_1)_{m} (a_2)_{n+p} (b_1)_{m} (b_2)_{n} (b_3)_p}{(c_1)_{m+n+p}} \frac{x^m y^n z^p}{m! n! p!}
\nonumber\\
\\
\hat{F}_{13} &=& F_T(a_1,a_2,a_2,b_1,b_2,b_1;c_1,c_1,c_1;x,y,z) = 
\sum_{m=0,n=0,p=0}^\infty
\frac{(a_1)_{m} (a_2)_{n+p} (b_1)_{m+p} (b_2)_{n}}{(c_1)_{m+n+p}} \frac{x^m y^n z^p}{m! n! p!}
\nonumber\\
\end{eqnarray}

There are three more triple hypergeometric functions, the Srivastava functions \cite{SRIKARL},
\begin{eqnarray}
\hat{H}_{A}(a,b,b';c,c';x,y,z) &=& \sum_{m=0,n=0,p=0}^\infty
\frac{(a)_{m+p} (b)_{m+n} (b')_{n+p}}{(c)_{m} { (c')} _{n+p} } \frac{x^m y^n z^p}{m! n! p!}
\\
\hat{H}_{B}(a,b,b';c_1,c_2,c_3;x,y,z) &=& \sum_{m=0,n=0,p=0}^\infty
\frac{(a)_{m+p} (b)_{m+n} (b')_{n+p}}{(c_1)_{m} (c_2)_{n} (c_3)_p} \frac{x^m y^n z^p}{m! n! p!}
\\
\hat{H}_{C}(a,b,b';c;x,y,z) &=& \sum_{m=0,n=0,p=0}^\infty
\frac{(a)_{m+p} (b)_{m+n} (b')_{n+p}}{(c)_{m+n+p}} \frac{x^m y^n z^p}{m! n! p!}.
\end{eqnarray}
These functions are given in the literature in different forms. The following identities hold comparing to 
Ref.~\cite{SRIKARL}:
{%
\begin{eqnarray}
\hat F_4(a_1,b_1,b_2;c_1;x,y,z)            &\to& f_{20a}(a_1,b_2,b_1;c_1;z,y,x) \\
\hat F_{14}(a_1,b_1,b_2;c_1;x,y,z)         &\to& f_{21a}(a_1,b_1,b_2;c_1,c_2;x,z,y) \\
\hat F_8(a_1,b_1,b_2,b_3;c_1,c_2;x,y,z)    &\to& f_{18a}(a_1,b_3,b_2,b_1;c_2,c_1;z,y,x) \\
\hat F_3(a_1,a_2,b_1,b_2;c_1,c_2,c_3;x,y,z)&\to& f_{10a}(a_2,b_1,b_2,a_1;c_2,c_3,c_1;y,z,x) \\
\hat F_{11}(a_1,a_2,b_1,b_2;c_1,c_2;x,y,z) &\to& f_{11a}(a_2,b_1,b_2,a_1;c_2,c_1,y,z,x) \\
\hat F_6(a_1,a_2,a_3,b_1,b_2;c_1,c_2;x,y,z)&\to& f_{6a}(b_1,a_1,a_3,a_2,b_1;c_1,c_2;x,z,y)  \\
\hat F_{12}(a_1,a_2,b_1,b_2;c_1,c_2;x,y,z) &\to& f_{12a}(b_1,a_1,a_2,b_2;c_2,c_1;y,x,z) \\
\hat F_{10}(a_1,a_2,b_1,b_2;c_1,c_2;x,y,z) &\to& f_{9a}(b_1,a_1,a_2,b_2;c_1,c_2;x,z,y)  \\
\hat F_7(a_1,a_2,a_3,b_1,b_2,b_3;c_1;x,y,z)&\to& f_{7a}(a_2,b_2,b_3,b_1,a_1;c_1;y,z,x)  \\
\hat F_{13}(a_1,a_2,b_1,b_2;c_1;x,y,z)     &\to& f_{13a}(a_2,b_1,b_2,a_1;c_1;y,z,x) \\
\hat H_A(a,b,b_1;c,c_1;x,y,z)              &\to& f_{15a}(b_1,a,b;c_1{,c};y,z,x) \\
\hat H_B(a,b,b_1;c_1,c_2,c_3;x,y,z)        &\to& f_{14a}(b,b_1,a;c_1,c_2,c_3;x,y,z) \\
\hat H_C(a,b,b_1;c;x,y,z)                  &\to& f_{16a}(b,b_1,a;c;x,y,z).
\end{eqnarray}
} 
A comprehensive list of triple hypergeometric series $f_{1b}$ to  $f_{62b}$ has been given in 
Ref.~\cite{SRIKARL,TRIPLE}.

The quadruple hypergeometric functions by Exton \cite{EXTON72a} are\footnote{Note various typographical errors in the 
literature.}
\begin{eqnarray}
K_{1} &=& \sum_{l,m,n,p = 0}^\infty \frac{(a)_{l+m+n+p} (b_1)_{l+m+n} (b_2)_p}{(c_1)_{l+p} (c_2)_m (c_3)_n}
\frac{x^l y^m z^n t^p}{l! m! n! p!}
\\
K_{2} &=& \sum_{l,m,n,p = 0}^\infty \frac{(a)_{l+m+n+p} (b_1)_{l+m+n} (b_2)_p}{(c_1)_{l} (c_2)_m (c_3)_n (c_4)_p}
\frac{x^l y^m z^n t^p}{l! m! n! p!}
\\
K_{3} &=& \sum_{l,m,n,p = 0}^\infty \frac{(a)_{l+m+n+p} (b_1)_{l+m} (b_2)_{n+p}}{(c_1)_{l+p} (c_2)_{m+n}}
\frac{x^l y^m z^n t^p}{l! m! n! p!}
\\
K_{4} &=& \sum_{l,m,n,p = 0}^\infty \frac{(a)_{l+m+n+p} (b_1)_{l+m} (b_2)_{n+p}}{(c_1)_{l+p} (c_2)_{m} (c_3)_n}
\frac{x^l y^m z^n t^p}{l! m! n! p!}
\\
K_{5} &=& \sum_{l,m,n,p = 0}^\infty \frac{(a)_{l+m+n+p} (b_1)_{l+m} (b_2)_{n+p}}{(c_1)_{l} (c_2)_m (c_3)_n (c_4)_p }
\frac{x^l y^m z^n t^p}{l! m! n! p!}
\\
K_{6} &=& \sum_{l,m,n,p = 0}^\infty \frac{(a)_{l+m+n+p} (b_1)_{l+m} (b_2)_{n} (b_3)_p}{(c_1)_{l} (c_2)_{m+n+p}}
\frac{x^l y^m z^n t^p}{l! m! n! p!}
\\
K_{7} &=& \sum_{l,m,n,p = 0}^\infty \frac{(a)_{l+m+n+p} (b_1)_{l+m} (b_2)_{n} (b_3)_p}{(c_1)_{l+n} (c_2)_{m+p}}
\frac{x^l y^m z^n t^p}{l! m! n! p!}
\\
K_{8} &=& \sum_{l,m,n,p = 0}^\infty \frac{(a)_{l+m+n+p} (b_1)_{l+m} (b_2)_{n} (b_3)_p}{(c_1)_{l+n} (c_2)_{m} (c_3)_{p}}
\frac{x^l y^m z^n t^p}{l! m! n! p!}
\\
{K_{9}} &=& 
{
\sum_{l,m,n,p = 0}^\infty \frac{(a)_{l+m+n+p} (b_1)_{l+m} (b_2)_{n} (b_3)_{p}}{(c_1)_l (c_2)_m (c_3)_{n+p}}
\frac{x^l y^m z^n t^p}{l! m! n! p!}
}
\\
K_{10} &=& \sum_{l,m,n,p = 0}^\infty \frac{(a)_{l+m+n+p} (b_1)_{l+m} (b_2)_{n} (b_3)_p}{(c_1)_{l} (c_2)_{m} (c_3)_{n} (c_4)_p}
\frac{x^l y^m z^n t^p}{l! m! n! p!}
\\
K_{11} &=& \sum_{l,m,n,p = 0}^\infty \frac{(a)_{l+m+n+p} (b_1)_{l} (b_2)_{m} (b_3)_n (b_4)_p}{(c_1)_{l+m+n} (c_2)_{p}}
\frac{x^l y^m z^n t^p}{l! m! n! p!}
\\
K_{12} &=& \sum_{l,m,n,p = 0}^\infty \frac{(a)_{l+m+n+p} (b_1)_{l} (b_2)_{m} (b_3)_n (b_4)_p}{(c_1)_{l+m} (c_2)_{n+p}}
\frac{x^l y^m z^n t^p}{l! m! n! p!}
\\
K_{13} &=& \sum_{l,m,n,p = 0}^\infty \frac{(a)_{l+m+n+p} (b_1)_{l} (b_2)_{m} (b_3)_n (b_4)_p}{(c_1)_{l+m} (c_2)_{n} (c_3)_p}
\frac{x^l y^m z^n t^p}{l! m! n! p!}
\\
K_{14} &=& \sum_{l,m,n,p = 0}^\infty \frac{(a)_{l+m+n} (b_1)_{p} (b_2)_{l+p} (b_3)_m (b_4)_n}{({c})_{l+m+n+p}}
\frac{x^l y^m z^n t^p}{l! m! n! p!}
\\
K_{15} &=& \sum_{l,m,n,p = 0}^\infty \frac{(a)_{l+m+n} (b_1)_{p} (b_2)_{l} (b_3)_m (b_4)_n (b_5)_p}{(c_1)_{l+m+n+p}}
\frac{x^l y^m z^n t^p}{l! m! n! p!}
\\
K_{16} &=& \sum_{l,m,n,p = 0}^\infty \frac{({a_1})_{l+m} ({a_2})_{l+n} ({a_3})_{m+p} ({a_4})_{n+p}}{({c})_{l+m+n+p}}
\frac{x^l y^m z^n t^p}{l! m! n! p!}
\\
K_{17} &=& \sum_{l,m,n,p = 0}^\infty \frac{(a)_{l+m} (b_1)_{l+n} (b_2)_{m+n} (b_3)_{p} (b_4)_p}{({c})_{l+m+n+p}}
\frac{x^l y^m z^n t^p}{l! m! n! p!}
\\
K_{18} &=& \sum_{l,m,n,p = 0}^\infty \frac{(a)_{l+m} (b_1)_{l+p} (b_2)_{m+n} (b_3)_{n} (b_4)_p}{(c_1)_{l+m+n+p}}
\frac{x^l y^m z^n t^p}{l! m! n! p!}
\\
K_{19} &=& \sum_{l,m,n,p = 0}^\infty \frac{(a)_{l+m} (b_1)_{l+n} (b_2)_{m} (b_3)_{n} (b_4)_p (b_5)_p}{(c_1)_{l+m+n+p}}
\frac{x^l y^m z^n t^p}{l! m! n! p!}
\\
K_{20} &=& {
\sum_{l,m,n,p = 0}^\infty \frac{(a_1)_{l+m} (a_2)_{n+p} (b_1)_{l} (b_2)_{m} (b_3)_{n} (b_4)_p}{(c_1)_{l+m+n+p}}
\frac{x^l y^m z^n t^p}{l! m! n! p!}
}
\\
K_{21} &=& {
\sum_{l,m,n,p = 0}^\infty \frac{(a)_{l+m} (b_1)_{l} (b_2)_{m} (b_3)_{n} (b_4)_p (b_5)_n (b_6)_p}{(c_1)_{l+m+n+p}}
\frac{x^l y^m z^n t^p}{l! m! n! p!}.
}
\end{eqnarray}

Furthermore, there are the multivariate Lauricella functions
\cite{Lauricella:1893} 
\begin{eqnarray}
F_A^{(n)}(a,b_1,...,b_n;c_1,...,c_n;x_1,...,x_n) 
&=& \sum_{m_1,...,m_n=0}^\infty \frac{(a)_{m_1+...+m_n} (b_1)_{m_1} ... 
(b_n)_{m_n}}{(c_1)_{m_1} ... 
(c_n)_{m_n}}
\frac{x_1^{m_1} ... x_n^{m_n}}{m_1! ... m_n!}
\nonumber\\
\\
F_B^{(n)}(a_1 ... a_n,b_1,...,b_n;c_1,...,c_n;x_1,...,x_n) 
&=& \sum_{m_1,...,m_n=0}^\infty \frac{(a_1)_{m_1} ... (a_n)_{m_n} 
(b_1)_{m_1} ... (b_n)_{m_n}}{(c_1)_{m_1} ... (c_n)_{m_n}}
\frac{x_1^{m_1} ... x_n^{m_n}}{m_1! ... m_n!}
\nonumber\\ 
\\
F_C^{(n)}(a,b;c_1,...,c_n;x_1,...,x_n) 
&=& \sum_{m_1,...,m_n=0}^\infty \frac{
(a)_{m_1+ ... m_n} 
(b)_{m_1+ ... m_n}}{(c_1)_{m_1} ... (c_n)_{m_n}}
\frac{x_1^{m_1} ... x_n^{m_n}}{m_1! ... m_n!}
\\
F_D^{(n)}(a,b_1 ... b_n;c;x_1,...,x_n) 
&=& \sum_{m_1,...,m_n=0}^\infty \frac{
(a)_{m_1+ ... m_n} 
(b_1)_{m_1} ... (b_n)_{m_n}}{(c)_{m_1 + ... m_n}}
\frac{x_1^{m_1} ... x_n^{m_n}}{m_1! ... m_n!}.
\nonumber\\
\end{eqnarray}
The file {\tt cases.m} gives an even more
extensive computer--readable list of these functions. 
\section{Mapping conditions to the Pochhammer case}
\label{sec:B}

\vspace*{1mm}
\noindent
The parameters of the general representations of the (partial) differential equations given in Section~\ref{sec:delist},
leading to product solutions  obey the well--know Pochhammer solutions, if they apply a number of relations.
Here we present some typical examples for these relations.

\vspace*{2mm}
\noindent
\underline{{\sf The case of two variables:}}\\
\begin{eqnarray}
\label{eq:F1a}
F_1 &:& \{a\}\to \alpha  \beta ,\{b\}\to \alpha +\beta +1,\{c,c_1\}\to -\gamma ,\{d,g,d_1,g_1\}\to -1,\{e,h,e_1,h_1\}\to 1,
\nonumber\\&&
\{f\}\to \beta ,\{j,j_1\}\to 0,\{a_1\}\to \alpha  \beta _1,\{b_1\}\to \alpha +\beta _1+1,\{f_1\}\to \beta _1\}\\
H_1 &:& \{a\}\to \alpha  \beta ,\{b\}\to \alpha +\beta +1,\{c\}\to -\delta ,\{d,j,g_1\}\to -1,\{e,d_1,e_1,h_1\}\to 1,
\nonumber\\&&
\{f\}\to \alpha -\beta -1,\{g,h,j_1\}\to 0,\{a_1\}\to \beta  \gamma ,\{b_1\}\to \beta +\gamma +1,\{c_1\}\to 1-\alpha ,
\nonumber\\&&
\{f_1\}\to \gamma \\
S_1 &:&
\{ a \} \to \alpha \alpha_1 \beta;
\{ b \} \to (1 + \alpha_1) (1 + \beta) + \alpha (1 + \alpha_1 +
\beta);
\{ c \} \to -(\gamma \delta);
\nonumber\\&&
\{ d \} \to -1 - \gamma - \delta;
\{ e \} \to 3 + \alpha + \alpha_1 + \beta;
\{ f \} \to (1 + \alpha + \alpha_1) \beta;
\{ g \} \to -\delta;
\nonumber\\&&
\{ h \} \to 3 + \alpha + \alpha_1 + 2 \beta;
\{ j \} \to \beta;
\{ l, q, d_1, g_1 \} \to -1;
\{ p, s, e_1, j_1 \} \to 1;
\nonumber\\&&
\{ r, h_1 \} \to 2;
\{ a_1 \} \to \alpha \alpha_1;
\{ b_1, f_1 \} \to 1 + \alpha + \alpha_1;
\{ c_1 \} \to -\gamma. 
\end{eqnarray}

\noindent
{\underline{\sf The case of three variables, \cite{SRIKARL}:}}\\
\begin{eqnarray}
f_{1b} &:&
\{ A \} \to b_1 b_2;
\{ B_0 \} \to -c;
\{ B_1 \} \to 1 + b_1 + b_2;
\{ C_1 \} \to -b_2;
\{ D_1 \} \to -b_1;
\nonumber\\&&
\{ E_0, H_1, L_1, H'_2, L''_2 \} \to -1;
\{ E_1, S_1, F'_0, F'_1, G''_0, G''_1 \} \to 1;
\{ F_1, G_1, H_0, L_0, B'_1, D'_1,
\nonumber\\&&
E'_1, G'_1, H'_1, L'_1, S'_0, S'_1, B''_1, C''_1, E''_1, F''_1, H''_1, L''_1, S''_1, S''_2 \} \to 0;
\{ A' \} \to a_1 a_2;  
\nonumber\\&&
\{ C'_0 \} \to 1 - b_1;
\{ C'_1 \} \to 1 + a_1 + a_2;
\{ A'' \} \to a_3 a_4;
\{ D''_0 \} \to 1 - b_2;
\nonumber\\&&
\{ D''_1 \} \to 1 + a_3 + a_4 \\
f_{21a} &:&
\{ A, A' \} \to a_1 a_2;
\{ B_0 \} \to -c_1;   
\{ B_1, C_1, B'_1, C'_1 \} \to 1 + a_1 + a_2;
\{ D_1, D'_1 \} \to a_2;
\nonumber\\&&
\{ E_0, F'_0, S'_0, G''_0, S''_2 \} \to -1;
\{ E_1, F_1, L_1, S_1, E'_1, F'_1, L'_1, S'_1, G''_1, L''_1, S''_1 \} \to 1;
\nonumber\\&&
\{ G_1, H_0, L_0, G'_1, H'_2, E''_1, F''_1, H''_1, L''_2 \} \to 0;
\{ H_1, H'_1 \} \to 2;
\{ C'_0, D''_0 \} \to -c_2;
\nonumber\\&&
\{ A'' \} \to a_1 a_3;
\{ B''_1, C''_1 \} \to a_3;
\{ D''_1 \} \to 1 + a_1 + a_3 \\
f_{27b} &:&
\{ A \} \to a_1 a_2;
\{ B_0, C'_0 \} \to 1 - b; 
\{ B_1 \} \to 1 + a_1 + a_2;
\{ C_1 \} \to a_2;
\{ D_1, F_1, G_1, 
\nonumber\\&&
L_1, S_1, D'_1, E'_1, G'_1, L'_1, S'_1, L''_2, S''_2 \} \to 0;
\{ E_0, E_1, H_0, H_1, F'_0, F'_1, H'_1, H'_2, E''_1, F''_1 \} \to 1;
\nonumber\\&&
\{ L_0, S'_0 \} \to -2;
\{ A' \} \to a_1 a_3;
\{ B'_1 \} \to a_3;
\{ C'_1 \} \to 1 + a_1 + a_3;
\{ A'' \} \to b (1 + b);
\nonumber\\&&
\{ B''_1, C''_1 \} \to -2 b;
\{ D''_0 \} \to -c;
\{ D''_1 \} \to 2 (3 + 2 b);
\nonumber\\&&
\{ G''_0 \} \to -1;   
\{ G''_1 \} \to 4;
\{ H''_1 \} \to 2; 
\{ L''_1, S''_1 \} \to -4. 
\end{eqnarray}

\noindent
{\underline{\sf The case of four variables, \cite{EXTON72a,EXTON1}:}}\\
\begin{eqnarray}
K_1 &:&
\{ A, A', A'' \} \to a b_1;
\{ B_0, E'''_0 \} \to -c_1;
\{ B_1, C_1, D_1, B'_1, C'_1, D'_1, B''_1, C''_1,
\nonumber\\&&
D''_1 \} \to 1 + a + b_1;
\{ E_1, E'_1, E''_1 \} \to b_1;
\{ F_0, P_0, G'_0, H''_0, L'''_0, P'''_2 \} \to -1;
\nonumber\\&&
\{ F_1, G_1, H_1, P_1, R_1, S_1, F'_1, G'_1, H'_1, P'_1, R'_1, S'_1, F''_1, G''_1, H''_1, P''_1, R''_1, S''_1, L'''_1, P'''_1,
\nonumber\\&&
R'''_1, S'''_1 \} \to 1;
\{ L_1, M_0, N_0, L'_1, M'_2, Q'_0, R'_0, L''_1, N''_2, Q''_2, S''_0, F'''_1, G'''_1, H'''_1, M'''_1,
\nonumber\\&&
N'''_1, Q'''_1, R'''_2, S'''_2 \} \to 0;
\{ M_1, N_1, Q_1, M'_1, N'_1, Q'_1, M''_1, N''_1, Q''_1 \} \to 2;
\{ C'_0 \} \to -c_2;
\nonumber\\&&
\{ D''_0 \} \to -c_3;
\{ A''' \} \to a b_2;
\{ B'''_1, C'''_1, D'''_1 \} \to b_2;
\{ E'''_1 \} \to 1 + a + b_2 \\
{
K_{21} }&:&
\{ A \} \to a b_ 1;
\{ B_ 0, C'_ 0, D''_ 0, E'''_ 0 \} \to -c_ 1;
\{ B_ 1 \} \to 1 + a + b_ 1;
\{ C_ 1 \} \to b_ 1;
\nonumber\\&&
\{ D_ 1, E_ 1, G_ 1, H_ 1, L_ 1, N_ 1, P_ 1, Q_ 1, R_ 1, S_ 1, D'_ 1, 
E'_ 1, F'_ 1, H'_ 1, L'_ 1, N'_ 1, P'_ 1, Q'_ 1, R'_ 1, S'_ 1, B''_ 
1, C''_ 1, 
\nonumber\\&&
E''_ 1, F''_ 1, G''_ 1, L''_ 1, M''_ 1, N''_ 1, P''_ 1, 
Q''_ 1, R''_ 1, S''_ 1, B'''_ 1, C'''_ 1, D'''_ 1, F'''_ 1, G'''_ 1, 
H'''_ 1, M'''_ 1, N'''_ 1, P'''_ 1, Q'''_ 1, 
\nonumber\\&&
R'''_ 1, S'''_ 1 \} \to 0;
\{ F_ 0, M_ 0, N_ 0, P_ 0, G'_ 0, M'_ 2, Q'_ 0, R'_ 0, H''_ 0, N''_ 
2, Q''_ 2, S''_ 0, L'''_ 0, P'''_ 2, R'''_ 2, S'''_ 2 \} \to -1;
\nonumber\\&&
\{ F_ 1, M_ 1, G'_ 1, M'_ 1, H''_ 1, L'''_ 1 \} \to 1;
\{ A' \} \to a b_ 2;
\{ B'_ 1 \} \to b_ 2;
\{ C'_ 1 \} \to 1 + a + b_ 2;
\nonumber\\&&
\{ A'' \} \to b_ 3 b_ 5;
\{ D''_ 1 \} \to 1 + b_ 3 + b_ 5;
\{ A''' \} \to b_ 4 b_6;
\{ E'''_ 1 \} \to 1 + b_ 4 + b_6 \}.
\end{eqnarray}
The complete list of relations is given in computer readable form in the attachment {\tt Mconditions.m} to this paper.

\section{A brief descriptions of the commands of {\tt HypSeries}} 
\label{sec:C}

\vspace*{1mm} 
\noindent
In the following we describe the commands available in the {\tt Mathematica} package {\tt HypSeries}.
To execute this package requires also the packages {\tt Sigma}, {\tt EvaluateMultiSums} 
\cite{Ablinger:2010pb,Blumlein:2012hg,Schneider:2013zna,Schneider:19},
and {\tt 
HarmonicSums} as well as other packages, see Appendix~\ref{sec:F}.
The user has to provide $n$  (partial) differential equations in the $n$--variable case. The commands
\[ 
{\tt solveDE1},~~
{\tt solveDE2},~~ 
{\tt solveDE3},~~ 
{\tt solveDE4}
\]
check whether the corresponding set of one to four variables has product solutions by
consulting internal lists of cases. If applicable, the corresponding product solution for the expansion coefficients $f[m]$ to $f[m,n,p,q]$ are 
provided. In the two--variate case subclasses are dealt with individually. The command is e.g.
\[{\tt {solveDE4}[\{eq1==0,eq2==0,eq3==0,eq4==0\},\{x,y,z,t\},\{m,n,p,q\}]}.\]

More general solutions are possible by using the command {\tt DEProductSolution}. One has to provide
the required $n$ differential equations in the list 
\[ {\tt sys = \{eq1==0,...,eqn==0\}}.\] 
Then 
\[
{\tt DEProductSolution[sys,\{x,y,...\},\{m,n,...\}]}
\]
returns the respective expansion coefficient {\tt f[m,n,p,q]}. Here the tools described in Section~\ref{sec:solveProd} are utilized.

If one has, on the reverse, a Pochhammer ratio {\tt A = f[m,n,p,q]} the command
\[
{\tt {findDE}[A,\{x,y,...\},\{m,n,...\}]}
\]
returns the system of differential equations obeyed by
\[
f(x,y,...) = \sum_{m,n,...\ge =0}^\infty f[m,n,...] \, x^m y^n \cdots
\]
Given a differential equation {\tt equ} in $n$ variables the command {\tt findRE} 
\[
{\tt findRE[eq==0,\{x,y,\ldots\},\{m,n,\ldots\}]}
\]
returns a corresponding recurrence for {\tt f[m,n,...]}. The last two commands implement the techniques presented in the beginning of Section~\ref{sec:recsol}.

To prepare for the expansion in the dimensional parameter $\ep$, which frequently occurs in the
parameters of the differential and difference equations, one usually needs to check for the convergence 
domain of the corresponding solution, to be able to perform the respective limit $N \rightarrow \infty$
in the sums involved. 
Generally it is assumed that $\{x,y,z,t\} \in ]-1,1[$. However,
often stronger conditions are needed in the multivariate cases. Internal Tables, cf.~\cite{SRIKARL},
allow to check for this in the two-- and three--variate cases using the commands {\tt findCond2} and  
{\tt findCond3}. One first has to determine the corresponding function label {\tt fcn}
via {\tt classifier2}, {\tt classifier3}, as e.g.
\[
\tt classifier3[f[m,n,p], \{x,y,z\}, \{m,n,p\}]
\]
returning {\tt fcn}. Then 
\[
 {\tt findCond3[fcn,\{x,y,z\}}]           
\]
returns the convergence conditions, which are in some cases given in implicit form.
The $\ep$--expansion is performed using algorithms implemented in {\tt Sigma}. The attached 
notebooks  {\tt ExHypSeries.nb} and {\tt ExSolvePartialLDE.nb} give a more detailed explanation on this.

In the cases the $\ep$--expansion of the considered higher transcendental functions can be 
performed to a certain power $O(\ep^k)$ one may want to check, whether the solution satisfies the 
corresponding differential equations. This is provided by the command {\tt CheckDE[sol,eq]}, where
{\tt sol} denotes the solution up to the corresponding degree in $\ep$ and {\tt eq} the differential 
equation 
\[{\tt  CheckDE[sol,eq]}\] 
returns then a result, which is of higher order in $\ep$.

\section{A brief descriptions of the commands of {\tt solvePartialLDE}}
\label{sec:D}

\vspace*{1mm}
\noindent
The {\tt Mathematica} package {\tt SolvePLDE.m} implements the 
aforementioned 
algorithms for solving partial linear difference equations. It requires {\tt Sigma} and {\tt HarmonicSums} 
to be loaded. Additionally, the software {\tt Singular} \cite{DGPS} must be installed, and made available 
through the interface given in~\cite{SingularInterface}. The installation path of {\tt Singular} can be set 
using the command appropriate for the user's system, e.g. \\[2mm]
{\tt <\,<Singular.m\\
SingularCommand = "{\it (path to)}/Singular-4.1.3-x86\_64-Linux/bin/Singular"
}.
\\[2mm]

The functions available are
\begin{itemize}
	\item $\mathtt {spread[p, q, \{n, k, ...\} (, \{eps,...\} ) ]}$: this function calculates the spread of the polynomials $p$ and $q$, in the variables $n,k,\ldots$. The symbols in the optional list are treated as an extension to the field over which the polynomials are defined. If the polynomials $p$ and $q$ contain symbolic parameters other than $n,k,\ldots$, such as for instance the dimensional regulator $\varepsilon$, they must be declared in the second list.
	\item $\mathtt {dispersion[p, q, \{n, k, ...\} (, \{eps,...\} ) ]}$: this function calculates the dispersion (it is the maximum of the spread) of the polynomials $p$ and $q$ in the variables $n,k,\ldots$. The second optional list has the same function as in the function {\tt spread}.
	\item $\mathtt {SolvePLDE[eq==rhs,f[n,k,...] ,(options) ] }$. This command solves the linear partial difference equation. It has the following available options:
	\begin{itemize}
		\item {\tt UseObject $\rightarrow$} {\em list of Harmonic sums and/or Pochhammer symbols}
			\\ Allows to define a list of harmonic sums and Pochhammer symbols to be searched in the numerator of the solution.

		\item {\tt PLDEdegBound $\rightarrow$} {\em d}
			\\ Allows to choose the total degree $d$ of the ansatz for the numerator of the solution. Defaults to 0.
		\item {\tt InsertDenFactor $\rightarrow$} {\em factors}
			\\ In the case the periodic denominator bound was 
not 
complete, the user may force the search to include {\em factors} in the denominator.
		\item {\tt PLDESymbols $\rightarrow$} {\em list}
			\\ Any symbols appearing other than the shift variables must be declared in {\em list}.	
		\item {\tt InitialValues $\rightarrow$} {\em list}
			\\ A list of initial values in the form $\{\{var_1\rightarrow val_1, var_2\rightarrow val_2,\ldots, initial value\},\ldots\}$
		\item {\tt SymbolDegree $\rightarrow$} {\em d}
			\\ When initial conditions are provided, a linear combination of the homogeneous solutions is built, having as coefficients rational functions in the symbols. This option sets the maximum total degree of the numerator and denominator of those rational functions.
	\end{itemize}
	\item $\mathtt { SolveExpand[eq==rhs,f[n,k,...], PLDEExpandIn\rightarrow\{\varepsilon,\ell_{min},\ell_{max}\} , }$
	\\
	$ \mathtt{ InitialValues\rightarrow\{\ldots\} ,(options) ] }$ : this command solves the PLDE in a series expansion in a parameter. The options are the same as for {\tt SolvePLDE}.
	\item $\mathtt { expandHypergPref[eq==rhs, f[n,k,...], fac] }$. This command derives a new equation whose solution has the hypergeometric factor {\tt fac} removed, as described in Section \ref{sec:hypergeometric-prefactor}.
\end{itemize}

\section{A constant}
\label{sec:E}

\vspace*{1mm}
\noindent
We calculate the constant $C$ given in~\eqref{eq:C1}.
The two contributions to this infinite sum do both diverge, while
\begin{eqnarray}
C \approx 2.759413418790153909406713643175.
\end{eqnarray}
Unfortunately (\ref{eq:C1}) cannot simply be 
written in terms of a hypergeometric function of main argument 1, since this diverges, which is easily 
seen by applying Gau\ss{}' formula. However, one can define it as the following limit
\begin{eqnarray}
C &=& \lim_{\ep \rightarrow 0}
\left\{1 -\, _2F_1\left(\frac{1}{2}-\frac{i
   \sqrt{3}}{2},\frac{1}{2}+\frac{i
   \sqrt{3}}{2};1;1-\ep\right)-\frac{\cosh
   \left[\frac{\sqrt{3} \pi }{2}\right] \ln (\ep)}{\pi
   }\right\}.
\end{eqnarray}
Before expanding in $\ep$ one should map the main argument as, cf.~e.g.~\cite{SLATER1},
\begin{eqnarray}
1 - \ep \rightarrow \frac{1}{\ep} - 1.
\end{eqnarray}
Furthermore, the relation for the digamma function 
\begin{eqnarray}
\psi(1-z) = \psi(z) + \pi {\rm cot}(\pi z)
\end{eqnarray}
shall be used. One finally obtains~\eqref{eq:C3}.
In deriving (\ref{eq:C3}) one obtains first an 
expression both containing the real and the imaginary part of 
$\psi\left(\frac{1}{2}+\frac{i\sqrt{3}}{2}\right)$. Those obey the following integral representations
\begin{eqnarray}
\Re\left[\psi\left(\frac{1}{2}+\frac{i\sqrt{3}}{2}\right)\right] 
&=&
\int_0^1 dt \left[\frac{ t^{-1/2} \cos\left[\frac{\sqrt{3}}{2} \ln(t) \right]}{t-1} - \frac{1}{\ln(t)}\right]
\\
\Im\left[\psi\left(\frac{1}{2}+\frac{i\sqrt{3}}{2}\right)\right] 
&=&
\int_0^1 \frac{dt}{\sqrt{t}(t-1)} \sin\left[\frac{\sqrt{3}}{2} \ln(t) \right]
= \frac{\pi}{2} 
\tanh\left[\frac{\sqrt{3}}{2} \pi \right].
\end{eqnarray}
While the imaginary part of 
$\psi\left(\frac{1}{2}+\frac{i\sqrt{3}}{2}\right)$ evaluates to a basic function, a further 
simplification seems not to be possible.

The new entities emerging here are therefore
\begin{eqnarray}
e^{\pi \sqrt{3}/2},~~~~~~~
         \text{and}~~~~~~~
\psi\left(\frac{1}{2}+\frac{i\sqrt{3}}{2}\right).
\end{eqnarray}
In the expansion of multi--variate series of the Pochhammer-type one expects quite new classes of special 
numbers to emerge, which will become a potential research topic in the future.

\section{Software required}
\label{sec:F}

\vspace*{1mm}
\noindent
The ancillary files cover the {\tt Mathematica} notebooks 
{\tt ExHypSeries.nb}
and {\tt ExSolvePartialLDE.nb}~.
Their execution require the following software packages, which can be downloaded from 
the software site of the RISC institute:

\vspace*{1mm}
\noindent
{\tt EvaluateMultiSums.m}~~~ 
{\tt https://risc.jku.at/sw/evaluatemultisums/}
\\
{\tt Guess.m, \tt LinearSystemSolver.m}
\\{\tt 
https://www3.risc.jku.at/research/combinat/software/ergosum/RISC/Guess.html} 
\\
{\tt HarmonicSums.m}~~~{\tt https://risc.jku.at/sw/harmonicsums/} 
\\
{\tt Sigma.m}\\
{\tt 
https://www3.risc.jku.at/research/combinat/software/Sigma/index.php} 
\\
{\tt Singular.m}~~~{\tt http://www.singular.uni-kl.de} 

\vspace*{4mm}
\noindent
The ancillary files contain the following input files for the example notebooks:

\vspace*{1mm}
\noindent
{\tt cases.m}
\\
{\tt converg.m}
\\
{\tt HypSeries.m}
\\
{\tt Mconditions.m}
\\
{\tt SolvePartialLDE.m}
\\
{\tt ExHypSeries.nb}
\\
{\tt ExSolvePartialLDE.nb}~.

\vspace*{1mm}\noindent
It is recommended to download all precomputed tables for {\tt 
HarmonicSums}. The {\tt Mathematica} notebooks {\tt ExHypSeries.nb} and {\tt 
ExsolvePartialLDE.nb} contain the information from where the correct version
of the used packages can be downloaded.

While the notebook {\tt ExsolvePartialLDE.nb} needs a few minutes computation time only,
{\tt ExHypSeries.nb} needs 1.32 days.
{\footnotesize

\end{document}